\documentclass[11pt, a4paper]{article}
\pdfoutput=1
\usepackage[utf8]{inputenc}

\usepackage{jheppub,amsmath,amssymb,slashed,url,bm,textgreek,upgreek,hyperref}
\usepackage{jheppub}  
\usepackage{tikz,lipsum,lmodern}
\usepackage[most]{tcolorbox}
\usepackage{amssymb} 
\usepackage{amsmath}
\usepackage{mathtools}
\usepackage{amsfonts}    
\usepackage{dsfont}
\usepackage{pdfpages}
\usepackage{verbatim}
\usepackage{tensor}
\usepackage{mathrsfs}
\usepackage{textgreek} 
\usepackage[mathscr]{euscript}
\usepackage[normalem]{ulem}
\usepackage{tikz}
\usepackage{makecell}
\usepackage[verbose]{placeins}
\usepackage{subcaption}
\usetikzlibrary{3d, arrows.meta, decorations.pathreplacing, decorations.markings,calc,shapes.misc,decorations.pathmorphing,patterns.meta, math}
\usepackage{pgfplots}
\pgfplotsset{compat=1.18}

\newcommand{\beq}{\begin{equation}}
\newcommand{\eeq}{\end{equation}}
\newcommand{\no}{\nonumber\\} 
\newcommand{\bea}{\begin{eqnarray}}
\newcommand{\ea}{\end{eqnarray}}
\newcommand{\barr}{\begin{array}}
\newcommand{\earr}{\end{array}}

\def\be{\begin{equation}}
\def\ee{\end{equation}}
\def\ba#1\ea{\begin{align}#1\end{align}}
\def\bg#1\eg{\begin{gather}#1\end{gather}}
\def\bm#1\em{\begin{multline}#1\end{multline}}
\def\bmd#1\emd{\begin{multlined}#1\end{multlined}}

 \def\L{\ell}

\def\vp{\chi}

\def\wg{\wedge}

\def\no{\nonumber}

\def\({\left(}
\def\){\right)}
\def\[{\left[}
\def\]{\right]}
\def\<{\langle}
\def\>{\rangle}

\def\bea{\begin{eqnarray}}
\def\eea{\end{eqnarray}}

\newcommand{\tr}{\operatorname{tr}}

\newcommand{\zb}{{\bar z}}

\def\nn{\nonumber}

\begin{document}

\global\long\def\aad{(a\tilde{a}+a^{\dagger}\tilde{a}^{\dagger})}%

\global\long\def\ad{{\rm ad}}%

\global\long\def\bij{\langle ij\rangle}%

\global\long\def\df{\coloneqq}%

\global\long\def\bs{b_{\alpha}^{*}}%

\global\long\def\bra{\langle}%

\global\long\def\dd{{\rm d}}%

\global\long\def\dg{{\rm {\rm \dot{\gamma}}}}%

\global\long\def\ddt{\frac{{\rm d^{2}}}{{\rm d}t^{2}}}%

\global\long\def\ddg{\nabla_{\dot{\gamma}}}%

\global\long\def\del{\mathcal{\delta}}%

\global\long\def\Del{\Delta}%

\global\long\def\dtau{\frac{\dd^{2}}{\dd\tau^{2}}}%

\global\long\def\ul{U(\Lambda)}%

\global\long\def\udl{U^{\dagger}(\Lambda)}%

\global\long\def\dl{D(\Lambda)}%

\global\long\def\da{\dagger}%

\global\long\def\id{{\rm id}}%

\global\long\def\ml{\mathcal{L}}%

\global\long\def\mm{\mathcal{\mathcal{M}}}%

\global\long\def\mf{\mathcal{\mathcal{F}}}%

\global\long\def\ket{\rangle}%

\global\long\def\kpp{k^{\prime}}%

\global\long\def\lr{\leftrightarrow}%

\global\long\def\lf{\leftrightarrow}%

\global\long\def\ma{\mathcal{A}}%

\global\long\def\mb{\mathcal{B}}%

\global\long\def\md{\mathcal{D}}%

\global\long\def\mbr{\mathbb{R}}%

\global\long\def\mbz{\mathbb{Z}}%

\global\long\def\mh{\mathcal{\mathcal{H}}}%

\global\long\def\mi{\mathcal{\mathcal{I}}}%

\global\long\def\ms{\mathcal{\mathcal{\mathcal{S}}}}%

\global\long\def\mg{\mathcal{\mathcal{G}}}%

\global\long\def\mfa{\mathcal{\mathfrak{a}}}%

\global\long\def\mfb{\mathcal{\mathfrak{b}}}%

\global\long\def\mfb{\mathcal{\mathfrak{b}}}%

\global\long\def\mfg{\mathcal{\mathfrak{g}}}%

\global\long\def\mj{\mathcal{\mathcal{J}}}%

\global\long\def\mk{\mathcal{K}}%

\global\long\def\mmp{\mathcal{\mathcal{P}}}%

\global\long\def\mn{\mathcal{\mathcal{\mathcal{N}}}}%

\global\long\def\mq{\mathcal{\mathcal{Q}}}%

\global\long\def\mo{\mathcal{O}}%

\global\long\def\qq{\mathcal{\mathcal{\mathcal{\quad}}}}%

\global\long\def\ww{\wedge}%

\global\long\def\ka{\kappa}%

\global\long\def\nn{\nabla}%

\global\long\def\nb{\overline{\nabla}}%

\global\long\def\pathint{\langle x_{f},t_{f}|x_{i},t_{i}\rangle}%

\global\long\def\ppp{p^{\prime}}%

\global\long\def\qpp{q^{\prime}}%

\global\long\def\we{\wedge}%

\global\long\def\pp{\prime}%

\global\long\def\sq{\square}%

\global\long\def\vp{\varphi}%

\global\long\def\ti{\widetilde{}}%

\global\long\def\wg{\widetilde{g}}%

\global\long\def\te{\theta}%

\global\long\def\tr{{\rm Tr}}%

\global\long\def\ta{{\rm \widetilde{\alpha}}}%

\global\long\def\sh{{\rm {\rm sh}}}%

\global\long\def\ch{{\rm ch}}%

\global\long\def\Si{{\rm {\rm \Sigma}}}%

\global\long\def\sch{{\rm {\rm Sch}}}%

\global\long\def\vol{{\rm {\rm {\rm Vol}}}}%

\global\long\def\reg{{\rm {\rm reg}}}%

\global\long\def\zb{{\rm {\rm |0(\beta)\ket}}}%

\title{Quantum fate of the Choptuik naked singularity
\vspace{-0.8cm}}

\author[a]{Chih-Hung Wu}

\affiliation[a]{Department of Physics, University of Washington, Seattle, WA 98195, USA}

\emailAdd{chwu29@uw.edu}

\abstract{Classical critical collapse provides a dynamical route from smooth initial data to a naked singularity, representing a sharper violation of predictability than ordinary black hole singularities.  We argue that this distinction is erased by quantum backreaction.  Building on the semiclassical interior analysis, where quantum self-energy of the collapsing matter generates a universal growing mode and a finite mass gap, we study the exterior naked singularity region that determines global visibility in the Einstein-scalar system.  We analyze controlled exterior models in both $2+1$ and $3+1$ dimensions. In the former, smooth matching and physical boundary conditions analytically select a vacuum polarization state, whose backreaction cloaks the classically naked region by a quantum trapped branch. In the latter, numerical horizon tracing shows that near a quantum-shifted threshold the exterior develops finite-mass marginally trapped surfaces rather than a zero-mass naked endpoint. These results suggest a global quantum picture in which the Choptuik naked singularity shares the fate of an ordinary black hole singularity: quantum effects push the putative Cauchy horizon behind a quantum-generated horizon, thereby reducing the loss of predictability to the standard black hole evaporation problem.}

\maketitle

\newpage

\section{Introduction}

Singularities mark the breakdown of the classical description of spacetime.  This breakdown is not equally severe in all circumstances.  In an ordinary black hole spacetime, the singularity is hidden behind an event horizon, so that the loss of predictability is confined to the black hole interior.  By contrast, a naked singularity would be visible to distant observers and would represent a more direct failure of classical determinism.  The \emph{weak cosmic censorship conjecture} was proposed precisely to exclude such visible singularities in generic gravitational collapse~\cite{Penrose:1969pc, Hawking:1994ss}.  The question of whether classical evolution can produce naked singularities therefore probes one of the sharpest tensions in gravitational physics.

Critical gravitational collapse provides perhaps the cleanest dynamical route to this problem.  In Choptuik's original numerical study of the spherically symmetric Einstein-scalar system, one considers a one-parameter family of smooth asymptotically flat initial data, labeled by $p$, interpolating between dispersion and black hole formation \cite{Choptuik:1992jv}.  At a critical value $p=p^\ast$, the evolution approaches a universal discretely self-similar critical solution.  For marginally supercritical data, the black hole mass obeys the scaling law
\begin{equation}
    M_{\rm BH} \sim (p-p^\ast)^\gamma ,
\end{equation}
where the exponent $\gamma$ is universal within a given matter model. For a massless scalar field in $3+1$ dimensions, $\gamma\simeq 0.37$.  At the threshold itself, the limiting spacetime contains a naked singularity and a Cauchy horizon emanating from the critical endpoint.

This phenomenon is now understood as Type II critical collapse.  In this case, the black hole mass plays the role of an order parameter: it vanishes continuously at the threshold, in close analogy with a second-order phase transition.  This should be contrasted with Type I collapse, where the critical solution is stationary or periodic in time and the black hole mass exhibits a finite gap.  The self-similar critical solutions relevant for Type II collapse may exhibit either discrete self-similarity (DSS), as in Choptuik's scalar collapse, or continuous self-similarity (CSS), as in Christodoulou's analytic construction of scalar field naked singularities \cite{Christodoulou:1994hg}.  CSS solutions are especially useful because they often allow much greater analytic control.  We refer to recent reviews for a broad account of the classical theory and its various extensions \cite{Gundlach:2025yje}.

The \emph{Choptuik naked singularity}, by which we broadly mean the zero-mass naked singularity arising at a Type II black hole threshold, is therefore special in two related ways.  First, it arises dynamically from smooth initial data after fine-tuning to the black hole threshold, rather than being imposed by hand.  Second, it is associated with a universal critical solution and is therefore not an arbitrary pathology of a special exact metric.  In this sense, critical collapse appears to give a counterexample to the weak cosmic censorship.  At the same time, the singularity forms in a regime of ever-increasing curvature and ever-decreasing length scale, so quantum effects from the collapsing matter are generally expected to intervene before a full theory of quantum gravity is required.  The central question is then not whether the classical singularity is ultimately resolved, but whether semiclassical effects already change the threshold dynamics.

\begin{figure}[t]
\centering

\begin{subfigure}[t]{0.48\textwidth}
\centering
\includegraphics[width=\textwidth]{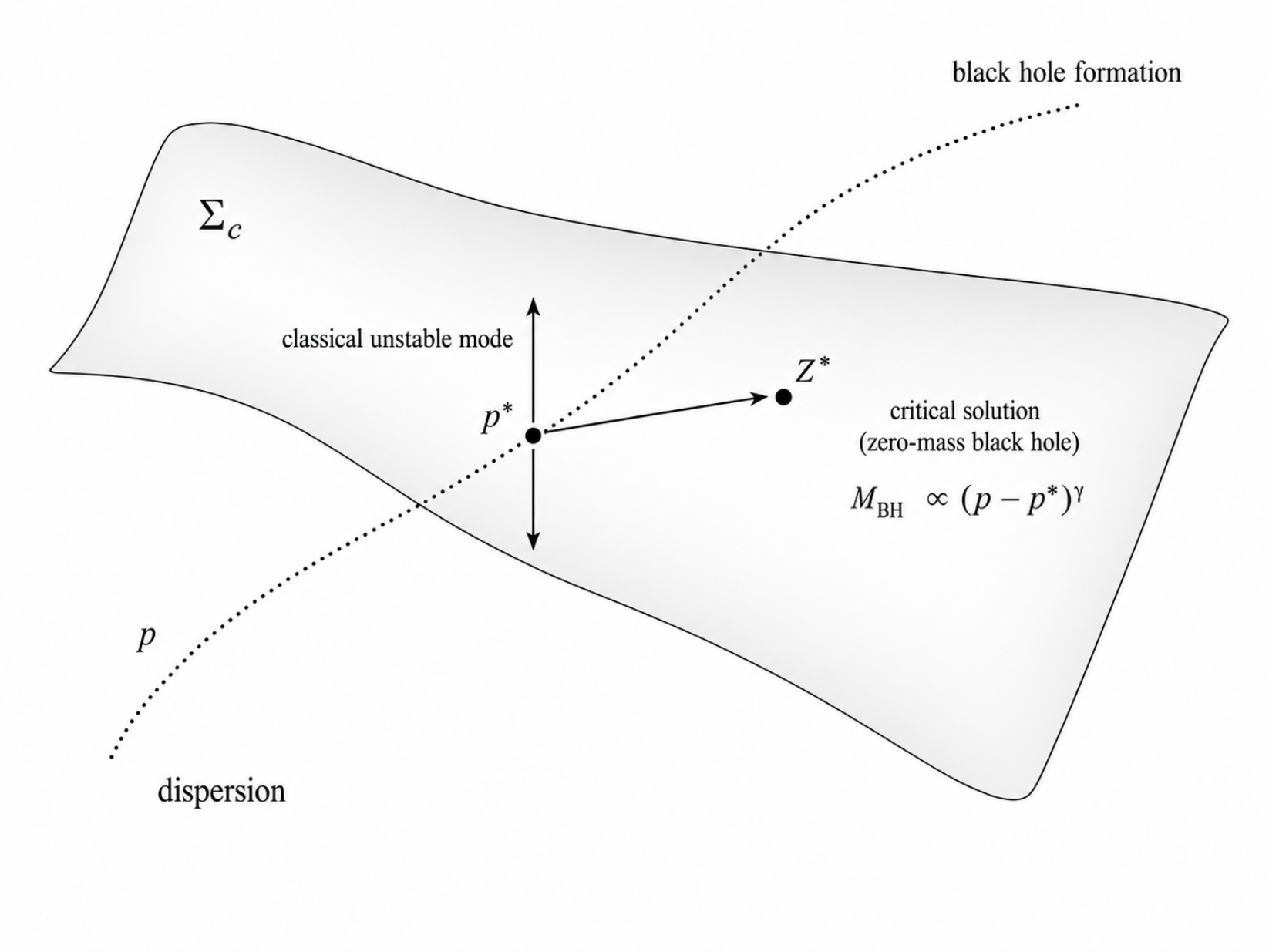}
\caption{a. Classical critical collapse.}
\label{fig:phase_classical}
\end{subfigure}
\hfill
\begin{subfigure}[t]{0.48\textwidth}
\centering
\includegraphics[width=\textwidth]{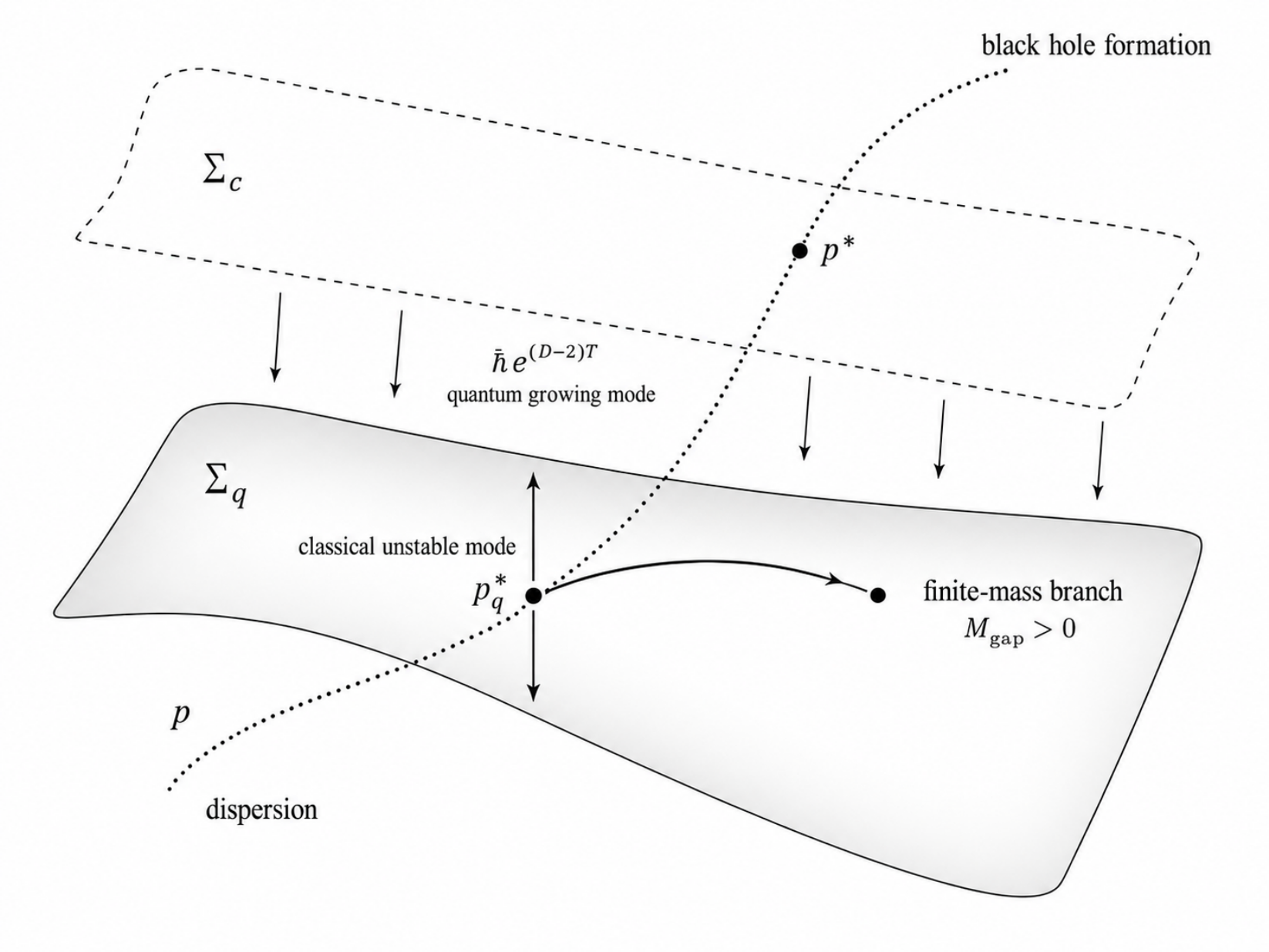}
\caption{b. Quantum critical collapse.}
\label{fig:phase_quantum}
\end{subfigure}

\caption{
Schematic phase space picture of classical and quantum critical collapse.
In the classical Type II scenario, the codimension-one critical surface
$\Sigma_c$ separates initial data that disperse from those that form
black holes. A one-parameter family of initial data $p$ intersects this
surface at the critical value $p^\ast$, whose evolution approaches
the critical solution $Z^\ast$. The classical unstable mode moves the
solution away from $\Sigma_c$, leading to either dispersion or black hole
formation, with the near-critical mass scaling
$M_{\rm BH}\propto (p-p^\ast)^\gamma$ in the supercritical regime.
In the quantum theory, the quantum growing mode
$\langle T_{\mu \nu} \rangle \propto \hbar e^{(D-2)T}$ shifts the effective threshold from $\Sigma_c$ to
$\Sigma_q$, so that the critical value becomes $p_q^\ast$.
The classical zero-mass critical endpoint is replaced by a finite-mass
branch with $M_{\rm gap}>0$.
}
\label{fig:phase_space_classical_quantum}
\end{figure}

\paragraph{Quantum effects near criticality.} This expectation has motivated many attempts to incorporate quantum effects into critical collapse, but no clear consensus has emerged.  Depending on the model, approximation scheme, and choice of quantum state, previous studies have found results ranging from the persistence of Type II scaling, the emergence of a mass gap, or even the breakdown of critical behavior altogether \cite{Strominger:1993tt, Zhou:1995zj, Bose:1996pi, Peleg:1996ce, Ayal:1997ab, Brady:1998fh, Chiba, Husain:2008tc, Ziprick:2009nd, Benitez:2020szx, Benitez:2021zjs, Berczi:2020nqy, Guenther:2020kro, Berczi:2021hdh, Hoelbling:2021axl, Varnhorst:2023dew, Moitra:2022umq, Zahn:2025tnh}.  A more detailed account of these issues can be found in \cite{Tomasevic:2025clf}.  The difficulty is not merely computational.  Critical spacetimes are time-dependent, self-similar backgrounds without a preferred static vacuum, and the physically relevant quantum source should come from self-energy of the collapsing matter rather than from an externally imposed Hawking-like flux.  A self-consistent treatment must therefore determine the quantum state and the renormalized stress tensor $\langle T_{\mu \nu} \rangle$ from regularity and boundary conditions intrinsic to the collapse problem itself.  This was the starting point of our previous analyses.

In previous work, we addressed this question in the self-similar region of analytically tractable critical spacetimes in the Einstein-scalar system \cite{Tomasevic:2025kqy, Tomasevic:2025clf}.  Since non-spherical perturbations of the scalar critical solution decay, the dominant dynamics near the threshold is captured by the $s$-wave sector.  Using a one-loop effective theory adapted to this sector, we found that regularity in the physical self-similar domain selects an asymptotically Minkowskian, Boulware-like state describing vacuum polarization of the collapsing scalar field.  The corresponding renormalized stress tensor contains a universal \emph{quantum growing mode},
\be
\langle T_{\mu\nu}^{(D)}\rangle
\sim
\hbar e^{(D-2)T}F_{\mu\nu}(x^i),
\ee
where $T$ is the logarithmic self-similar time.  This scaling defines a quantum Lyapunov exponent $\omega_q=D-2$.  When this quantum mode is included together with the classical unstable mode, the critical threshold is shifted and a finite mass gap appears. See Figure~\ref{fig:phase_space_classical_quantum} for an illustration of the resulting phase space structure.  The quantum growing exponent is kinematical in origin and therefore not tied to the detailed coefficients of a particular model: the quantum mode reflects how a one-loop stress tensor scales in a self-similar background.  Equivalently, the classical Type II behavior is replaced, within the admissible semiclassical regime, by a quantum-modified threshold resembling a Type I scenario, where a trapped surface forms even under arbitrary fine-tuning.  This provides a concrete mechanism by which vacuum polarization can enforce horizon formation before the naked critical endpoint is reached.

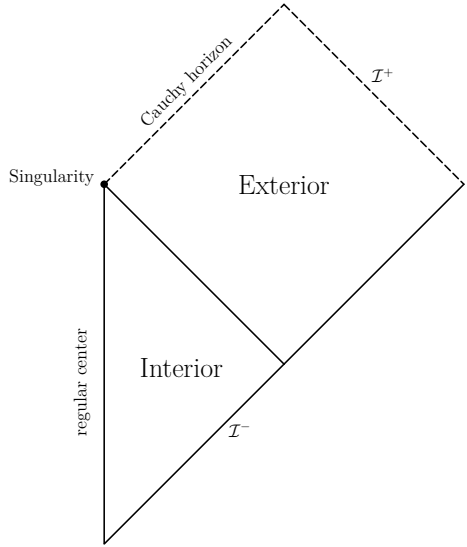
\begin{figure}[t]
\centering
\resizebox{0.40\textwidth}{!}{%
\begin{tikzpicture}[
  line cap=round,
  line join=round,
  every node/.style={inner sep=1.5pt},
boundary/.style={draw=black, line width=1.25pt},
horizon/.style={draw=black, line width=1.25pt, dash pattern=on 6pt off 3pt}
]

\def\L{5.0}

\coordinate (A) at (0,0);        
\coordinate (B) at (0,-10);      
\coordinate (D) at (\L,-\L);     
\coordinate (T) at (\L,\L);      
\coordinate (R) at (2*\L,0);     


\draw[horizon] (A) -- (T);

\draw[horizon] (T) -- (R);

\draw[boundary] (R) -- (D);

\draw[boundary] (A) -- (B) -- (D);

\draw[boundary] (A) -- (D);

\fill (A) circle[radius=3.0pt];


\node[font=\fontsize{22}{26}\selectfont]
  at ($(A)!0.5!(R)$)
  {Exterior};

\node[font=\fontsize{22}{26}\selectfont]
  at (2.15,-5.10)
  {Interior};

\node[font=\Large, anchor=west]
  at ($(A)-(2.70,-0.20)$)
  {Singularity};

\node[font=\Large, rotate=90]
  at ($(A)!0.55!(B)+(-0.62,0)$)
  {regular center};

\node[font=\Large, rotate=45, anchor=south]
  at ($(A)!0.48!(T)+(0.10,0.25)$)
  {Cauchy horizon};

\node[font=\Large, sloped, above=-15pt]
  at ($(B)!0.72!(D)+(0.18,-0.16)$)
  {$\mathcal I^-$};

\node[font=\Large, sloped, above=15pt]
  at ($(T)!0.55!(R)$)
  {$\mathcal I^+$};

\end{tikzpicture}%
}
\caption{
Penrose diagram illustrating the decomposition of a naked singularity spacetime
into an interior region and an exterior region. The dashed future light cone of
the singularity is the Cauchy horizon, while the dashed future null infinity
$\mathcal I^+$ indicates incompleteness.
}
\label{fig:ExteriorNakedness}
\end{figure}

\paragraph{The exterior nakedness problem.} However, this is not yet the full global story. The past light cone of the critical endpoint separates the spacetime into an interior region and an exterior naked singularity region. The former explains how the singularity arises from regular collapse, while the latter determines whether the singularity is visible to distant observers. For a naked singularity, the failure of predictability is not only the local divergence of curvature, but also the incompleteness of future null infinity caused by null rays emanating from the singular endpoint.\footnote{This separation is implicit in the classical constructions of naked singularities and has been made particularly explicit in recent mathematical treatments~\cite{Rodnianski:2019ylb,Shlapentokh-Rothman:2022byc, Cicortas:2024hpk, Cicortas:2026xdw}. Recent work on obstructions to the global visibility of singularities also emphasizes the same point from a classical causality perspective. That is, local nakedness does not by itself imply global visibility, since the exterior null generators may be obstructed before reaching future null infinity \cite{Jay:2026huw}.} We refer to Figure~\ref{fig:ExteriorNakedness} for an explicit Penrose diagram. Our semiclassical analysis was deliberately focused on the interior self-similar region. 

The exterior problem is therefore the missing half of the problem, and completing its semiclassical picture is not merely a technical refinement but a conceptual completion. In the interior, vacuum polarization enforces horizon formation and removes the local naked endpoint. The problem addressed in this paper is whether this quantum trapped branch persists into the exterior region, and whether the classically visible Cauchy horizon is hidden behind a horizon once one-loop backreaction is included. Moreover, while the exterior region inherits self-similar data near the past light cone, exact self-similarity is gradually lost as one evolves toward the far exterior. The one-loop analysis is therefore more challenging than in the interior.

If the answer is affirmative, then the fate of the Choptuik singularity is directly linked to that of the endpoint of Hawking evaporation~\cite{Hawking:1975vcx}: quantum effects place the Cauchy horizon inside the resulting black hole region, so that from the outside the geometry resembles an evaporating black hole. The Choptuik singularity then shares the fate of an ordinary black hole singularity, and the information loss problem~\cite{Hawking:1976ra} is therefore not enlarged by critical collapse. It is the familiar problem of an evaporating black hole, now reached dynamically from the threshold of collapse.

\paragraph{Exterior models and main results.} We work in the Einstein-scalar system and focus on two CSS exterior models: the $2+1$-dimensional Garfinkle spacetime~\cite{Garfinkle:2000br} and the $3+1$-dimensional Roberts spacetime~\cite{Roberts:1989sk, Brady:1994xfa, Oshiro:1994hd, Brady:1994aq}. Their roles are different. In $2+1$ dimensions, the scalar critical solution observed numerically is CSS~\cite{Pretorius:2000yu, Husain:2000vm}, and the Garfinkle family gives an analytic description of this critical behavior in the self-similar regime~\cite{Garfinkle:2002vn, Cavaglia:2004mt, Jalmuzna:2015hoa}.  In $3+1$ dimensions, the true Choptuik solution is DSS, while the Roberts solution should instead be viewed as an analytically tractable CSS proxy whose perturbation structure is closely tied to the breaking of CSS toward DSS~\cite{Frolov:1997uu, Frolov:1998tq, Frolov:1999fv}. In our previous interior analysis, these models allowed us to isolate the one-loop quantum growing mode and the associated horizon formation mechanism.  The present work instead tests whether the same mechanism survives in the exterior region controlling global visibility.

The three-dimensional example admits an analytic treatment.  The Garfinkle family provides an exact CSS background with a well-understood interior structure.  In numerical simulations of $2+1$-dimensional scalar collapse with negative cosmological constant, the critical solution is well approximated by a member of the Garfinkle family in the near-singularity region.  The cosmological constant is essential for defining the asymptotic problem, but it becomes subleading in the small-scale self-similar regime where critical behavior is locally valid~\cite{Cavaglia:2004mt, Jalmuzna:2015hoa}.  However, once one asks for an exterior completion, the cosmological constant again becomes important. It supplies the structure needed to embed the self-similar critical region into a larger spacetime and affects how the continued exterior should be interpreted.  In this case, the classical exterior can be obtained by a null continuation beyond the past light cone, following the construction of \cite{Jalmuzna:2015hoa}.  Starting from this null-continued exterior, we impose semiclassical matching conditions across the light cone.  Smooth matching, absence of Hawking-like asymptotic flux, and regularity of the quantum-corrected geometry select the exterior continuation of the same vacuum polarization state sourced by the collapsing scalar field.  Solving the one-loop equations then gives an explicit quantum-corrected exterior spacetime.  Horizon tracing in this geometry shows that the classically naked Cauchy horizon region is cloaked by a quantum trapped branch.  In this analytic model, the mechanism by which vacuum polarization removes exterior nakedness can therefore be seen explicitly.

The four-dimensional problem is more subtle. The Roberts solution is a simple exact CSS solution of the Einstein-scalar system and captures several features of near-critical collapse. In particular, Roberts exhibits complex growing modes that already encode the oscillatory structure expected when a CSS background is driven toward a DSS attractor \cite{Frolov:1997uu,Frolov:1998tq}, and previous analyses show that generic growing perturbations evolve away from Roberts toward the DSS critical solution \cite{Frolov:1999fv}.  In our previous work, we further argued that the one-loop quantum growing mode is kinematical in origin and should survive the replacement of the CSS proxy by the true DSS critical solution~\cite{Tomasevic:2025kqy, Tomasevic:2025clf}.  The role of the Roberts example is to understand whether the semiclassical shielding mechanism holds in an analytic setting that captures the CSS-breaking route toward the true Choptuik problem.

In the Roberts interior, the one-loop calculation again gives a clean analytic quantum growing mode. However, a globally satisfactory exterior continuation with all desired physical properties is not available in closed form.  We therefore formulate the problem as an exterior finite-strip construction in double-null coordinates. The strip is seeded by Roberts-like data near the past light cone and matched to weak-field exterior behavior without introducing a null shell or artificial incoming flux.  We then propagate the classical and quantum perturbation sectors into the exterior and trace marginally trapped surfaces. The numerical analysis reveals a threshold ribbon in the space of classical and quantum amplitudes. Along its centerline, the exterior develops a finite-mass marginally trapped surface near the quantum-shifted threshold, rather than approaching a zero-mass naked endpoint. The numerical results indicate that the quantum horizon structure found in the interior is not lost in the exterior region.

The main conclusion of the paper is therefore the first step toward a global quantum picture of critical collapse.  In representative models of classically naked critical geometries, quantum vacuum polarization can generate or maintain a quasi-local trapped branch in the exterior region.  The classically visible Cauchy horizon and naked singularity structure is replaced by a quantum-shifted, finite-mass horizon.  Within the one-loop, near-critical, and linear regime, the Choptuik naked singularity is therefore driven toward the same category as an ordinary black hole singularity. It is hidden behind a horizon and ultimately tied to the physics of evaporation.

Let us also emphasize the scope of the result.  We do not prove a full nonlinear event horizon theorem for the exact Choptuik spacetime.  Our analysis is perturbative, one-loop, and restricted to the dominant $s$-wave sector.  The horizons we identify are quasi-local trapped surfaces, not teleological event horizons.  Nevertheless, the restricted setting is also what makes the calculation sharp: the quantum source is fixed by self-energy of the collapsing field, the exterior matching conditions are explicit, and the horizon criterion is geometric.  The agreement between the analytic $2+1$ construction and the numerical $3+1$ exterior analysis suggests that the mechanism is not an artifact of the interior calculation.

The paper is organized as follows.  In Section~\ref{sec:critical collapse and nakedness}, we briefly review critical collapse and formulate the exterior nakedness problem.  We set up the Einstein-scalar system, summarize the one-loop framework, and state the light cone matching and horizon tracing criteria used throughout the paper.  In Section~\ref{sec:2+1d_Garfinkle}, we analyze the $2+1$-dimensional exterior problem.  We review the Garfinkle interior and its quantum horizon, construct the exterior null continuation, compute the quantum backreaction beyond the light cone, and trace the resulting quantum trapped branch.  In Section~\ref{sec:3+1d_Roberts}, we turn to the $3+1$-dimensional problem.  We review the Roberts interior result, construct a shell-free exterior strip, propagate the classical and quantum perturbation sectors, and present the numerical evidence for a threshold ribbon and finite-mass exterior horizons.  We conclude in Section~\ref{sec:Discussion} with the implications for cosmic censorship, the relation to black hole evaporation, and directions toward a full global treatment of quantum critical collapse.

\section{Critical collapse and the exterior nakedness problem}
\label{sec:critical collapse and nakedness}

This section collects the classical and semiclassical ingredients needed for the exterior analysis. We first review the aspects of critical collapse that are relevant for the nakedness of the singularity. The exterior problem is then formulated as a matching problem across the past light cone of the singularity. We also set up the Einstein-scalar system and the one-loop framework, before stating the matching and horizon tracing criteria used in the analytic $2+1$ and numerical $3+1$ models.

\subsection{From local horizons to global nakedness}
\label{sec2:visibility}

Critical gravitational collapse is most naturally described as a dynamical systems problem in the space of initial data. Consider a one-parameter family of smooth asymptotically flat initial data labeled by $p$, crossing the boundary between the dispersion and black hole formation. There exists a critical value $p^\ast$ separating the two basins of attraction. Near this threshold, the evolution approaches a universal critical solution before eventually departing either toward dispersion or toward black hole formation. Figure~\ref{fig:phase_classical} is a generic picture for the classical story.

The canonical example is the spherically symmetric massless scalar field in $3+1$ dimensions, where the critical solution is the Choptuik spacetime \cite{Choptuik:1992jv}. More generally, the black hole threshold is governed by a codimension-one critical surface in phase space. The critical solution has precisely one physical growing perturbation transverse to this surface, while the amplitudes of all other perturbations decay as the solution approaches the critical spacetime. In the scalar field case, this unique growing mode is spherically symmetric. Nonspherical perturbations decay, so the spherical sector captures the dominant near-threshold instability~\cite{Gundlach:1995kd, Gundlach:1996eg, Martin-Garcia:1998zqj}. We refer to the review \cite{Gundlach:2025yje} and to our previous discussions in \cite{Tomasevic:2025kqy,Tomasevic:2025clf} for more details.

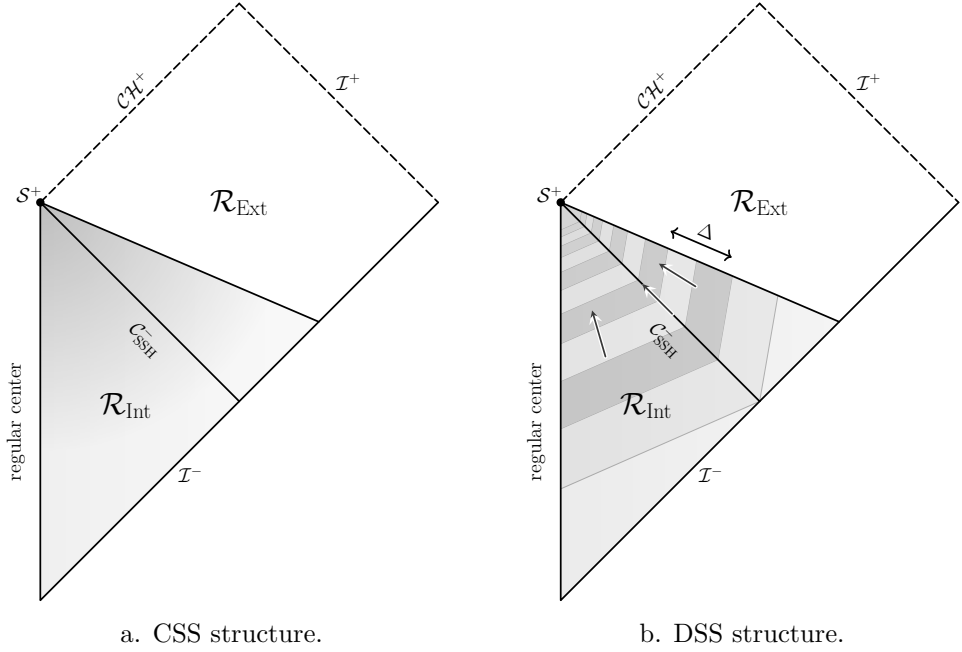
\begin{figure}[t]
\centering

\begin{subfigure}[t]{0.40\textwidth}
\centering
\resizebox{0.95\textwidth}{!}{%
\begin{tikzpicture}[
  line cap=round,
  line join=round,
  every node/.style={inner sep=1.5pt},
  boundary/.style={draw=black, line width=1.25pt},
  horizon/.style={draw=black, line width=1.25pt, dash pattern=on 6pt off 3pt}
]

\def\L{5.0}

\coordinate (A) at (0,0);        
\coordinate (B) at (0,-10);      
\coordinate (D) at (\L,-\L);     
\coordinate (T) at (\L,\L);      
\coordinate (R) at (2*\L,0);     
\coordinate (E) at ($(D)!0.4!(R)$);

\fill[white] (A) -- (T) -- (R) -- (D) -- cycle;

\begin{scope}
  \clip (A) -- (E) -- (D) -- (B) -- cycle;

  \shade[
    shading=radial,
    inner color=black!88,
    outer color=white
  ]
    ($(A)+(0.05,-0.03)$) circle[radius=6.2];

  \shade[
    shading=axis,
    shading angle=-32,
    left color=black!28,
    right color=white,
    opacity=0.32
  ]
    (-0.6,0.5) rectangle (10.5,-10.5);
\end{scope}

\begin{scope}
  \clip (A) -- (E) -- (D) -- (B) -- cycle;

  \shade[
    shading=axis,
    shading angle=-31,
    left color=black!5,
    right color=white,
    opacity=0.68
  ]
    (-0.65,0.65) rectangle (10.6,-10.6);
\end{scope}

\draw[horizon] (A) -- (T);       
\draw[horizon] (T) -- (R);       
\draw[boundary] (R) -- (D);
\draw[boundary] (A) -- (B) -- (D);
\draw[boundary] (A) -- (E);      
\draw[boundary] (A) -- (D);      

\fill (A) circle[radius=3.0pt];

\node[font=\fontsize{22}{26}\selectfont]
  at ($(A)!0.5!(R)$)
  {$\mathcal{R}_{\rm Ext}$};

\node[font=\fontsize{22}{26}\selectfont]
  at (2.15,-5.10)
  {$\mathcal{R}_{\rm Int}$};

\node[font=\Large, anchor=east]
  at ($(A)+(0.10,0.20)$)
  {$\mathcal{S}^{+}$};

\node[font=\Large, rotate=90]
  at ($(A)!0.55!(B)+(-0.62,0)$)
  {regular center};

\node[font=\Large, rotate=-45, anchor=north]
  at ($(A)!0.54!(D)+(0.18,-0.50)$)
  {$\mathcal{C}^{-}_{\rm SSH}$};

\node[font=\Large, rotate=45, anchor=south]
  at ($(A)!0.48!(T)+(0.10,0.25)$)
  {$\mathcal{CH}^{+}$};

\node[font=\Large, sloped, above=-15pt]
  at ($(B)!0.72!(D)+(0.18,-0.16)$)
  {$\mathcal I^-$};

\node[font=\Large, sloped, above=15pt]
  at ($(T)!0.55!(R)$)
  {$\mathcal I^+$};

\end{tikzpicture}%
}
\caption{a. CSS structure.}
\label{fig:IntExtCSS_panel}
\end{subfigure}
\hspace{1.5em}
\begin{subfigure}[t]{0.40\textwidth}
\centering
\resizebox{0.95\textwidth}{!}{%
\begin{tikzpicture}[
line cap=round,
line join=round,
every node/.style={inner sep=1.5pt},
boundary/.style={draw=black, line width=1.25pt},
horizon/.style={
  draw=black,
  line width=1.25pt,
  dash pattern=on 6pt off 3pt
},
crystal/.style={
  draw=black!78,
  line width=1.25pt,
  ->,
  >=stealth,
  preaction={draw=white, line width=3.1pt}
}
]

\def\L{5.0}

\coordinate (A) at (0,0);        
\coordinate (B) at (0,-10);      
\coordinate (D) at (\L,-\L);     
\coordinate (T) at (\L,\L);      
\coordinate (R) at (2*\L,0);     
\coordinate (E) at ($(D)!0.4!(R)$);

\fill[white] (A) -- (T) -- (R) -- (D) -- cycle;

\begin{scope}
\clip (A) -- (E) -- (D) -- (B) -- cycle;

\fill[gray!10] (-1,1) rectangle (11,-11);

\coordinate (Cbase) at ($(A)!0.72!(B)$);
\coordinate (Hbase) at (D);
\coordinate (Obase) at ($(A)!0.78!(E)$);

\foreach \so/\si/\c in {
1.000/0.790/gray!20,
0.790/0.624/gray!47,
0.624/0.493/gray!18,
0.493/0.389/gray!44,
0.389/0.307/gray!20,
0.307/0.243/gray!41,
0.243/0.192/gray!22,
0.192/0.152/gray!38,
0.152/0.120/gray!24,
0.120/0.095/gray!35
}{
  \path[fill=\c]
    ($(A)!\so!(Cbase)$) --
    ($(A)!\so!(Hbase)$) --
    ($(A)!\so!(Obase)$) --
    ($(A)!\si!(Obase)$) --
    ($(A)!\si!(Hbase)$) --
    ($(A)!\si!(Cbase)$) -- cycle;

  \draw[black!38, line width=0.42pt]
    ($(A)!\si!(Cbase)$) --
    ($(A)!\si!(Hbase)$) --
    ($(A)!\si!(Obase)$);
}

\draw[black!38, line width=0.42pt]
  (Cbase) -- (Hbase) -- (Obase);

\path[fill=gray!32]
  (A) --
  ($(A)!0.095!(Cbase)$) --
  ($(A)!0.095!(Hbase)$) --
  ($(A)!0.095!(Obase)$) -- cycle;

\shade[
  shading=axis,
  shading angle=0,
  left color=black!18,
  right color=white,
  opacity=0.20
]
  (-0.10,-10.2) rectangle (7.2,0.2);

\end{scope}

\draw[horizon] (A) -- (T);       
\draw[horizon] (T) -- (R);       
\draw[boundary] (R) -- (D);
\draw[boundary] (A) -- (B) -- (D);
\draw[boundary] (A) -- (E);      
\draw[boundary] (A) -- (D);      

\fill (A) circle[radius=3.0pt];


\coordinate (Pint) at ($(B)!0.45!(D)$);
\draw[crystal]
  ($(A)!0.50!(Pint)$) --
  ($(A)!0.35!(Pint)$);

\draw[crystal]
  ($(A)!0.56!(D)$) --
  ($(A)!0.41!(D)$);

\coordinate (Pext) at ($(D)!0.58!(E)$);
\draw[crystal]
  ($(A)!0.55!(Pext)$) --
  ($(A)!0.40!(Pext)$);

\coordinate (DelL) at ($(A)!0.38!(E)+(0.10,0.42)$);
\coordinate (DelR) at ($(A)!0.60!(E)+(0.10,0.42)$);

\draw[<->, line width=1.25pt] (DelL) -- (DelR)
  node[midway, sloped, above=3pt, font=\Large] {$\Delta$};



\node[font=\fontsize{22}{26}\selectfont]
  at ($(A)!0.5!(R)$)
  {$\mathcal{R}_{\rm Ext}$};

\node[font=\fontsize{22}{26}\selectfont]
  at (2.15,-5.10)
  {$\mathcal{R}_{\rm Int}$};

\node[font=\Large, anchor=east]
  at ($(A)+(0.10,0.20)$)
  {$\mathcal{S}^{+}$};

\node[font=\Large, rotate=90]
  at ($(A)!0.55!(B)+(-0.62,0)$)
  {regular center};

\node[font=\Large, rotate=-45, anchor=north]
  at ($(A)!0.54!(D)+(0.18,-0.50)$)
  {$\mathcal{C}^{-}_{\rm SSH}$};

\node[font=\Large, rotate=45, anchor=south]
  at ($(A)!0.48!(T)+(0.10,0.25)$)
  {$\mathcal{CH}^{+}$};

\node[font=\Large, sloped, above=-15pt]
  at ($(B)!0.72!(D)+(0.18,-0.16)$)
  {$\mathcal I^-$};

\node[font=\Large, sloped, above=15pt]
  at ($(T)!0.55!(R)$)
  {$\mathcal I^+$};

\end{tikzpicture}%
}
\caption{b. DSS structure.}
\label{fig:IntExtDSS_panel}
\end{subfigure}

\caption{
Comparison of the global causal structure and self-similar swept region in the
CSS and DSS cases. In both panels, the critical spacetime
is decomposed into an interior region $\mathcal{R}_{\rm Int}$ and an
exterior region $\mathcal{R}_{\rm Ext}$. The null line
$\mathcal{C}^{-}_{\rm SSH}$ denotes the past light cone/self-similarity horizon,
while the dashed null line $\mathcal{CH}^{+}$ denotes the Cauchy horizon and the
dashed $\mathcal I^+$ indicates incompleteness. For CSS spacetimes, the self-similar
region is continuous and is represented by a smooth gradient
accumulating toward the future curvature singularity $\mathcal{S}^+$. In the DSS panel, the alternating shaded regions schematically
represent successive fundamental domains related by the same
discrete rescaling of period $\Delta$. Their detailed boundaries
are coordinate dependent and should not be interpreted as physical
or null hypersurfaces. The arrows indicate that the crystal translation direction is timelike 
in $\mathcal{R}_{\rm Int}$, null on
$\mathcal{C}^{-}_{\rm SSH}$, and spacelike in the self-similar
part of $\mathcal{R}_{\rm Ext}$.
}
\label{fig:IntExt_CSS_DSS}
\end{figure}

The critical solutions relevant for Type II collapse are self-similar. In a CSS spacetime, there is a homothetic vector field $\xi$ such that, with our convention
\begin{equation}
    \mathcal{L}_{\xi} g_{\mu\nu}=2 g_{\mu\nu}, \qquad \xi=-\frac{\partial}{\partial T}.
\end{equation}
Equivalently, in coordinates $(T,x^i)$ adapted to the self-similarity, one may write
\begin{equation}
    g_{\mu\nu}(T,x^i) \propto e^{-2T}\tilde g_{\mu\nu}(x^i),
\end{equation}
where increasing $T$ corresponds to probing smaller spacetime scales. In a DSS spacetime, the rescaled metric is periodic rather than independent of $T$,
\begin{equation}
    g_{\mu\nu}(T,x^i)\propto e^{-2T}\tilde g_{\mu\nu}(T,x^i),
    \qquad
    \tilde g_{\mu\nu}(T+\Delta,x^i)=\tilde g_{\mu\nu}(T,x^i),
\end{equation}
with echoing period $\Delta$. The Choptuik solution is DSS, whereas many analytically tractable models, including the Christodoulou, Garfinkle, and Roberts-type solutions used as admissible proxies, are CSS. The distinction between CSS and DSS is important technically, but the geometric lesson is common: the self-similar region focuses the collapse toward a singular endpoint at $T\to\infty$. See Figure~\ref{fig:IntExt_CSS_DSS} for Penrose diagrams of both CSS and DSS critical spacetimes.

In Type II critical collapse, the black hole mass plays the role of an order parameter. For marginally supercritical data,
\begin{equation}
    M_{\rm BH}\sim (p-p^\ast)^\gamma,
    \label{eq:classical_choptuik_scaling}
\end{equation}
where $\gamma$ is universal within a given matter model. This contrasts with Type I collapse, where the black hole mass exhibits a finite gap at the threshold. In our previous work, we showed that one-loop backreaction converts the classical Type II threshold into a quantum-modified threshold with a finite mass gap, making the semiclassical transition resemble a Type I scenario~\cite{Tomasevic:2025kqy,Tomasevic:2025clf}. This picture is illustrated in Figure~\ref{fig:phase_quantum}.

The exact critical spacetime itself contains a curvature singularity at the accumulation point, which we denote by $\mathcal{S}^+$. In self-similar coordinates this is the endpoint $T\to\infty$ reached at finite proper time from regular points in its past. The past light cone of $\mathcal{S}^+$, denoted schematically by $\mathcal{C}_{\rm SSH}^-$, plays a special role. It is often a self-similarity horizon (SSH), where the critical geometry remains regular there, but the homothetic vector becomes null. To its future, the outgoing null boundary emanating from $\mathcal{S}^+$ behaves as a Cauchy horizon $\mathcal{CH}^+$. A continuation beyond this Cauchy horizon is not uniquely determined by the classical data in the domain of dependence. This is the geometric origin of the predictability problem associated with the critical naked singularity.

The past light cone $\mathcal{C}^{-}_{\rm SSH}$ naturally divides the naked singularity problem into two regions,
\begin{equation}
    \mathcal{R}_{\rm Int}
    \qquad \text{and} \qquad
    \mathcal{R}_{\rm Ext}.
\end{equation}
The interior region $\mathcal{R}_{\rm Int}$ lies inside the past light cone and describes the dynamical formation of the singularity from regular collapse. The exterior region $\mathcal{R}_{\rm Ext}$ lies outside the past light cone and extends toward future null infinity. This exterior region determines whether the singularity is actually visible to distant observers. Equivalently, it determines whether the future null infinity $\mathcal{I}^+$ is complete or whether it terminates on the future light cone of the singularity. In this sense, the nakedness property is an exterior property, while the interior region supplies the regular dynamical origin of the singularity.

This division has been made explicit in recent mathematical treatments of exterior nakedness, interior fill-ins, and instability of locally naked singularities \cite{Rodnianski:2019ylb,Shlapentokh-Rothman:2022byc,Cicortas:2024hpk, Cicortas:2026xdw}. It is especially useful for the present problem because the two regions have different physical roles. The trapped branch found in our previous interior analysis shows that quantum effects can modify the local horizon formation \cite{Tomasevic:2025kqy,Tomasevic:2025clf}, but it does not by itself determine whether outgoing null rays associated with the would-be Cauchy horizon still reach $\mathcal{I}^+$. Conversely, an exterior construction that reaches future null infinity can diagnose nakedness even before one has completely solved the regular interior problem. A complete censorship statement requires both sides.\footnote{It is important to note that this exterior problem is not a second computation of the
critical exponent.  In Type II collapse, the mass scaling exponent $\gamma$ and the echoing
period are determined by the interior self-similar attractor and its single unstable
mode.  The exterior evolution can affect non-universal data, such as radiation
profiles, Bondi-type masses, and the global causal completion, but it does not
change the local eigenvalue problem that gives the critical scaling law.}

This is why local horizon formation and global nakedness must be distinguished. In spherical symmetry, a marginally trapped surface can be diagnosed quasi-locally by the vanishing of an appropriate null expansion, or equivalently by the geometric condition
\begin{equation}
    (\nabla r)^2=0,
    \label{eq:horizon_condition_preview}
\end{equation}
where $r$ is the areal radius. This criterion will be discussed in detail in Section~\ref{sec2:horizontracing}. It is well suited for tracing apparent horizons or marginally outer trapped surfaces. However, the event horizon is teleological: its location depends on the entire future development. Therefore, a horizon found in the interior does not automatically settle the exterior problem. One must follow the corrected geometry into $\mathcal{R}_{\rm Ext}$ and ask whether the exterior Cauchy horizon structure is cloaked by a trapped branch before it can define an incomplete visible null boundary.

The exterior problem is also technically different from the interior one. The interior calculation benefits from the self-similar structure near the critical endpoint. By contrast, an exterior spacetime that connects to a physically meaningful asymptotic region cannot remain globally self-similar. Exact global self-similarity is generally incompatible with ordinary asymptotic flatness, or with the boundary conditions needed to define the exterior problem. Thus $\mathcal{R}_{\rm Ext}$ should be viewed as an interpolating region, where it inherits self-similar data near $\mathcal{C}^{-}_{\rm SSH}$ but gradually loses exact self-similarity as one moves toward the far exterior, which can be seen explicitly in Figure~\ref{fig:IntExt_CSS_DSS}. In the semiclassical problem this makes the state selection and matching conditions more delicate. The renormalized stress tensor $\langle T_{\mu \nu} \rangle$ must be compatible with the interior quantum state near the light cone, while also avoiding artificial Hawking-like fluxes or distributional sources in the exterior.

The goal of this paper can now be stated geometrically. The previous interior analysis showed that one-loop effects generate a quantum trapped branch in the self-similar formation region. The present work asks whether this branch extends into the exterior region $\mathcal{R}_{\rm Ext}$.

\subsection{Einstein-scalar collapse and one-loop vacuum polarization}
\label{sec2:Einsteinscalaroneloop}

We now specify the Einstein-scalar system and the semiclassical framework used throughout the paper. Consider Einstein gravity in $D=d+1$ spacetime dimensions minimally coupled to $N$ free massless scalar fields $f_i$, with $i=1,\ldots,N$.\footnote{The large $N$ limit gives a controlled semiclassical expansion in which matter loops dominate over graviton loops. In practice, $N$ is also a useful modeling parameter. For sufficiently large $N$, the quantum source can be made visible while the geometry remains within the linear and semiclassical regimes. This point applies equally to the horizon tracing analyses in \cite{Tomasevic:2025kqy,Tomasevic:2025clf}. A concrete setting where an effectively large number of matter modes arises is discussed in Appendix~D of \cite{Shi:2026wnk}, for magnetically charged black holes in the presence of electrically charged fermions and Landau levels.} The classical action is
\be \label{eq:Einsteinscalar}
S=\int d^D x \sqrt{-g} \bigg[\frac{1}{16 \pi G_N}(R-2 \Lambda)-\frac{1}{2}\sum_{i=1}^N(\nabla f_i)^2 \bigg].
\ee
From now on we set $c=1$ and $8\pi G_N=1$,  while keeping explicit factors of $\hbar$ in the quantum source. Note that the one-loop source scales linearly with the number of matter fields. Equivalently, the semiclassical expansion parameter is the combination $N\hbar$. However, we assume that the classical profiles are scaled so that the background stress tensor
is held fixed as $N$ is varied, with $N$ only to control the size of the one-loop source. With this convention, our goal is to solve the semiclassical Einstein equation
\be \label{eq:semiclassicalEinstein}
G_{\mu \nu}+\Lambda g_{\mu \nu}
=
T_{\mu \nu}
+
\frac{N\hbar}{\ell^{D-2}}\langle T_{\mu\nu}\rangle ,
\ee
where the classical sector satisfies
\be
T_{\mu \nu}=\sum_{i=1}^N\bigg(\nabla_\mu f_i \nabla_\nu f_i-\frac{1}{2}g_{\mu \nu}(\nabla f_i)^2\bigg),
\qquad
\Box f_i=0.
\ee
Here $\ell$ denotes the characteristic length scale. The quantum part $\langle T_{\mu\nu}\rangle$ will be obtained from the effective theory that captures the $s$-wave sector of the system. We reserve Greek indices $\mu,\nu,\ldots$ for $D$-dimensional quantities and Latin indices $a,b,\ldots$ for two-dimensional quantities. Dimensionality will be explicitly indicated whenever there is potential ambiguity. In spherical symmetry we write
\be \label{eq:sphericalmetric}
ds^2_{(D)}=g_{ab}(x^c)dx^a dx^b+r^2(x^c)d \Omega^2_{D-2},
\ee
where $r(x^c)$ is the areal radius. We denote background quantities, such as the unperturbed metric, with a bar, e.g. $\bar{g}_{\mu\nu}$, and distinguish interior and exterior quantities by the subscripts $\mathrm{int}$ and $\mathrm{ext}$, respectively.

The $s$-wave sector is captured by a two-dimensional dilaton gravity model.\footnote{The focus on the $s$-wave sector is both physically motivated and technically controlled. The unique physical growing mode of the scalar critical solution is spherically symmetric, while nonspherical perturbations decay once the gauge invariant constraints and regularity conditions are imposed. This statement is kinematical and does not depend on whether the source is treated classically or semiclassically. More about these points can be found in \cite{Tomasevic:2025clf}.} Consider a $D=d+1$ spacetime factorized into a two-dimensional spacetime with metric $g_{ab}$ and an internal sphere. We adopt the ansatz
\be \label{eq:sphericalansatz}
ds^2_{(D)}
=
g_{ab} dx^a dx^b
+
\ell^2 e^{-\frac{4 \phi}{D-2}} d \Omega^2_{D-2}.
\ee
The exponential parametrization of the dilaton $\phi$ ensures positivity of the areal radius, and the normalization is chosen such that the angular volume factor contributes universally as $e^{-2\phi}$. After imposing the ansatz, the Lagrangian densities that capture the spherical sector of the Einstein-scalar system are\footnote{The dilaton kinetic term in the gravitational action can be removed by a Weyl rescaling of $g_{ab}$. In that frame the coefficients appearing in the trace anomaly and in the anomaly-induced effective action are correspondingly reshuffled; see, for example, the discussion in~\cite{Shi:2026wnk}.  In this paper we keep the same conformal frame and normalization conventions as in~\cite{Tomasevic:2025clf,Tomasevic:2025kqy}.}
\be \label{eq:gravityaction}
\mathcal{L}_{\text{grav}}
=
\frac{\Omega_{D-2} \ell^{D-2}}{16 \pi G_N }
\sqrt{-g_{(2)}}e^{-2 \phi}
\bigg[
R^{(2)}
+
\frac{4(D-3)}{D-2}(\nabla \phi)^2
-
\frac{1}{\ell^2}(D-2)(D-3)e^{\frac{4\phi}{D-2}}
\bigg],
\ee
\be \label{eq:matteraction}
\mathcal{L}_{\text{matter}}
=
-\frac{\Omega_{D-2} \ell^{D-2}}{2}
\sqrt{-g_{(2)}} e^{-2\phi}
\sum_{i=1}^N(\nabla f_i)^2 .
\ee
Since the matter fields couple to the dilaton, they are not free conformal scalars on the metric $g_{ab}$. Nevertheless, the matter action retains Weyl symmetry, and its one-loop trace anomaly is fixed to be~\cite{Mukhanov:1994ax, Bousso:1997cg, Kummer:1998dc, Kummer:1999zy, Balbinot:2000iy, Fabbri:2003vy, Hofmann:2004kk}
\be \label{eq:traceanomaly}
\langle T^a_{\ a}\rangle
=
\frac{1}{24\pi}
\left[
R-6(\nabla\phi)^2+6\Box\phi
\right].
\ee
This anomaly follows from standard path integral quantization using the heat kernel formalism. It is one-loop exact, state independent, and insensitive to the choice of regularization scheme. More detailed discussions of this construction, and its applications to black hole and critical spacetimes can be found in \cite{Wu:2023uyb,Tomasevic:2025clf,Shi:2026wnk}.

We define the covariant renormalized stress tensor from the one-loop effective action by
\be \label{eq:def_stresstensor}
\langle T_{ab}\rangle \equiv -\frac{2}{\sqrt{-g}}\frac{\delta \Gamma_{\text{1-loop}}}{\delta g^{ab}}.
\ee
The trace anomaly fixes the anomaly-induced part of the action, but only up to Weyl-invariant terms and local counterterms,
\be \label{eq:fulloneloop}
\Gamma_{\text{1-loop}}
=
\Gamma_{\text{anom}}
+
\Gamma_{\text{W}}
+
\Gamma_{\text{ct}}.
\ee
Here $\Gamma_{\text{anom}}$ is a particular solution of the functional differential equation whose trace reproduces \eqref{eq:traceanomaly}. In the conventions above,
\be \label{eq:oneloopanom}
\Gamma_{\text{anom}}
=
-\frac{1}{96\pi}
\int d^2x \sqrt{-g}
\bigg[
R\frac{1}{\Box}R
-
12(\nabla\phi)^2\frac{1}{\Box}R
+
12\phi R
\bigg].
\ee
The first term is the Polyakov action for conformal scalars~\cite{Christensen:1977jc, Polyakov:1981rd}, while the remaining terms are required by the dilaton coupling and reflect the higher-dimensional origin of the $s$-wave sector. The counterterm $\Gamma_{\text{ct}}$ is local and state independent. The term $\Gamma_{\text{W}}$ denotes Weyl-invariant nonlocal functionals characterizing the most general solution not fixed by the trace anomaly. It contributes to the normal ordered part of the quantum stress tensor and can carry state-dependent information. In the quantum states relevant to our problem, we choose the subtraction prescription so that the vacuum contribution of $\Gamma_{\text{W}}$ is absorbed into the definition of the state, in close analogy with normal ordering. With this prescription, $\Gamma_{\text{W}}$ does not obstruct the state-selection criteria we will impose.

Diffeomorphism invariance of the combined metric-dilaton system implies the modified conservation law
\be \label{eq:modified_conservation_law}
\nabla^a\langle T_{ab}\rangle
-
\frac{1}{\sqrt{-g}}
\frac{\delta \Gamma_{\text{1-loop}}}{\delta \phi}
\nabla_b\phi
=
0.
\ee
This is the two-dimensional expression of the higher-dimensional conservation equation $\nabla^\mu \langle T^{(D)}_{\mu\nu}\rangle=0$ for the $s$-wave stress tensor. To evaluate the effective action and its associated stress tensor, it is convenient to introduce two auxiliary fields $\chi_1$ and $\chi_2$, whose role is to make the action local. More precisely,
\be \label{eq:Gamma_ext}
\Gamma_{\rm anom}[g,\phi]
=
\operatorname*{ext}_{\chi_1,\chi_2}
\Big(
\Gamma_{\chi_1}[g,\phi,\chi_1]
+\Gamma_{\chi_2}[g,\phi,\chi_2]
+\Gamma_{\phi}[g,\phi]
\Big).
\ee
The corresponding auxiliary equations take the form
\be \label{eq:aux}
\Box \chi_1= \lambda_1 R+ \lambda_2 (\nabla \phi)^2, \qquad
\Box \chi_2=- \mu_1 R-\mu_2 (\nabla \phi)^2,
\ee
with coefficients fixed by reproducing \eqref{eq:oneloopanom}. Solving these auxiliary equations with appropriate boundary conditions is equivalent to choosing the Green's functions in the nonlocal action. The homogeneous parts of $\chi_1$ and $\chi_2$ encode the quantum state and the normal ordering prescription. We will use this standard prescription~\cite{Wu:2023uyb,Tomasevic:2025clf,Shi:2026wnk} in the explicit $2+1$ and $3+1$ calculations below.

The physical state relevant for critical collapse is fixed by the collapse problem itself. It should be a Hadamard state satisfying the Wald axioms for the renormalized stress tensor~\cite{Wald:1977up, Wald1978, Wald:1978ce}. In general, the Hadamard condition requires that the short distance singularity structure be the universal Minkowski one, so that local UV divergences can be removed by covariant counterterms. In the self-similar interior, two further requirements are sufficient. First, the stress tensor must be regular in the physical domain of the critical spacetime, away from the limiting singularity at $T\to\infty$. Second, the state should not introduce incoming or outgoing quantum fluxes that are not sourced by the collapsing matter. These two conditions select the Boulware-like, asymptotically Minkowskian state found in \cite{Tomasevic:2025kqy,Tomasevic:2025clf}. It is Boulware-like not because the critical spacetime is static, but because it describes genuine vacuum polarization and has vanishing flux in the asymptotic region relevant for the incoming collapse~\cite{Boulware:1974dm}.

This choice is also motivated by the physical time ordering of the critical collapse problem. Standard Hawking radiation is associated with the later evolution of a geometry containing a trapped region, whereas the universal critical scaling is determined by the earliest MOTS, or equivalently by the apparent horizon when it first forms. The present state choice is therefore intended to isolate the vacuum polarization backreaction relevant before this first trapping event.

With this state fixed, the corresponding higher-dimensional stress tensor develops a universal growing behavior
\be \label{eq:quantum_growing_mode_general}
\langle T_{\mu\nu}^{(D)}\rangle
\sim
\hbar\, e^{(D-2)T}F_{\mu\nu}(x^i).
\ee
Here $T$ is the logarithmic self-similar time, and $F_{\mu\nu}(x^i)$ is a dimensionless tensor profile depending only on the spatial similarity coordinates $x^i$. In this sense, the one-loop stress tensor behaves as a universal quantum perturbation with Lyapunov exponent
\be
\omega_q=D-2.
\ee
Two distinctions will be useful in what follows. First, the Boulware-like state in the self-similar interior is not selected by the absence of outgoing flux alone. It follows from regularity together with collapse boundary conditions that remove independently specified state fluxes. In the explicit solutions, the resulting higher-dimensional stress tensor is asymptotically Minkowskian. The exterior problem is different. Local matching across $\mathcal{C}_{\rm SSH}^-$ fixes only the continuation of the same state near the light cone. Additional exterior boundary conditions, discussed in Sections~\ref{sec:2+1d_Garfinkle} and~\ref{sec:3+1d_Roberts} are required to determine its continuation away from the matching surface. Second, once the reduced stress tensor respects self-similarity and contains no additional exponential $T$-dependence, the factor $e^
{(D-2)T}$  follows from the areal-radius factor in the spherical uplift. The detailed profile $F_{\mu \nu}$ remains state dependent. We emphasize that showing that quantum effects introduce a growing, dynamically relevant mode does not determine, by dimensional analysis alone, whether a mass gap forms. This requires the explicit backreaction and horizon-tracing analysis. See also the detailed analysis in~\cite{Tomasevic:2025clf}.

In a DSS spacetime, $F_{\mu\nu}$ is replaced by a bounded function periodic in $T$ with the echoing period, while the overall growth exponent remains $\omega_q=D-2$. The explicit profiles were obtained in \cite{Tomasevic:2025clf} and will be reviewed for the Garfinkle and Roberts spacetimes in Sections~\ref{sec3:Garfinkleinterior} and~\ref{sec4:Robertsinterior}. In these two examples the one-loop backreaction can be solved analytically in the interior.

The scaling in \eqref{eq:quantum_growing_mode_general} is kinematical. It follows from self-similarity and from the areal radius factor that lifts the two-dimensional stress tensor to the higher-dimensional $s$-wave stress tensor~\cite{Tomasevic:2025clf}. This is why this scaling is expected to hold more broadly than in the particular CSS examples used for analytic control. For the same reason, higher-loop terms do not acquire an additional self-similar enhancement relative to the one-loop term; they share the same kinematical scaling but are suppressed by extra powers of $\hbar$. Thus the one-loop stress tensor is the leading quantum source capable of competing with the classical unstable mode in the admissible semiclassical regime. In the examples studied in~\cite{Tomasevic:2025kqy,Tomasevic:2025clf}, it is this competition that shifts the critical threshold and produces a finite mass gap, replacing the classical Type II zero-mass endpoint by a quantum-modified threshold resembling Type I behavior.

The exterior problem requires matching across the past light cone $\mathcal{C}_{\rm SSH}^-$ of the critical endpoint, which imposes two additional requirements. Third, the exterior stress tensor must be the continuation of the same quantum state across $\mathcal{C}_{\rm SSH}^-$. Fourth, the matching should not create a distributional stress tensor or a null shell unless such a shell is explicitly part of the model. For critical collapse, no such shell is expected either classically or semiclassically, since the past light cone is an artificial division used to organize the calculation, not a physical layer of matter. Classically, this means that the metric should be sufficiently regular across the matching surface, or more generally that the Barrab\'es-Israel null shell data vanish~\cite{Israel:1966rt,Barrabes:1991ng,Poisson:2002nv}. Equivalently, the curvature should contain no delta function supported on $\mathcal{C}_{\rm SSH}^-$.  These local junction conditions ensure that the state is locally Hadamard across $\mathcal{C}_{\rm SSH}^-$ and, together with regularity and the exterior boundary conditions, define the smooth state continuation used in the models below.

This should be distinguished from an Unruh-like state~\cite{Unruh:1976db, Davies:1976ei, Hiscock:1980ze, Hiscock:1981xb}. In a black hole spacetime formed from collapse, with a future horizon already present and outgoing Hawking radiation at future null infinity, an Unruh state is the appropriate late-time effective description. The present problem concerns the earlier near-critical regime and the first trapped surface.\footnote{This is the standard object in critical collapse, where the critical scaling is associated with the first marginally trapped surface or first apparent horizon. Later horizon evolution, including additional matter accumulation, is generally not part of the universal critical regime.} There the relevant source is the self-energy of the collapsing scalar field itself, not a prescribed Hawking flux from a black hole that has already formed. If a global future horizon subsequently develops, a later transition from the Boulware-like description to an Unruh-like evaporation description may be required. That transition belongs to the global evaporation problem and is not needed for locating the first quantum trapped branch. This separation does not exclude the possibility of pre-Hawking radiation before trapping, which remains an important open question and is discussed further in Section~\ref{sec:Discussion}.

\subsection{Light cone matching and exterior horizon tracing}
\label{sec2:horizontracing}

We now formulate the geometric criteria used to diagnose horizon formation in the exterior. The discussion is independent of the detailed form of the $2+1$ and $3+1$ models below, and applies to any spherically symmetric semiclassical geometry of the form~\eqref{eq:sphericalmetric}. In a double-null patch, we write
\be
ds^2
=
-e^{2\rho(u,v)}du\,dv
+
r^2(u,v)d\Omega_{D-2}^2 .
\label{eq:double_null_general}
\ee
The future directed radial null directions are generated by vectors proportional to $\partial_v$ and $\partial_u$. The corresponding null expansions are proportional to
\be
\theta_+ \propto \partial_v r,
\qquad
\theta_- \propto \partial_u r.
\label{eq:null_expansions}
\ee
A marginal surface is obtained when one of the radial null expansions vanishes. For our case, the branch of interest is the marginally outer trapped one, namely
$\theta_+=0$ with the ingoing expansion keeping its expected sign, $\theta_-<0$. In spherical symmetry, the union of the two possible marginal branches can be detected by the coordinate invariant condition
\be
(\nabla r)^2
=
g^{ab}\nabla_a r \nabla_b r
=
0 .
\label{eq:geometric_horizon_condition}
\ee
In the double-null coordinates \eqref{eq:double_null_general}, we have$(\nabla r)^2\propto \partial_u r\,\partial_v r$,
so \eqref{eq:geometric_horizon_condition} detects the locus where either radial expansion vanishes. The branch selection is made by checking the sign of the nonvanishing expansion. In the perturbative regimes studied in this paper, $\theta_-$ remains bounded away from zero, while the outgoing expansion can change sign and produce a marginally outer trapped surface. Since \eqref{eq:geometric_horizon_condition} is coordinate invariant, we will evaluate it in adapted self-similar coordinates $(T,x^i)$, where the classical and quantum growing modes are most transparent.

A useful point is that \eqref{eq:geometric_horizon_condition} is nonlinear in the corrected geometry. This is precisely why horizon formation can be located even when the metric perturbations themselves remain small. We will write the corrected areal radius schematically by expanding around the background $\bar{r}$ as
\be
r
=
\bar{r}
+
\delta r_{\rm c}
+
\delta r_{\rm q},
\ee
incorporating the classical and quantum perturbations. The horizon tracing function $(\nabla r)^2$ contains both perturbations through the nonlinear geometric combination \eqref{eq:geometric_horizon_condition}. A root of $(\nabla r)^2=0$ can therefore appear in a perturbative regime, provided the root occurs before the perturbative expansion or the semiclassical approximation breaks down.

Here we make the notions more precise. A marginally outer trapped surface (MOTS) is a local codimension two surface satisfying the appropriate vanishing expansion condition discussed above. An apparent horizon (AH) is a hypersurface foliated by such marginal surfaces and is tied to a choice of slicing. In critical collapse, the universal scaling is usually associated with the earliest MOTS (EMOTS), or equivalently with the AH when it first forms. This object depends only weakly on the slicing compared with the later evolution of the AH after additional matter has fallen in. On the other hand, the \emph{global} EMOTS, and the associated critical scaling law are interior notions.

We characterize the size of a trapped surface by a quasi-local mass. We consider the general such function known as the Hawking mass~\cite{Hawking:1968qt, Hayward:1993ph}, or more specifically the Misner-Sharp-Hernandez mass in spherically symmetric spacetimes~\cite{Misner:1964je, Hernandez:1966zia}. Since the two examples we consider have different dimensions and asymptotics, we spell out the definitions separately. In the $2+1$ dimensional asymptotically AdS case, we use
\be
M_{\rm H}^{(3)}
\equiv
\frac{r^2}{\ell^2}
-
(\nabla r)^2
\quad
\xrightarrow{\,(\nabla r)^2=0\,}
\quad
M_{\rm H}^{(3)}
=
\frac{r^2}{\ell^2},
\label{eq:Hawking_mass_3d}
\ee
where $\ell$ is the AdS length. In the $3+1$ dimensional asymptotically flat case, we use
\be
M_{\rm H}^{(4)}
\equiv
\frac{r}{2}\left[1-(\nabla r)^2\right]
\quad
\xrightarrow{\,(\nabla r)^2=0\,}
\quad
M_{\rm H}^{(4)}
=
\frac{r}{2}.
\label{eq:Hawking_mass_4d}
\ee
These quasi-local masses will be used to determine whether the quantum-generated trapped surface has finite mass near the shifted threshold.

We will call an exterior horizon \emph{controlled} when the following local conditions are satisfied.  First, the candidate horizon must be a root of $(\nabla r)^2=0$ within the exterior patch where the classical continuation and the one-loop construction are valid. Second, the curvature at the root must remain weak in Planck units, so that we have effective field theory control.  Generically, one evaluates the curvature invariants using the full truncated metric retained in the calculation. A convenient diagnostic is
\be
\epsilon_{\rm curv}
\equiv
\max_{\rm MOTS}
\left(
|R|,
\left|R_{\mu\nu}R^{\mu\nu}\right|^{1/2},
\left|R_{\mu\nu\rho\sigma}R^{\mu\nu\rho\sigma}\right|^{1/2}
\right)
\ll 1,
\label{eq:curvature_control}
\ee
where one may include the Riemann-squared invariant because tidal curvature can become large even when Ricci invariants are small. Third, the one-loop source must remain small compared with the curvature scale of the same truncated geometry.  We quantify this by the source-to-curvature ratio, where we use the invariant scalar diagnostic
\be
\epsilon_{\rm src}
\equiv
\left.
\frac{
\frac{N\hbar}{\ell^{D-2}}
\left|
\langle T_{\mu\nu}\rangle
\langle T^{\mu\nu}\rangle
\right|^{1/2}
}{
\max\left(
|R|,
\left|R_{\mu\nu}R^{\mu\nu}\right|^{1/2},
\left|R_{\mu\nu\rho\sigma}R^{\mu\nu\rho\sigma}\right|^{1/2}
\right)
}
\right|_{\rm MOTS}
\ll1 .
\label{eq:source_curvature_control}
\ee
All contractions are evaluated using the full truncated metric.  This condition is the invariant version of requiring the one-loop stress tensor to remain a perturbative source for the geometry.\footnote{We only use the coordinate component ratio $\delta g/\bar{g}$ as a diagnostic.  Such a ratio is gauge-dependent, and in self-similar coordinates, some background components can become small precisely where the horizon equation is most sensitive.} Finally, the root must be stable within the approximation scheme. It must not be produced by constraint violation, numerical boundary artifacts, or a mismatch at $\mathcal{C}_{\rm SSH}^-$. It must also not be an artifact of perturbation truncation, where omitted terms are parametrically smaller and do not affect the existence of MOTS.

In the analytic $2+1$ model, the corresponding checks can be carried out directly using the closed-form exterior stress tensor and corrected geometry. In the numerical $3+1$ model, similar criteria are implemented as diagnostics: the root of $(\nabla r)^2$ must be stable under changes of the exterior strip, insensitive to the outer boundary within the chosen control window, and accompanied by small constraint residuals and curvature. The finite Hawking mass of the root is then the exterior analog of the finite mass gap found in the interior analysis.

\section{Quantum critical exterior spacetime in $2+1$ dimensions}
\label{sec:2+1d_Garfinkle}

We first analyze the exterior problem in a case where the full construction can be carried out analytically. The $2+1$ dimensional Garfinkle spacetime provides a CSS model of scalar critical collapse with closed-form interior quantum backreaction. Its exterior is obtained by a null continuation across the past light cone of the singularity, following~\cite{Jalmuzna:2015hoa}. This allows us to formulate the matching problem explicitly, and horizon tracing reveals a quantum trapped branch.

\subsection{The Garfinkle interior and its quantum horizon}
\label{sec3:Garfinkleinterior}

Critical collapse in $2+1$ dimensions is special because black holes require a negative cosmological constant, as in the Ba\~{n}ados-Teitelboim-Zanelli (BTZ) solution~\cite{Banados:1992wn}, while the Type II self-similar regime is controlled by arbitrarily small scales where $\Lambda$ becomes dynamically irrelevant.  The Garfinkle family captures precisely this local CSS regime: the true numerical critical solution for scalar collapse in asymptotically AdS$_3$ is CSS near the singularity~\cite{Pretorius:2000yu, Husain:2000vm}, and is well approximated there by a member of the Garfinkle family~\cite{Garfinkle:2000br,Garfinkle:2002vn,Jalmuzna:2015hoa}.\footnote{We note that the classical circularly symmetric Einstein-scalar system with $\Lambda<0$ has recently been shown to satisfy weak cosmic censorship for generic $C^k$ data, with a mass gap and blueshift instability playing central roles~\cite{Cicortas:2026xdw}. Our use of the Garfinkle spacetime is different, and should be distinguished from classical genericity results. We focus on the fine-tuned critical solution and ask how one-loop vacuum polarization modifies the naked region.} The local critical solution can be studied in the $\Lambda=0$ approximation, while the exterior completion still remembers the negative cosmological constant needed to embed the critical region into an asymptotically AdS spacetime; see Section~4.1 of~\cite{Tomasevic:2025clf} for a detailed review.

The value of the Garfinkle model is therefore twofold. It is closely related to the numerical 2+1-dimensional CSS critical solution, and its analytic form allows the one-loop stress tensor, semiclassical backreaction, and exterior horizon tracing to be treated in closed form. Nevertheless, a natural next step is to repeat the analysis directly on the numerical critical spacetime. In this subsection we review the $\Lambda=0$ interior solution and its one-loop quantum horizon. The exterior null continuation and the role of $\Lambda$ will be addressed in Section~\ref{sec3:exteriornull}.

In adapted self-similar coordinates $(T,x)$, the background interior is represented by a one-parameter family labeled by $n$,
\be \label{eq:Garfinklemetric}
ds^2_{\rm int}
=
\ell^2 e^{-2T}
\bigg[
e^{2\bar{\rho}(x,n)}
\bigg(dx-\frac{x}{2n}dT\bigg)dT
+
\bar{r}^2(x,n)d\theta^2
\bigg],
\ee
with
\be
e^{2\bar{\rho}(x,n)}
=
2n
\bigg(\frac{1+x^n}{2}\bigg)^{4(1-\frac{1}{2n})},
\qquad
\bar{r}(x,n)=\frac{1-x^{2n}}{2}.
\ee
The coordinate range for the interior region is 
\be
T\in(-\infty,\infty),
\qquad
x\in[0,1].
\ee
The center is at $x=1$, while the past light cone of the singularity is at $x=0$.  The background scalar field supporting the geometry is
\be
\bar{f}(T,x)
=
\sqrt{\frac{2n-1}{2n}}
\bigg[
T
-
2\ln\bigg(\frac{1+x^n}{2}\bigg)
\bigg].
\ee
The curvature singularity is reached at the self-similar accumulation point $T\to\infty$ at fixed $x$.  For the solution to be analytic at both the center and the light cone, we take $n$ to be a positive integer $n \in \mathbb{Z}_{>0}$.  The case $n=4$ gives the best approximation to the numerical critical solution once the cosmological constant and the null continuation are included~\cite{Jalmuzna:2015hoa}, although most formulas below will be kept for general positive integer $n$.

\begin{figure}[t]
\centering
\resizebox{0.30\textwidth}{!}{%
\begin{tikzpicture}[
  line cap=round,
  line join=round,
  every node/.style={inner sep=1.5pt},
  boundary/.style={draw=black, line width=1.25pt},
  outerlabel/.style={font=\Large, align=center},
  innercoord/.style={font=\large, align=center, fill=gray!8, inner sep=1.1pt}
]

\def\L{5.0}

\coordinate (A) at (0,0);        
\coordinate (B) at (0,-10);      
\coordinate (D) at (\L,-\L);     

\fill[gray!8] (A) -- (B) -- (D) -- cycle;

\draw[boundary] (A) -- (B);   
\draw[boundary] (B) -- (D);   
\draw[boundary] (A) -- (D);   

\fill (A) circle[radius=3.0pt];

\node[font=\Large, anchor=east, align=right]
  at ($(A)+(-0.35,0.25)$)
  {Singularity\\[2pt]$T\to\infty$\\[2pt]$(u\to0^-,\,v\to0^-)$};


\node[outerlabel, rotate=90]
  at ($(A)!0.56!(B)+(-0.50,0)$)
  {regular center};

\node[outerlabel, rotate=-45]
  at ($(A)!0.56!(D)+(0.42,0.42)$)
  {past light cone};

\node[outerlabel, rotate=0]
  at ($(B)!0.58!(D)+(0.42,-0.42)$)
  {$\mathcal I^-$};


\node[innercoord, rotate=90]
  at ($(A)!0.56!(B)+(0.55,0)$)
  {$x=1~~ (u=v)$};

\node[innercoord, rotate=-45]
  at ($(A)!0.53!(D)+(-0.28,-0.28)$)
  {$x=0~~ (v=0)$};

\node[innercoord, rotate=45]
  at ($(B)!0.56!(D)+(-0.30,0.30)$)
  {$T\to-\infty~~ (u\to-\infty)$};

\end{tikzpicture}%
}
\caption{
Penrose diagram of the Garfinkle interior region, where the corresponding adapted
and null coordinate limits are indicated.}
\label{fig:GarfinkleInterior}
\end{figure}
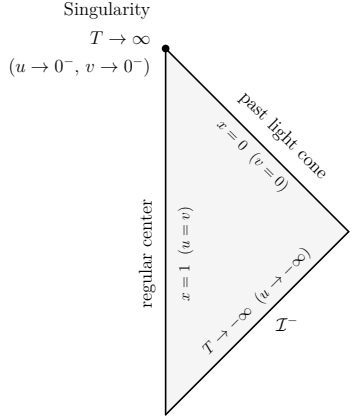

It is often useful to introduce double-null coordinates $(u,v)$ related to the adapted coordinates by
\be \label{eq:Garfinkletcoord}
x=\frac{v}{u},
\qquad
T=-2n\ln(-u),
\qquad
u=-e^{-\frac{T}{2n}},
\qquad
v=ux=-e^{-\frac{T}{2n}}x .
\ee
In the interior region we have
\be
u\in(-\infty,0],
\qquad
v\in(-\infty,0],
\qquad
|v|\leq |u|.
\ee
The metric becomes
\be \label{eq:Garfinkle_uv_metric}
ds^2_{\rm int}=-e^{2\bar{\rho}}du\,dv+\bar{r}^2d\theta^2,
\ee
where
\be
e^{2\bar{\rho}}
=
4n^2
\bigg[
\frac{(-u)^n+(-v)^n}{2}
\bigg]^{4(1-\frac{1}{2n})},
\qquad
\bar{r}
=
\frac{1}{2}
\left[
(-u)^{2n}-(-v)^{2n}
\right].
\ee
The scalar field is
\be
\bar{f}
=
-2\sqrt{\frac{2n-1}{2n}}
\ln
\bigg[
\frac{(-u)^n+(-v)^n}{2}
\bigg].
\ee
In these coordinates, the center is at $u=v$, past null infinity corresponds to $u\to-\infty$, and the past light cone of the singularity is $v=0$ at finite negative $u$.  The singular endpoint is reached by taking $u\to0^-$ with $v=ux\to0^-$. See Figure~\ref{fig:GarfinkleInterior} for a Penrose diagram of the Garfinkle interior.  Although the critical solutions of interest have positive integer $n$, the formal value $n=\frac12$ is a useful sanity check: the scalar field vanishes and the metric reduces to flat Minkowski spacetime written in self-similar coordinates.

Let us recall the relevant one-loop result in the interior~\cite{Tomasevic:2025clf}.  Applying the one-loop effective theory of Section~\ref{sec2:Einsteinscalaroneloop} to the Garfinkle background, regularity at the center and at the past light cone fixes the homogeneous auxiliary field data and selects the Boulware-like, asymptotically Minkowskian state associated with the self-energy of the collapsing scalar field.  In the conventions of Section~\ref{sec2:Einsteinscalaroneloop}, the required local counterterms can be chosen as
\be \label{eq:Garfinkle_counterterms}
\Gamma_{\rm ct}
=
\int d^2x\sqrt{-g}
\left[
\alpha_1\phi R+\alpha_2 fR+\alpha_3(\nabla f)^2
\right],
\ee
with
\be
\alpha_1=\frac{1}{16\pi},
\qquad
\alpha_2=\frac{1}{24\pi}\sqrt{1-\frac{1}{2n}},
\qquad
\alpha_3=\frac{17n-4}{192n\pi}.
\ee
These counterterms remove spurious local divergences while preserving the physical state criteria. In double-null coordinates, the regular quantum stress tensor is
\bea
\langle T^{(3)}_{uu}\rangle_{\rm int}
&=&
-\frac{(2n-1)n}{16\pi^2 u^2 (u^{2n}-v^{2n})^3(u^n+v^n)^2}
\bigg[
u^{6n}
-8u^{5n}v^n
-8u^{4n}v^{2n}
+12u^{3n}v^{3n}
\no\\
&\quad&
+9u^{2n}v^{4n}
-4u^n v^{5n}
-2v^{6n}
+8u^{4n}(u^n+v^n)^2
\ln\bigg(\frac{2u^n}{u^n+v^n}\bigg)
\bigg],
\label{eq:Garfinkle_Tuu_uv}
\\
\langle T^{(3)}_{vv}\rangle_{\rm int}
&=&
\frac{(2n-1)n}{16\pi^2 v^2 (u^{2n}-v^{2n})^5}
\bigg[
-3u^{6n}v^{2n}
+10u^{5n}v^{3n}
-u^{4n}v^{4n}
-20u^{3n}v^{5n}+11u^{2n}v^{6n}
\no\\
&\quad&
+10u^n v^{7n}
-7v^{8n}
+8(2u^{2n}v^{6n}-v^{8n}-u^{4n}v^{4n})
\ln\bigg(\frac{2u^n}{u^n+v^n}\bigg)
\bigg],
\label{eq:Garfinkle_Tvv_uv}
\\
\langle T^{(3)}_{uv}\rangle_{\rm int}&=&0,
\label{eq:Garfinkle_Tuv_uv}
\\
\langle T^{(3)}_{\theta\theta}\rangle_{\rm int}
&=&
-\frac{(2n-1)4^{-1-\frac{1}{n}}}{n\pi^2 u}
\frac{(u^n+v^n)^{-5+\frac{2}{n}}}{u^n-v^n}
v^{n-1}
\bigg[
3u^{3n}
-4u^{2n}v^n
+3u^n v^{2n}-2v^{3n}
\no\\
&\quad&
-8u^{2n}v^n
\ln\bigg(\frac{2u^n}{u^n+v^n}\bigg)
\bigg].
\label{eq:Garfinkle_Ttt_uv}
\eea
Here and below the powers are understood in the same real branch as the interior Garfinkle patch.  The expression is cumbersome in $(u,v)$ coordinates, but it makes the state interpretation transparent. The stress tensor is regular at both the center $u=v$ and the past light cone $v=0$, vanishes in the flat limit $n=\frac12$, and carries no artificial asymptotic flux. Explicitly, $\lim_{u\to-\infty}\langle T^{(3)}_{vv}\rangle=0$ and, formally, $\lim_{v\to\infty}\langle T^{(3)}_{uu}\rangle=0$.  The first limit shows that there is no imposed incoming flux from $\mathcal I^-$.  The second should be interpreted with care, since the Garfinkle interior by itself does not define the future infinity. Nevertheless, it is the sense in which the state is asymptotically Minkowskian and describes self-energy of the collapsing matter without Hawking-like flux.

The simple structure becomes manifest after transforming back to the adapted coordinates $(T,x)$.  One obtains
\bea
\langle T^{(3)}_{TT}\rangle_{\rm int}
&=&
e^T
\frac{2n-1}{64n\pi^2(x^{2n}-1)^3}
\bigg[
1-10x^n+14x^{2n}-10x^{3n}+5x^{4n}
\no\\
&\quad&
+8(x^{4n}+1)\ln\bigg(\frac{2}{1+x^n}\bigg)
\bigg],
\label{eq:Garfinkle_TTT_Tx}
\\
\langle T^{(3)}_{xT}\rangle_{\rm int}
&=&
e^T
\frac{(2n-1)x^{2n-1}}{32\pi^2(x^{2n}-1)^3}
\bigg[
-3+10x^n-7x^{2n}
-8x^{2n}\ln\bigg(\frac{2}{1+x^n}\bigg)
\bigg],
\label{eq:Garfinkle_TxT_Tx}
\\
\langle T^{(3)}_{xx}\rangle_{\rm int}
&=&
e^T
\frac{(2n-1)n x^{2n-2}}{16\pi^2(x^{2n}-1)^3}
\bigg[
3-10x^n+7x^{2n}
+8x^{2n}\ln\bigg(\frac{2}{1+x^n}\bigg)
\bigg],
\label{eq:Garfinkle_Txx_Tx}
\\
\langle T^{(3)}_{\theta\theta}\rangle_{\rm int}
&=&
e^T
\frac{(2n-1)4^{-1-\frac{1}{n}}x^{n-1}(1+x^n)^{-5+\frac{2}{n}}}{n\pi^2(x^n-1)}
\bigg[
3-4x^n+3x^{2n}-2x^{3n}
\no\\
&\quad&
-8x^n\ln\bigg(\frac{2}{1+x^n}\bigg)
\bigg].
\label{eq:Garfinkle_Ttt_Tx}
\eea
Thus the stress tensor organizes into the homogeneous form
\be \label{eq:Garfinkle_quantum_stress_scaling}
\langle T^{(3)}_{\mu\nu}\rangle_{\rm int}
=
e^T F_{\mu\nu}(x,n),
\ee
where $F_{\mu\nu}(x,n)$ is real analytic on $x\in[0,1]$ for positive integer $n$.  This is the explicit realization of the universal quantum growing mode discussed around~\eqref{eq:quantum_growing_mode_general}.  The quantum Lyapunov exponent is
\be
\omega_q=1,
\ee
which is positive and therefore grows toward the singular endpoint $T\to\infty$.

This growth is not a pathology of the state.  Regularity means being regular at finite $T$ in the physical self-similar domain, including the center and the past light cone. It is not a claim that the classical singular endpoint at $T\to\infty$ is nonsingular.  On the contrary, the growth of $\langle T^{(3)}_{\mu\nu}\rangle$ with $T$ is precisely the physical signal that quantum effects become increasingly important as the singular regime is approached.  Our semiclassical analysis is always performed at finite, moderately large $T$, where the one-loop expansion, linearized backreaction, and the horizon criteria of Section~\ref{sec2:horizontracing} remain valid.

The corresponding one-loop backreaction can be solved exactly in the interior.  We write the semiclassical metric in the same adapted form as the classical solution,
\be
ds^2_{\rm int}
=
\ell^2 e^{-2T}
\bigg[
F(T,x)
\bigg(dx-\frac{x}{2n}dT\bigg)dT
+
r^2(T,x)d\theta^2
\bigg],
\ee
with
\be
F(T,x)
=
\bar{F}(x)
+
\frac{N \hbar}{\ell}F_q(x)e^T,
\qquad
r(T,x)
=
\bar{r}(x)
+
\frac{N \hbar}{\ell}r_q(x)e^T,
\ee
where $\bar{F}(x)=e^{2\bar{\rho}(x,n)}$. Solving the semiclassical Einstein equation at $O(\hbar)$ gives
\be \label{eq:quantumGarfinkle}
F_q(x)
=
-\frac{(2n-1)4^{\frac{1}{n}-4}(1+x^n)^{3-\frac{2}{n}}}{\pi^2(x^n-1)}
\bigg[
x^{2n}-1
+
4\ln\bigg(\frac{2}{1+x^n}\bigg)
\bigg],
\qquad
r_q=\frac{3}{64\pi^2}.
\ee
The corrected geometry is real analytic throughout the interior domain $x\in[0,1]$ for positive integer $n$, and the full semiclassical equations, including the angular component, are satisfied to $O(\hbar)$.  The fact that $r_q$ is a constant makes the interior calculation especially simple, but it is still nontrivial for horizon tracing because the invariant quantity $(\nabla r)^2$ is nonlinear in the corrected geometry.

For the semiclassical spacetime, the physical areal radius is $\ell e^{-T}r(T,x)$. To linear order in perturbations, including both the classical growing mode and the quantum backreaction, the horizon tracing function takes the form
\be
(\nabla r)^2_{\rm int}
\approx
\bar{h}(x)
+
(p-p^\ast)e^{(1-\frac{1}{2n})T}h_c(x)
+
\frac{N \hbar}{\ell}e^T h_q(x),
\label{eq:Garfinkle_interior_horizon_function}
\ee
where the unperturbed Garfinkle background contribution is
\be
\bar{h}(x)
=
2^{4-\frac{2}{n}}
x^{2n-1}
(1+x^n)^{\frac{2}{n}-4}.
\label{eq:Garfinkle_f0}
\ee
This background quantity vanishes at the past light cone $x=0$. Thus, in the pure $\Lambda=0$ CSS background, the light cone is already marginal. The second term in \eqref{eq:Garfinkle_interior_horizon_function} is the contribution from the dominant classical perturbation, with classical Lyapunov exponent $\omega_c=1-\frac{1}{2n}$, while the third term is the contribution from the one-loop quantum mode. The analytic forms of $h_c(x)$ and $h_q(x)$, as well as the detailed horizon tracing analysis for general $n$, were given in~\cite{Tomasevic:2025clf}.

The fact that the $\Lambda=0$ light cone is already marginal should not be confused with the exterior question. The cosmological constant and the null continuation are needed to determine the correct exterior structure. The important lesson is that the regular quantum state produces a growing one-loop stress tensor, the backreaction can be solved in closed form, and the quantum mode generates a trapped branch in the local self-similar region. 

The exterior analysis below imports precisely the data fixed in this interior calculation.  At the past light cone $v=0$, we keep the classical Garfinkle geometry, the scalar profile, and the quantum state selected by the interior regularity conditions.  The exterior quantum state is required to be the continuation of the same state.  In this sense, the interior Garfinkle calculation supplies the boundary data for the exterior problem. The continuation of a smooth light cone and a smooth semiclassical geometry will determine whether the classically naked Cauchy horizon region is cloaked by a quantum trapped branch.

\subsection{The exterior null continuation and classical perturbation modes}
\label{sec3:exteriornull}

The naive analytic continuation of the Garfinkle solution through the past light cone into $x<0$ produces trapped surfaces and a spacelike central singularity, and is therefore not the appropriate exterior critical geometry~\cite{Jalmuzna:2015hoa, Tomasevic:2025clf}.  Instead, following~\cite{Jalmuzna:2015hoa}, one matches the interior Garfinkle solution across $v=0$ to a distinct null continuation.  This continuation is a local exterior model beyond the past light cone of the critical endpoint.  It is exact at $\Lambda=0$, but the negative cosmological constant plays an essential role in the exterior problem and will be included below as a perturbation. Related issues can be found in~\cite{Jalmuzna:2015hoa} and are reviewed in Section~4.1 of~\cite{Tomasevic:2025clf}. This is the classical background on which we will impose the one-loop matching conditions in Section~\ref{sec3:quantumbackreaction}.

In the exterior domain
\be
u\leq 0,
\qquad
v\geq 0,
\qquad
x=\frac{v}{u}\leq 0,
\ee
the null continuation is defined by
\be
ds^2_{\rm ext}
=
\ell^2
\left[
-e^{2\bar{\rho}(u)}du\,dv
+
\bar{r}^2(u)d\theta^2
\right],
\ee
with
\be
e^{2\bar{\rho}(u)}
=
4n^2
\left[
\frac{(-u)^n}{2}
\right]^{4(1-\frac{1}{2n})},
\qquad
\bar{r}(u)
=
\frac{(-u)^{2n}}{2},
\ee
and
\be
\bar{f}
=
-2\sqrt{1-\frac{1}{2n}}
\ln
\left[
\frac{(-u)^n}{2}
\right].
\ee
Equivalently, in the adapted self-similar coordinates, the same continuation is represented by\footnote{The factor of $2n$ in $e^{2 \bar{\rho}}$ comes from converting the null continuation of~\cite{Jalmuzna:2015hoa} to the regular null coordinate at the light cone and to our adapted metric normalization.  In particular, we use the regular exterior null coordinate, denoted $\hat v$ in~\cite{Jalmuzna:2015hoa}, rather than the singular coordinate used in some intermediate expressions.  With this choice, the metric matches the interior convention used in Section~\ref{sec3:Garfinkleinterior}.}
\be
e^{2\bar{\rho}}
=
(2n)2^{-4(1-\frac{1}{2n})},
\qquad
\bar{r}=\frac12,
\qquad
\bar{f}(T,x)
=
\sqrt{1-\frac{1}{2n}}
\left[
T+2\ln2\right] .
\ee
Thus the exterior null continuation is still CSS at $\Lambda=0$.  Its defining feature is that the exterior fields are independent of $v$ in double-null coordinates, or equivalently that the leading similarity profiles are constant or simple functions of the exterior variable $x$.

The matching at $v=0$ is smooth enough for the present purpose but not analytic.  On the interior side, the dependence on the transverse coordinate near the light cone appears through powers such as $v^n$, or equivalently $x^n$. Schematically,
\be
\bar{\Psi}_{\rm int}(u,v)
=
\Psi_0(u)
+
\Psi_1(u)v^n
+
\cdots ,
\qquad
\bar{\Psi}_{\rm ext}(u,v)
=
\Psi_0(u),
\ee
for any of the background fields $\bar{\Psi}$.  Therefore derivatives up to order $n-1$ match across $v=0$, while the $n$th derivative generally jumps.  The null continuation is consequently $C^{n-1}$ but not $C^n$.  For positive integer $n\geq2$, the metric is at least $C^1$, so the curvature contains no delta function supported on the light cone.  In particular, for the case $n=4$ most closely related to the numerical critical solution, the background matching is $C^3$.  This is sufficient to avoid a classical null shell at the matching surface.

Geometrically, the null continuation should not be interpreted as a standard Bondi exterior~\cite{Bondi:1962px, Sachs:1962wk}.  Since $r_0(u)$ is independent of $v$, moving to large $v$ at fixed $u$ does not take one to large areal radius. Rather, it moves along a null direction of constant area radius.  This is the sense in which the continuation is Vaidya-like, or an ``onion'' of outgoing null layers~\cite{Vaidya:1951fdr}.  The surface $u=0$ is the future null boundary associated with the accumulation point and has the character of a Cauchy horizon for the local critical solution.  In a complete spacetime this local exterior would eventually have to be embedded into an asymptotically AdS geometry, and exact CSS would be lost.  In the present construction, we will incorporate the leading effect of this embedding by adding the first-order $\Lambda$-correction to the null continuation.\footnote{An exact continuation beyond the light cone with $\Lambda<0$ was also constructed in Appendix~L of~\cite{Jalmuzna:2015hoa}. That solution preserves the qualitative features observed numerically outside the light cone and perturbatively reduces to the null continuation plus the leading $\Lambda$ corrections used here. Treating this exact $\Lambda<0$ exterior semiclassically would require a separate analysis of the appropriate quantum state, including the role of the asymptotically AdS boundary conditions and the correct notion of vacuum polarization in that global spacetime. We use the perturbative form because it makes the matching to the interior CSS solution, the classical growing mode, and the one-loop stress tensor transparent. It is also sufficient since numerically, the exterior region probed by the near-critical analysis corresponds to very small negative values of the exterior coordinate, where the exact $\Lambda<0$ continuation cannot be distinguished from its perturbative expansion to the order used here.}

The null continuation is also the natural setting for continuing the classical perturbation modes from the interior~\cite{Jalmuzna:2015hoa,Tomasevic:2025clf}.  Linear perturbations around the CSS Garfinkle background are labeled by classical Lyapunov exponents
\be
\omega_c=\frac{m}{2n}.
\ee
The regular interior spectrum contains growing modes
\be
m=2,3,\ldots,n-1,
\qquad
m=2n-1,
\ee
together with decaying modes
\be
m<0,
\qquad
m\neq -(2k+1)n,
\qquad
k\in\mathbb{Z}_{\geq0}.
\ee
Thus the unmodified Garfinkle solution with parameter $n$ has $n-1$ growing modes.  For $n=4$, this would give three growing modes, apparently in tension with the usual critical collapse expectation of a single unstable mode.  The amended construction with $\Lambda<0$ and the null continuation was proposed to address this tension: it is conjectured that, in the true critical solution approximated by the amended Garfinkle geometry, the relevant physical instability is selected by the top growing mode~\cite{Jalmuzna:2015hoa}
\be
m=2n-1,
\qquad
\omega_c=1-\frac{1}{2n}.
\ee
This is the mode we keep in the exterior horizon tracing analysis.

We now write the exterior perturbative geometry in a form that includes both this top growing mode and the leading effect of the negative cosmological constant.  Since $\Lambda=-1/\ell^2$ introduces the infrared scale, its local effect in the self-similar region is controlled by the dimensionless ratio between the self-similar length scale and the AdS radius.  With the metric written as $\ell^2 e^{-2T}$ times a dimensionless CSS profile, this ratio appears as
\be
y\equiv e^{-2T}.
\ee
Thus the $\Lambda$ correction is a sourced decaying perturbation of the $\Lambda=0$ CSS equations, with effective classical Lyapunov exponent
\be
\omega_\Lambda=-2.
\ee
Equivalently, in solving the Einstein equations one keeps the cosmological term $\Lambda g_{\mu\nu}$ and expands the resulting fields to first order in $y=e^{-2T}$. This is an expansion in the small local curvature scale relative to the AdS scale, not an expansion in $T$ itself.

We therefore use the ansatz
\be
F(T,x)
=
\bar{F}
+
e^{-2T}F_\Lambda(x)
+
(p-p^\ast)e^{\omega_c T}F_c(x)
+
\cdots ,
\ee
\be
r(T,x)
=
\bar{r}
+
e^{-2T}r_\Lambda(x)
+
(p-p^\ast)e^{\omega_c T}r_c(x)
+
\cdots ,
\ee
with $\bar{F}=e^{2 \bar{\rho}}$, and
\be
f(T,x)
=
\sqrt{1-\frac{1}{2n}}
\left[
T+2 \ln{2}
+
e^{-2T}f_\Lambda(x)
+
(p-p^\ast)e^{\omega_c T}f_c(x)
+
\cdots
\right].
\ee
In the gauge where the correction vanishes at the light cone,
\be
F_\Lambda(x)
=
(2n)
\frac{
2^{-7+\frac{4}{n}}(1-8n)n^2
}{
(1-5n)(1-6n)
}
x,
\qquad
r_\Lambda(x)
=
\frac{
4^{\frac{1}{n}}n^2
}{
16(1-6n)
}
x,
\ee
and
\be
f_\Lambda(x)
=
-
\frac{
4^{\frac{1}{n}}n^3
}{
8(1-5n)(1-6n)
}
x.
\ee
For the top classical growing mode, the maximally differentiable exterior continuation gives
\be
F_c(x)
=
-2\bar{F}
\left[
\frac{C_b(1-n)+C_c n}{2n}
+
\frac{C_c(1-2n)}{4n}x
\right],
\qquad
r_c(x)
=
-\frac{C_c}{2}(1-x),
\ee
and\footnote{There appears to be a missing coefficient in the scalar perturbation written in~\cite{Jalmuzna:2015hoa}. For the maximally differentiable continuation, the constant $C_b$ term must be multiplied by $(n-1)/(2n-1)$ in order to satisfy the linearized scalar equation and the light cone constraint. We use this corrected scalar profile below. Since the correction is confined to the scalar perturbation, it does not affect the metric perturbations or the horizon tracing function.}
\be
f_c(x)
=
-
\left[
\frac{n-1}{2n-1}C_b
-
C_c
+
C_c x
\right],
\ee
where
\be
C_b
\equiv
\frac{\Gamma(\frac12-\omega_c)}
{\sqrt{\pi}\Gamma(1-\omega_c)},
\qquad
C_c
\equiv 
\frac{\Gamma(\omega_c-\frac12)}
{\sqrt{\pi}\Gamma(\omega_c)}.
\ee 
Note that the coefficient $c_2$ used in~\cite{Jalmuzna:2015hoa} is identified as
\be
c_2=-(p-p^\ast)
\ee
on both sides of the light cone.  We absorb this sign into the profiles, so that the physical amplitude in the horizon tracing function is $(p-p^\ast)$, with supercritical data corresponding to $(p-p^\ast)>0$. These profiles are the exterior continuation of the dominant interior growing perturbation.

The $\Lambda$ correction is subleading in the local self-similar interior, but it is indispensable in the exterior.  In the pure $\Lambda=0$ null continuation, the areal radius is independent of $v$, so every outgoing layer is marginal:
\be
\partial_v \bar{r}=0,
\qquad
\partial_u \bar{r}\neq0 .
\ee
The apparent horizon is therefore degenerate rather than a distinguished hypersurface.  The leading $\Lambda$ correction removes this degeneracy by eliminating the marginal surfaces of the background exterior.  The top classical growing mode can then restore a well-defined exterior apparent horizon for supercritical data, corresponding to $c_2<0$, or equivalently $(p-p^\ast)>0$ in our convention. This is the classical structure found in~\cite{Jalmuzna:2015hoa}, and we will return to its horizon tracing consequences in Section~\ref{sec3:horizontracing}.  In the next subsection we add the one-loop vacuum polarization state to this background and impose the semiclassical matching conditions across the light cone.

\subsection{Quantum backreaction beyond the light cone}
\label{sec3:quantumbackreaction}

We now compute the quantum stress tensor and its backreaction in the exterior.  The important simplification is that the background is more degenerate than the interior Garfinkle geometry.  For the local two-dimensional calculation, we strip off the overall length scale and work with dimensionless null coordinates. In this convention, the reduced exterior metric and dilaton are
\be
ds^2_{\rm ext}
=
-e^{2\bar{\rho}(u)}du\,dv,
\quad
e^{2\bar{\rho}(u)}
=
4n^2
\left[
\frac{(-u)^n}{2}
\right]^{4(1-\frac{1}{2n})},
\quad
\bar{\phi}
=
-\frac12
\ln
\left[
\frac{(-u)^{2n}}{2}
\right].
\ee
Since both $\bar{\rho}$ and $\bar{\phi}$ depend only on $u$, we have
\be
R=0,
\qquad
(\nabla \bar{\phi})^2=0,
\qquad
\Box \bar{\phi}=0.
\ee
Consequently the trace anomaly~\eqref{eq:traceanomaly} vanishes in the exterior,
\be
\langle T^a_{\ a}\rangle_{\rm ext}=0,
\ee
The sources in the auxiliary field equations~\eqref{eq:aux} vanish, and the auxiliary fields are purely homogeneous,
\be
\chi_1=C_1(v)+C_2(u),
\qquad
\chi_2=C_3(v)+C_4(u).
\ee
The functions $C_i$ encode the exterior choice of state.  They are not determined by the anomaly alone. They must be fixed by matching to the interior quantum state and by physical boundary conditions in the exterior.

The local matching at the light cone imposes the first constraint.  We use the same renormalization scheme and local counterterms as in the interior, and require that the renormalized stress tensor contain no delta function layer at $v=0$.  For the interior stress tensor in \eqref{eq:Garfinkle_Tuu_uv}--\eqref{eq:Garfinkle_Ttt_uv}, the light-cone limits are
\be
\langle T^{(3)}_{uu}\rangle_{\rm int}\big|_{v=0^-}
=
-\frac{n(2n-1)}{16\pi^2}
(1+8\ln2)
(-u)^{-2n-2},
\ee
while
\be
\langle T^{(3)}_{vv}\rangle_{\rm int}\to0,
\qquad
\langle T^{(3)}_{\theta\theta}\rangle_{\rm int}\to0
\qquad
(v\to0^-)
\ee
for $n\geq2$.  The case $n=1$ is too singular for this matching, since the $vv$ and angular components do not even have a continuous $C^0$ limit compatible with the exterior continuation.  For $n=2$ the stress tensor can be matched at $C^0$, but generically not at $C^1$.  For $n\geq3$, the matching can be made $C^1$ across the light cone.  We therefore restrict to $n\geq2$ below.

The $C^0$ matching of $\langle T_{uu}\rangle$ is nontrivial as it must reproduce the same $(-u)^{-2n-2}$ profile as the interior result and must not leave any dependence on the auxiliary field normalization.    This requirement fixes the $u$-dependent homogeneous modes to be constants, with vanishing derivatives, and gives
\be
C_2-C_4
=
\frac{1-5n+6(2n-1)\ln2}{24n\pi\lambda_2}.
\label{eq:Garfinkle_C2_C4_constraint}
\ee
Here $\lambda_2$ is a normalization parameter in the localized auxiliary field representation of the anomaly-induced action.  One may choose a basis in which the other coefficients in \eqref{eq:aux}, such as $\lambda_1,\mu_1,\mu_2$, are expressed in terms of $\lambda_2$. Since the original nonlocal effective action has no such free physical parameter, all final stress tensor components and backreacted metric functions must be independent of $\lambda_2$. 

After this local matching, the remaining state dependence is carried by the two functions of $v$, $C_1(v)$ and $C_3(v)$.  The exterior stress tensor still depends on these functions and their derivatives.  It is therefore useful to introduce
\be
\Delta(v)\equiv C_1(v)-C_3(v),
\qquad
\Sigma(v)\equiv C_1(v)+C_3(v).
\ee
In terms of these variables, the flux component depends on $\Delta$ as
\be
\langle T^{(3)}_{uu}\rangle_{\rm ext}
=
-\frac{n}{16\pi^2}(-u)^{-2n-2}
\left[
(2n-1)(1+8\ln2)-32n\pi\lambda_2\Delta(v)
\right],
\label{eq:Garfinkle_ext_Tuu_Delta}
\ee
while the other components involve $\Delta'$, $\Delta''$, $\Sigma'$, and $\Sigma''$.  Thus the local light cone matching fixes only the jet of these functions at $v=0$, not their global continuation.  Smooth matching of the nonzero $uu$ component, the vanishing of the angular component, and the absence of an artificial transverse kink require
\be
\Delta(0)=\Delta'(0)=\Delta''(0)=0.
\label{eq:Garfinkle_Delta_local_conditions}
\ee
The corresponding local constraints on $\Sigma$ are determined by the $vv$ component and by conservation.  These conditions ensure that the exterior state is the local continuation of the same Hadamard state selected in the interior, with no quantum surface layer at the light cone. However, they do not uniquely determine the global exterior state, as many choices of $\Delta(v)$ and $\Sigma(v)$ have the same light cone jets.  A further physical condition compatible with the semiclassical Einstein equations, is therefore required.

We impose that there be no persistent outgoing quantum flux at the far end of the null continuation.  This should not be interpreted as a Bondi condition at future null infinity, since the null continuation is not a standard Bondi exterior.  Rather, it is a boundary condition along the null direction of the exterior model, where the outgoing flux component should decay as one moves to large $v$ at fixed $u$.  Equivalently,
\be
\lim_{v\to\infty}
\langle T^{(3)}_{uu}\rangle_{\rm ext}
=
0.
\label{eq:far-end}
\ee
This fixes the asymptotic value of the state-dependent difference. We emphasize that this is a condition on the far end of this particular null continuation, not the assumption that the outgoing component vanishes throughout the exterior. Indeed, the resulting $\langle T^{(3)}_{uu}\rangle_{\rm ext}$ in~\eqref{eq:Garfinkle_ext_Tuu} is nonzero at finite $v$ and only decays as $v \to \infty$. The purpose of~\eqref{eq:far-end} is to exclude a persistent outgoing flux at the far boundary and to isolate a minimal continuation of the Boulware-like vacuum polarization state. It does not rule out particle creation in a different global state or in the later backreacted evolution.

It is convenient to write
\be
\Delta(v)=\frac{D(v)}{\lambda_2}.
\ee
The far end condition then requires
\be
D(\infty)
=
-\frac{(2n-1)(1+8\ln2)}{32n\pi}.
\ee
A simple analytic profile satisfying both the light cone junction conditions and the far end decay condition is
\be
D(v)
=
D_\infty
\left(1-e^{-\alpha v^3}\right),
\qquad
D_\infty
=
-\frac{(2n-1)(1+8\ln2)}{32n\pi},
\qquad
\alpha>0 .
\label{eq:Garfinkle_ext_D_profile}
\ee
The power $v^3$ is the minimal analytic choice that makes $D$, $D'$, and $D''$ vanish at the light cone.  The condition $\alpha>0$ is essential, as it ensures that the outgoing flux decays along the null continuation rather than approaching a nonzero value at the far end.\footnote{Other profiles are possible, but they are not arbitrary.  The local jets at $v=0$ must satisfy the junction conditions, the large $v$ limit must remove persistent outgoing radiation, and the remaining homogeneous data must be compatible with conservation and with the semiclassical Einstein equations.  Slower decaying or polynomial tail profiles can be made possible, but they introduce extra choices such as tail exponents and length scales.  The exponential profile is a minimal representative of the admissible class that is analytic at the light cone, rapidly localizes the state dressing along the null direction, gives uniform control of derivatives at large $v$, and keeps the final stress tensor independent of the auxiliary field normalization.}  This parameter only fixes the scale of the exterior null coordinate $v$, and by a normalization of $v$ we set $\alpha=1$ from now on.

The exterior stress tensor now takes the form
\bea
\langle T^{(3)}_{uu}\rangle_{\rm ext}
&=&
-\frac{n}{16\pi^2}
(-u)^{-2n-2}
\left[
(2n-1)(1+8\ln2)
+
32n\pi D(v)
\right],
\label{eq:Garfinkle_ext_Tuu_preSigma}
\\
\langle T^{(3)}_{vv}\rangle_{\rm ext}
&=&
-\frac{1}{\pi u^{2n}}
\left[
\frac{\Sigma'(v)D'(v)}{\lambda_2}
-
\frac{\Sigma''(v)}{8\pi\lambda_2}
+
\frac{D''(v)}{6}
\right],
\label{eq:Garfinkle_ext_Tvv_preSigma}
\\
\langle T^{(3)}_{uv}\rangle_{\rm ext}
&=&0,
\\
\langle T^{(3)}_{\theta\theta}\rangle_{\rm ext}
&=&
\frac{2^{1-\frac{2}{n}}}{n\pi}
u^{1-2n}D'(v).
\label{eq:Garfinkle_ext_Ttt_preSigma}
\eea
Thus $D(v)$ controls the outgoing flux sector, while $\Sigma(v)$ appears only in the $vv$ component. The latter is fixed by the remaining semiclassical Einstein constraint.

We now solve the metric backreaction.  We restore overall $\ell^2$ and use the ansatz
\be
ds^2_{\rm ext}
=
\ell^2
\left[
-
e^{2(\bar{\rho}+\frac{N \hbar}{\ell}A_q)}du\,dv
+
\left(\bar{r}+\frac{N \hbar}{\ell}r_q\right)^2d\theta^2
\right],
\label{eq:Garfinkle_ext_metric_q_uv}
\ee
and keep terms to $O(\hbar)$.  The semiclassical Einstein equations can be organized as
\bea
&&
\left[
(2n-1)
+
\frac{(1-2n)u}{n}\partial_u
+
\frac{u^2}{2n}\partial_u^2
\right] r_q(u,v)
-
u^{2n+1}\partial_u A_q(u,v)
\nonumber\\
&&\hspace{7cm}
=
\frac{n}{2\pi}D(v)
+
\frac{2n-1}{64\pi^2}(1+8\ln2),
\label{eq:Garfinkle_ext_uu_eq}
\\
&&
\partial_v^2 r_q(u,v)
=
\frac{1}{2\pi}
\left[
\frac{\Sigma'(v)D'(v)}{\lambda_2}
-
\frac{\Sigma''(v)}{8\pi\lambda_2}
+
\frac{D''(v)}{6}
\right],
\label{eq:Garfinkle_ext_vv_eq}
\\
&&
\partial_u\partial_v r_q(u,v)=0,
\label{eq:Garfinkle_ext_uv_eq}
\\
&&
\partial_u\partial_v A_q(u,v)
=
-\frac{n}{2\pi}u^{-2n-1}D'(v).
\label{eq:Garfinkle_ext_theta_eq}
\eea
These four equations are not independent.  The contracted Bianchi identity, together with the conservation of the renormalized stress tensor, implies that one component acts as a constraint once the remaining equations are solved.  We keep all four equations because this makes clear why $\Sigma(v)$ is not a freely specifiable state function.

We first solve the angular equation \eqref{eq:Garfinkle_ext_theta_eq}.  Its general solution is
\be
A_q(u,v)
=
\frac{D(v)}{4\pi}u^{-2n}
+
P(u)
+
Q(v).
\label{eq:Garfinkle_Aq_general}
\ee
The function $P(u)$ is fixed by matching to the interior quantum geometry at the light cone:
\be
P(u)
=
-\frac{(2n-1)(1-4\ln2)}{64n\pi^2}
(-u)^{-2n}.
\label{eq:Garfinkle_Pu}
\ee
The function $Q(v)$ is a residual homogeneous term.  Although it does not enter the $uu$ equation, it would correspond to an additional homogeneous distortion of the exterior quantum geometry not fixed by the matched quantum source.  We choose the smooth continuation compatible with the light cone matching and set $Q(v)=0$.

Next, we consider the areal-radius correction.  The $uv$ equation gives
\be
\partial_u\partial_v r_q=0
\qquad
\Longrightarrow
\qquad
r_q(u,v)=f(u)+g(v).
\ee
Substituting this form into the $uu$ equation shows that the $v$ dependence of the right-hand side is already fixed by $D(v)$ and by the solution for $A_q$.  The algebraic $r_q$ term then forbids an independent nontrivial $g(v)$. Equivalently, differentiating the resulting $uu$ equation with respect to $v$ gives $g'(v)=0$.  Thus the $v$-dependent part of $r_q$ is at most a constant. We have
\be
r_q(u,v)
=
\frac{3}{64\pi^2}
+
c_1(-u)^{2n-1}
+
c_2(-u)^{2n}
+
c_3.
\ee
Regularity as $u\to-\infty$ sets $c_1=c_2=0$.  Matching to the interior value at the light cone and excluding an additional constant radius shift set $c_3=0$.

We now return to the $vv$ equation.  Since $r_q$ is constant, the left-hand side of \eqref{eq:Garfinkle_ext_vv_eq} vanishes.  The $vv$ equation is therefore not an equation for a new metric function. It is the remaining constraint on the auxiliary field data:
\be
\frac{\Sigma'(v)D'(v)}{\lambda_2}
-
\frac{\Sigma''(v)}{8\pi\lambda_2}
+
\frac{D''(v)}{6}
=
0,
\ee
resulting in $\langle T^{(3)}_{vv}\rangle_{\rm ext}=0$. Once $D(v)$ has been fixed by local matching and the far-end condition, $\Sigma(v)$ is fixed by the $vv$ constraint required for the chosen stress tensor to solve the semiclassical Einstein equations, hence the explicit solution for $\Sigma(v)$ is not needed.

With the $vv$ constraint imposed, the exterior quantum stress tensor becomes
\bea
\langle T^{(3)}_{uu}\rangle_{\rm ext}
&=&
-\frac{n(2n-1)}{16\pi^2}
(1+8\ln2)
(-u)^{-2n-2}
e^{- v^3},
\label{eq:Garfinkle_ext_Tuu}
\\
\langle T^{(3)}_{vv}\rangle_{\rm ext}
&=&0,
\\
\langle T^{(3)}_{uv}\rangle_{\rm ext}
&=&0,
\label{eq:Garfinkle_ext_Tvv_Tuv}
\\
\langle T^{(3)}_{\theta\theta}\rangle_{\rm ext}
&=&
\frac{
3(2n-1)(1+8\ln2)
}{
32n^2\pi^2
}
2^{1-\frac{2}{n}}
(-u)^{1-2n}
v^2
e^{- v^3}.
\label{eq:Garfinkle_ext_Ttt}
\eea
The tensor is manifestly real for $u<0$ and positive integer $n$, vanishes in the flat limit $n\to\frac12$, and is covariantly conserved on the exterior background.  It has the required light cone limits,
\be
\lim_{v\to0^+}
\langle T^{(3)}_{uu}\rangle_{\rm ext}
=
\langle T^{(3)}_{uu}\rangle_{\rm int}\big|_{v=0^-},
\qquad
\lim_{v\to0^+}
\langle T^{(3)}_{\theta\theta}\rangle_{\rm ext}=0,
\ee
and it decays at the far end of the null continuation,
\be
\lim_{v\to\infty}
\langle T^{(3)}_{\mu\nu}\rangle_{\rm ext}
=
0
\qquad
(\text{at fixed }u<0).
\ee
All components also vanish as $u\to-\infty$.  As $u\to0^-$, the stress tensor grows.  This is not an imposed Hawking flux, but the exterior continuation of the same quantum state, and its growth reflects the RG flow toward the UV end of the self-similar region, now along the Cauchy-horizon-like boundary of the null continuation. The state remains Boulware-like in the same sense as in the interior.

In adapted coordinates $(T,x)$, the stress tensor becomes
\bea
\langle T^{(3)}_{TT}\rangle_{\rm ext}
&=&
-\frac{(2n-1)(1+8\ln2)}{64n\pi^2}
e^T
\exp\!\left(
x^3 e^{-\frac{3T}{2n}}
\right),
\label{eq:Garfinkle_ext_TTT}
\\
\langle T^{(3)}_{Tx}\rangle_{\rm ext}
&=&0,
\\
\langle T^{(3)}_{xx}\rangle_{\rm ext}
&=&0,
\label{eq:Garfinkle_ext_TTx_Txx}
\\
\langle T^{(3)}_{\theta\theta}\rangle_{\rm ext}
&=&
\frac{
3(2n-1)(1+8\ln2)
}{
n^2\pi^2
}
4^{-2-\frac{1}{n}}
x^2
e^{(1-\frac{3}{2n})T}
\exp\!\left(
x^3 e^{-\frac{3T}{2n}}
\right).
\label{eq:Garfinkle_ext_Ttt_Tx}
\eea
It is therefore not exactly CSS, since the state profile introduces the dimensionless combination $
v^3=-x^3e^{-\frac{3T}{2n}}$. This mild breaking of exact self-similarity is expected.  It implements the far-end condition that removes persistent outgoing radiation along the null continuation.

The resulting profiles for the semiclassical exterior geometry are
\be
A_q(u,v)
=
\frac{(2n-1)(-u)^{-2n}}{128n\pi^2}
\left[
(1+8\ln2)e^{-v^3}
-
3
\right],
\qquad
r_q(u,v)
=
\frac{3}{64\pi^2}.
\label{eq:Garfinkle_ext_backreaction_uv}
\ee
At $v=0$, this matches the light cone limit of the interior quantum geometry. Rewriting the metric in the adapted form
\be
ds^2_{\rm ext}
=
\ell^2e^{-2T}
\left[
F(T,x)
\left(dx-\frac{x}{2n}dT\right)dT
+
r^2(T,x)d\theta^2
\right],
\ee
we obtain
\be
F(T,x)
=
\bar{F}
+
\frac{N \hbar}{\ell}F_q(T,x),
\qquad
r(T,x)
=
\bar{r}
+
\frac{N\hbar}{\ell}r_q(T,x),
\ee
with
\be
F_q(T,x)
=
\frac{(2n-1)2^{\frac{2}{n}-9}}{\pi^2}
e^T
\left[
(1+8\ln2)
\exp\!\left(
x^3 e^{-\frac{3T}{2n}}
\right)
-
3
\right],
\quad
r_q(T,x)
=
\frac{3}{64\pi^2}e^T.
\label{eq:Garfinkle_ext_Fq_rq_Tx}
\ee
These expressions are real for real $T$ and $x\leq0$, have the same differentiability class across the light cone as the quantum stress tensor, and solve the semiclassical Einstein equations at $O(\hbar)$.  This analytic exterior quantum geometry is the input for the horizon tracing calculation in the next subsection.

\subsection{Horizon tracing and quantum trapped branch}
\label{sec3:horizontracing}

We now trace MOTS in the quantum corrected exterior geometry.  The horizon condition is the invariant equation $(\nabla r)^2=0$, where $r=\ell e^{-T}r(T,x)$ is the physical areal radius including perturbations. To make the expansion precise, define
\be
y\equiv e^{-2T},
\qquad
\eta\equiv p-p^\ast,
\qquad
\lambda\equiv \frac{N\hbar}{\ell}.
\ee
We therefore expand the horizon function to first order in the $\Lambda$ deformation, the top classical growing mode,\footnote{In fact, the classical lower growing modes found in the exterior do not contribute to the leading horizon function. Their exterior radius perturbations do not change the relevant derivative entering the MOTS condition, while the top mode does~\cite{Jalmuzna:2015hoa}.} and the one-loop source.  The calculation is not a raw $y\to0$ expansion at fixed $\eta$ and fixed $\lambda$, because the classical and quantum perturbations contain positive powers of $e^T$.  Instead, it is first-order perturbation theory in $\eta$ and $\lambda$, performed on an exterior background known to first order in $y$.  Therefore we keep the mixed terms $y\eta$ and $y\lambda$, where they describe how the leading $\Lambda$ correction to the exterior background modifies the linear classical and quantum perturbations.

With this convention, we decompose the horizon function as
\be
(\nabla r)^2
=
\bar{h}(T,x)
+
h_\Lambda(T,x)
+
h_c(T,x)
+
h_{\Lambda c}(T,x)
+
h_q(T,x)
+
h_{\Lambda q}(T,x).
\label{eq:Garfinkle_H_all_terms}
\ee
The subscripts indicate the origin of each contribution: $\Lambda$ the leading cosmological correction, $c$ the classical growing mode, and $q$ the one-loop quantum correction. Mixed subscripts such as $\Lambda c$ and $\Lambda q$ denote the corresponding linear perturbations evaluated on the $\Lambda$-deformed exterior background.  The full spatial profiles are lengthy and not illuminating, so we will only record the structural form needed for horizon tracing.

For the spacetime, one finds
\be
\bar{h}(T,x)=0,
\qquad
h_q(T,x)=0.
\label{eq:Garfinkle_f0_fq_zero}
\ee
The first equality is the horizon degeneracy of the $\Lambda=0$ null continuation discussed in the end of Section~\ref{sec3:exteriornull}. Equivalently, every outgoing null layer of the pure null continuation is marginal.  The leading $\Lambda$ correction $h_\Lambda (T,x)$ removes this degeneracy, while the classical growing mode and the quantum backreaction can restore a nontrivial MOTS branch. The second equality follows from the fact that the exterior quantum correction has constant radius profile at this order. Without the $\Lambda$ correction giving rise to $h_{\Lambda q}$, the quantum radius perturbation does not change $\partial_v  r$ and therefore does not contribute to $(\nabla r)^2$ at linear order. These are not drawbacks, but rather a structural feature of the $2+1$ dimensional exterior problem, indicating that $\Lambda$ correction is indispensable.

The nontrivial horizon function is therefore
\be
(\nabla r)^2
=
h_\Lambda(T,x)
+
h_c(T,x)
+
h_{\Lambda c}(T,x)
+
h_{\Lambda q}(T,x).
\label{eq:Garfinkle_H_effective}
\ee
Schematically, the corresponding terms scale with $T$ as
\be
h_\Lambda\sim e^{-2T},
\qquad
h_c\sim \eta e^{\omega_c T},
\qquad
h_{\Lambda c}\sim \eta e^{(\omega_c-2)T},
\qquad
h_{\Lambda q}\sim \lambda e^{-T}\,\mathcal Q(T,x),
\label{eq:Garfinkle_H_scaling}
\ee
with $\omega_c=1-\frac{1}{2n}$. The function $\mathcal Q(T,x)$ is a bounded profile in the near-light-cone, self-similar window. Away from this region it contains the exponential decay inherited from the exterior state profile.  In this exterior horizon equation, $h_\Lambda$ plays the role of the effective background term because $\bar{h}=0$.  Relative to this $\Lambda$-corrected background, multiplying the horizon equation by $e^{2T}$ gives the scalings
\be
e^{2T}h_\Lambda\sim O(1),
\quad
e^{2T}h_c\sim \eta e^{(\omega_c+2)T},
\quad
e^{2T}h_{\Lambda c}\sim \eta e^{\omega_c T},
\quad
e^{2T}h_{\Lambda q}\sim \lambda e^{T}\mathcal Q(T,x).
\label{eq:Garfinkle_effective_scalings}
\ee
Thus the effective exterior Lyapunov exponents differ from the interior analysis~\cite{Tomasevic:2025clf, Tomasevic:2025kqy}.  The classical top mode is stronger than the quantum mode, while the quantum piece is still a growing contribution relative to the $\Lambda$-corrected background.  This is the regime in which the exterior quantum branch should be understood: we are not sending $T$ arbitrarily large at fixed $\eta$ and $\lambda$, where one term would inevitably dominate and the perturbative expansion would fail.  Rather, we work in a near-critical window with moderate $T$ and small dressed amplitudes, where the retained terms can compete and the omitted nonlinear terms remain subleading.

\subsubsection*{Quantum horizon branch at the classical threshold $p=p^\ast$}

We first set $\eta \equiv p-p^\ast=0$.  The horizon function reduces to
\be
(\nabla r)^2_{\eta=0}
=
h_\Lambda(T,x)
+
h_{\Lambda q}(T,x)=0.
\label{eq:Garfinkle_H_eta0}
\ee
Thus the classical growing mode is absent, and any exterior MOTS branch is generated by the competition between the $\Lambda$-corrected background and the one-loop source.  This is the cleanest exterior analog of the interior quantum horizon mechanism.

The equation~\eqref{eq:Garfinkle_H_eta0} can be written in the compact form
\be
e^{-T}
=
\lambda \kappa_n
\left[
A\exp\left(
x^3 e^{-\frac{3T}{2n}}
\right)
-
3
\right],
\qquad
x\leq0,
\label{eq:Garfinkle_eta0_root_equation}
\ee
where $A\equiv 8\ln2+1$ and
$\kappa_n\equiv \frac{2n-1}{64\pi^2 n}$. Thus the location of the MOTS depends on the quantum source only through the ratio $\lambda=N\hbar/\ell$.  The separate values of $N$ and $\ell$ still matter for the curvature and source diagnostics discussed around~\eqref{eq:curvature_control} and \eqref{eq:source_curvature_control}, but the root equation itself only sees $\lambda$.

\begin{figure}[t]
    \centering
    \begin{minipage}{0.49\textwidth}
        \centering
        \includegraphics[width=\linewidth]{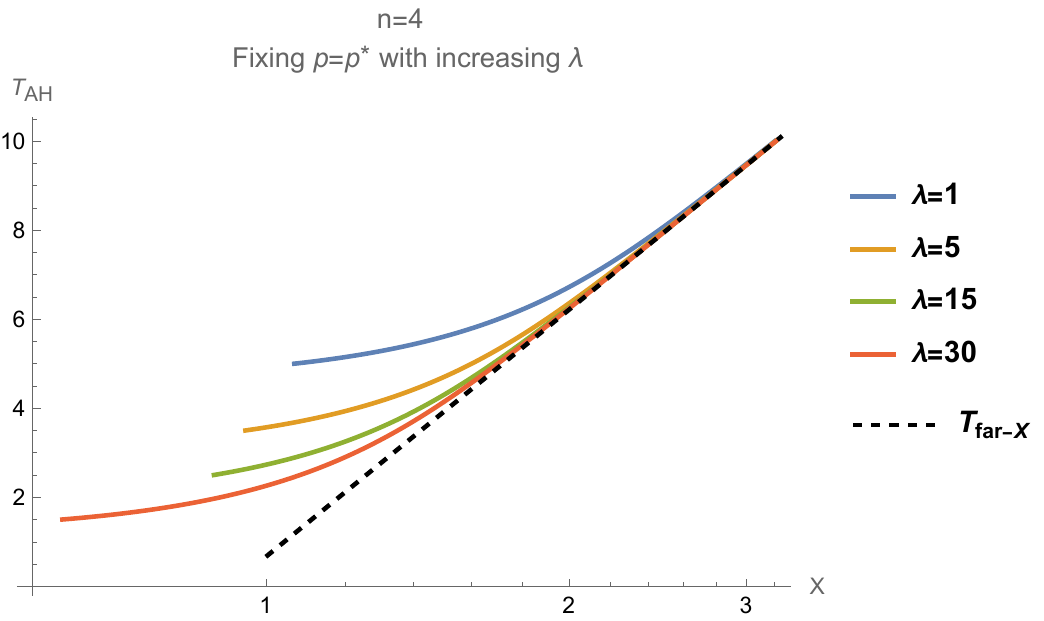}
    \end{minipage}
    \begin{minipage}{0.49\textwidth}
        \centering
        \includegraphics[width=\linewidth]{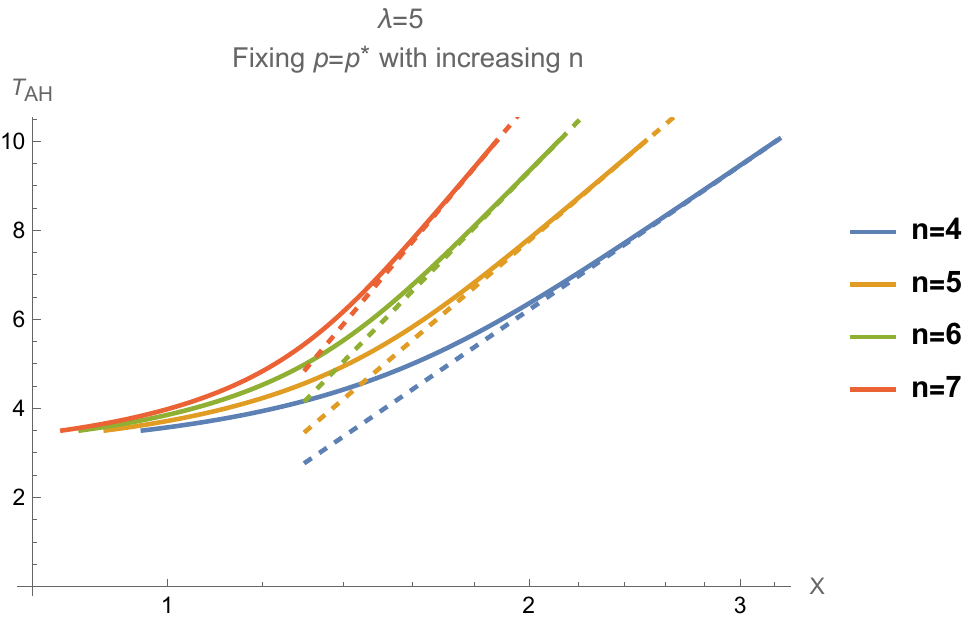}
    \end{minipage}
    \caption{Exterior MOTS branch at the classical threshold $p=p^\ast$, plotted as $T_{\rm AH}(X)$ with $X\equiv |x|=-x$ on a logarithmic scale.  The left panel fixes $n=4$ and varies $\lambda=N\hbar/\ell$: increasing $\lambda$ shifts the moderate-$X$ branch to smaller $T_{\rm AH}$, while all curves approach the same far-region asymptote \eqref{eq:Garfinkle_Tfar}, shown as a dashed line.  The right panel fixes $\lambda=5$ and varies $n$, showing that the existence of the exterior quantum MOTS branch is robust across the Garfinkle family.
}
    
    \label{fig:Garfinkle_critical_TAH}
\end{figure}

It is useful to analyze \eqref{eq:Garfinkle_eta0_root_equation} in two regimes.  In the moderate-$X$ regime, with
\be
X\equiv |x|,
\qquad
X^3 e^{-\frac{3T}{2n}}\ll1,
\label{eq:Garfinkle_moderate_condition}
\ee
we expand $\exp\left(
x^3 e^{-\frac{3T}{2n}}
\right)
=
\exp\left(
-X^3 e^{-\frac{3T}{2n}}
\right)
\simeq
1
-
X^3 e^{-\frac{3T}{2n}}$. The root equation then gives
\be
T_{\rm AH}^{\rm mod}
\simeq
\ln
\frac{1}{\lambda\kappa_n(A-3)}
+
\frac{A}{A-3}
X^3 e^{-\frac{3T_{\rm AH}}{2n}}
+
\cdots,
\label{eq:Garfinkle_Tmod}
\ee
for the apparent horizon formation. Since $A-3=8\ln2-2>0$, the first term gives the dominant location,
\be
T_{\rm AH}^{\rm mod}
\simeq
\ln
\frac{1}{\lambda\kappa_n(8\ln2-2)} .
\ee
The next term shows that $T_{\rm AH}$ increases mildly as one moves outward in $X$, as long as the moderate-$X$ condition \eqref{eq:Garfinkle_moderate_condition} remains valid.  Increasing $\lambda$ therefore moves the quantum exterior horizon to smaller $T$, as expected for a stronger one-loop source. This is shown in the left panel of Figure~\ref{fig:Garfinkle_critical_TAH}.

In the far-$X$ regime, the exponential profile in the exterior becomes important. We write
\be
s
\equiv
X^3e^{-\frac{3T}{2n}}
\label{eq:Garfinkle_s_definition} \implies e^{-T}
=
X^{-2n}s^{\frac{2n}{3}},
\ee
and the root equation becomes
\be
X^{-2n}s^{\frac{2n}{3}}
=
\lambda\kappa_n
\left[
Ae^{-s}-3
\right].
\label{eq:Garfinkle_far_root_s}
\ee
The right-hand side must be positive, so a root requires
\be
Ae^{-s}>3
\qquad
\Longleftrightarrow
\qquad
s<s_\ast,
\qquad
s_\ast\equiv \ln\frac{A}{3}.
\label{eq:Garfinkle_sstar}
\ee
Equivalently,
\be
T
>
2n\ln X
-
\frac{2n}{3}\ln s_\ast .
\ee
For large $X$, the left-hand side of \eqref{eq:Garfinkle_far_root_s} is small, so the bracket on the right-hand side must also become small.  Hence $s$ approaches $s_\ast$ from below.  The far-exterior root is therefore
\be
T_{\rm AH}^{\rm far}
=
2n\ln X
-
\frac{2n}{3}
\ln s_\ast
+
O\left(\lambda^{-1}X^{-2n}\right),
\label{eq:Garfinkle_Tfar}
\ee
The leading far-$X$ behavior is independent of $\lambda$.  The role of $\lambda$ is instead to determine where the curve leaves the moderate-$X$ regime and approaches this common asymptotic line. This behavior is visible in Figure~\ref{fig:Garfinkle_critical_TAH}: the curves separate at moderate $X$ but collapse onto the same dashed curve in the far region.

\begin{figure}[t]
    \centering
    \includegraphics[width=0.65\textwidth]{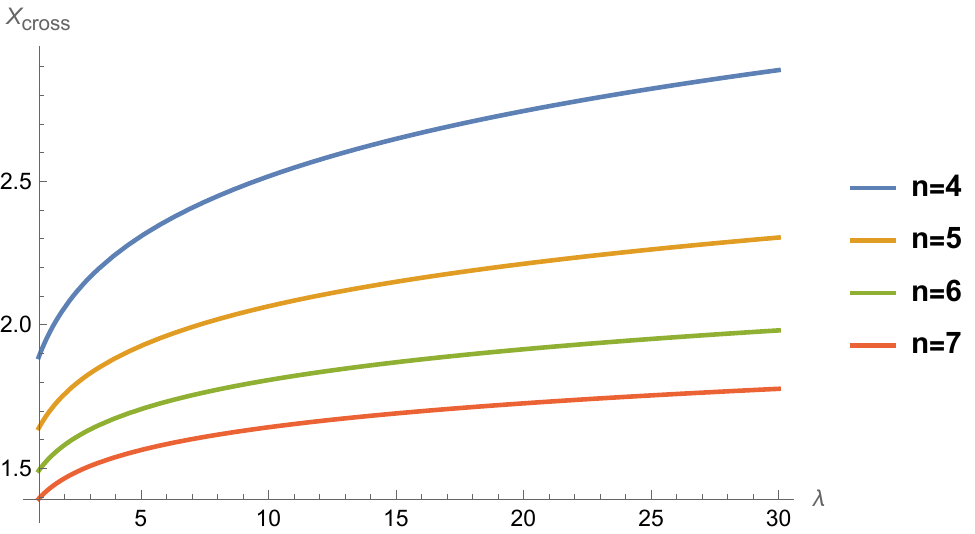}
    \caption{
    Crossover scale between the moderate-$X$ and far-$X$ regimes of the exterior MOTS branch at $p=p^\ast$. Increasing $\lambda$ lowers the moderate regime value of $T_{\rm AH}$ and therefore pushes the onset of the far-$X$ regime closer to the light cone.  The mild dependence on $n$ shows that the existence of a far-region branch is a structural feature of the analytic exterior construction.}
    \label{fig:Garfinkle_Xcross}
\end{figure}

\begin{figure}[t]
\centering
\begin{minipage}{0.48\textwidth}
\centering
\includegraphics[width=\linewidth]{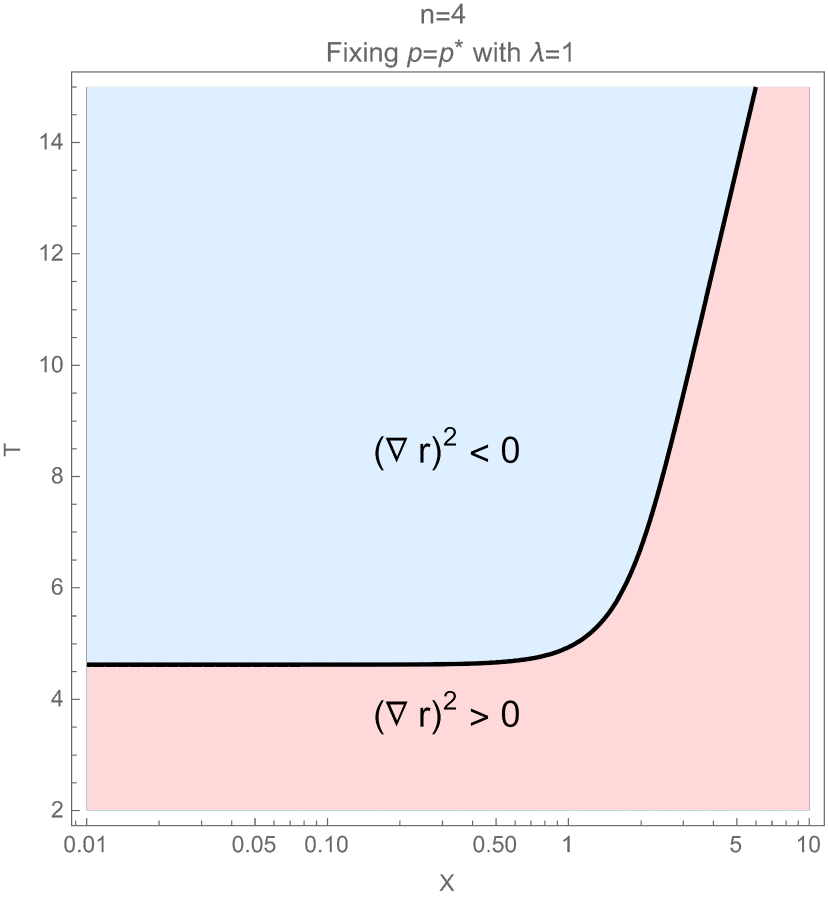}
\end{minipage}
\hfill
\begin{minipage}{0.48\textwidth}
\centering
\includegraphics[width=\linewidth]{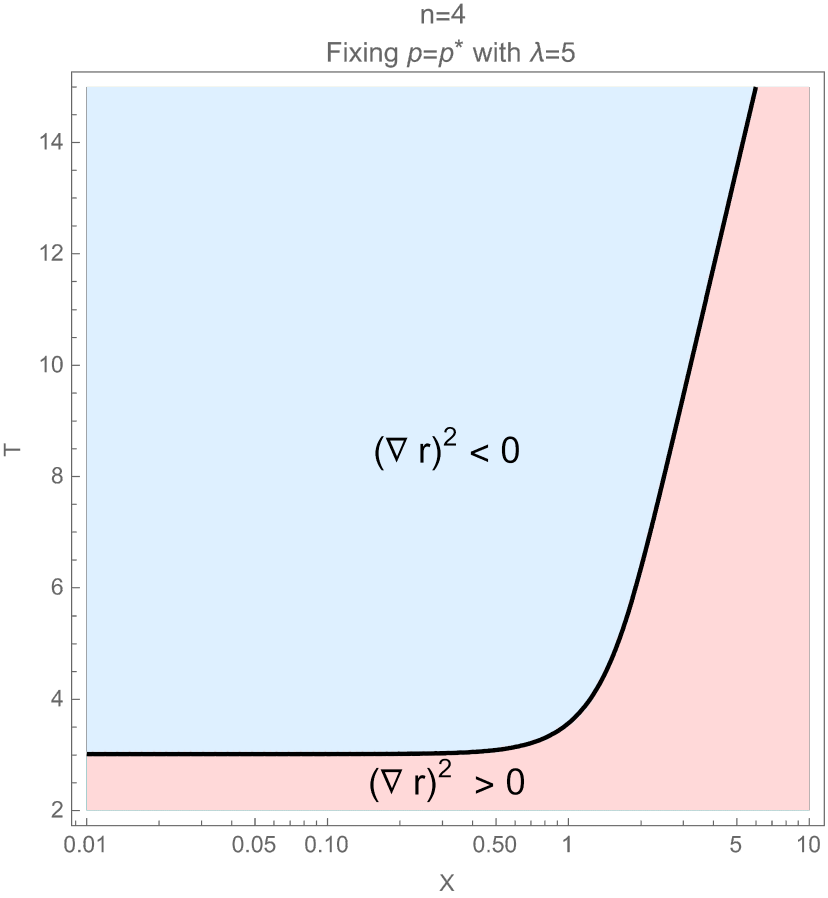}
\end{minipage}
\caption{
 Sign structure of the exterior horizon function at $p=p^\ast$ for $n=4$ at $\lambda=1$ and $\lambda=5$.  We plot the sign of $(\nabla r)^2$, with the black curve being the zero contour $(\nabla r)^2=0$, which gives the exterior MOTS branch.  The blue region has $(\nabla r)^2<0$ and is trapped, while the red region has $(\nabla r)^2>0$ and is untrapped.  At fixed $X$, the black contour gives the root $T_{\rm AH}(X)$; at fixed $T$, it gives the boundary in $X$ between trapped and untrapped regions.  Comparing $\lambda=1$ and $\lambda=5$ shows that increasing the one-loop strength shifts the moderate-$X$ MOTS branch to smaller $T$, in agreement with the analytic estimate \eqref{eq:Garfinkle_Tmod}, while preserving the same qualitative trapped region structure.
}
\label{fig:Garfinkle_signplot_critical}
\end{figure}

A rough estimate of the crossover between the two regimes is obtained by evaluating the far-regime variable $s$ at the moderate-regime root.  Let
\be
T_0
\equiv
\ln
\frac{1}{\lambda\kappa_n(A-3)} .
\ee
Then the crossover scale satisfies
\be
X_{\rm cross}^3
e^{-\frac{3T_0}{2n}}
\sim
s_\ast  \implies
X_{\rm cross}
\sim
s_\ast^{1/3}
\exp\left(\frac{T_0}{2n}\right).
\label{eq:Garfinkle_Xcross}
\ee
Thus larger $\lambda$ decreases $T_0$ and pushes the onset of the far-$X$ regime closer to the light cone, as illustrated in Figure~\ref{fig:Garfinkle_Xcross}.

Furthermore, Figure~\ref{fig:Garfinkle_signplot_critical} shows the sign of the full horizon function $(\nabla r)^2$ and overlays the zero contour.  At fixed $X$, the contour gives the same root $T_{\rm AH}(X)$ discussed above.  At fixed $T$, it instead identifies the interval in $X$ that lies inside the trapped region.  The two viewpoints are therefore consistent, where the MOTS curve is both the graph of $T_{\rm AH}(X)$ and the boundary between trapped and untrapped regions in the exterior patch.

The same equation also clarifies the relation to the dressed one-loop amplitude $\sim \lambda e^T$.  At the root,
\be
\lambda e^{T_{\rm AH}}
=
\frac{
1
}{
\kappa_n
\left[
A\exp\left(
x^3e^{-\frac{3T_{\rm AH}}{2n}}
\right)
-
3
\right]
}.
\label{eq:Garfinkle_dressed_amplitude_at_root}
\ee
In the moderate-$X$ regime, this quantity depends mainly on the constant $A-3=8\ln2-2$.  In the far-$X$ regime, where $T_{\rm AH}\sim 2n\ln X$, it grows as
\be
\lambda e^{T_{\rm AH}}
\sim
\lambda X^{2n}.
\ee
This growth does not by itself signal a breakdown of the horizon calculation.  The horizon equation is sensitive to the cancellation between the $\Lambda$-corrected background and the quantum source, and a large value of a coordinate-dependent dressed amplitude is not a gauge-invariant control criterion.  The appropriate checks are instead the invariant curvature scale $\epsilon_{\rm curv}$, the source-to-curvature ratio $\epsilon_{\rm src}$ defined in \eqref{eq:curvature_control} and \eqref{eq:source_curvature_control}, and the branch condition discussed below.

We should note the distinction between the ratio $\lambda$ that controls the MOTS equation and the individual scales $N$ and $\ell$ that control semiclassical validity in Planck units.  At $\eta=0$, the root equation depends only on $\lambda$. The control diagnostics, however, distinguish $N$ and $\ell$ separately. For fixed $\lambda$, increasing $N$ and $\ell$ together leaves the MOTS equation unchanged, but suppresses the physical curvature scale carried by the metric.  Schematically, we have
$\epsilon_{\rm curv}
\sim
\frac{\lambda}{\ell^2}$, whereas $\epsilon_{\rm src}$ is controlled mainly by the relative size of the quantum source compared with the curvature of the same truncated geometry.  Thus one may first choose $\lambda$ to obtain an admissible exterior MOTS and to keep $\epsilon_{\rm src}$ small, and then take the large-$\ell$, large-$N$ limit at fixed $\lambda$ to make $\epsilon_{\rm curv}$ small.  This separates the existence of the quantum MOTS, controlled by $\lambda$, from EFT control of the geometry, improved in the large-$\ell$, large-$N$ semiclassical regime, as in the interior analysis~\cite{Tomasevic:2025clf}.

\subsubsection*{Near-critical detuning and the quantum-shifted threshold}

We next turn on a small classical detuning $\eta=p-p^\ast$.  The full horizon equation is then
\be
(\nabla r)^2
=
h_\Lambda
+
h_c
+
h_{\Lambda c}
+
h_{\Lambda q}
=
0.
\label{eq:Garfinkle_H_eta}
\ee
The sign of $\eta$ has the usual critical collapse interpretation: positive $\eta$ corresponds to the supercritical side and helps form trapped surfaces, while negative $\eta$ corresponds to the subcritical side and works against horizon formation.  In the semiclassical problem, however, this classical language must be used with care. The interior analysis of~\cite{Tomasevic:2025clf,Tomasevic:2025kqy} showed that the one-loop growing mode shifts the threshold. Thus the relevant question is not whether the classical parameter is positive or negative by itself, but whether the combined classical and quantum perturbations produce a trapped branch.

\begin{figure}[t]
    \centering
    \includegraphics[width=0.68\textwidth]{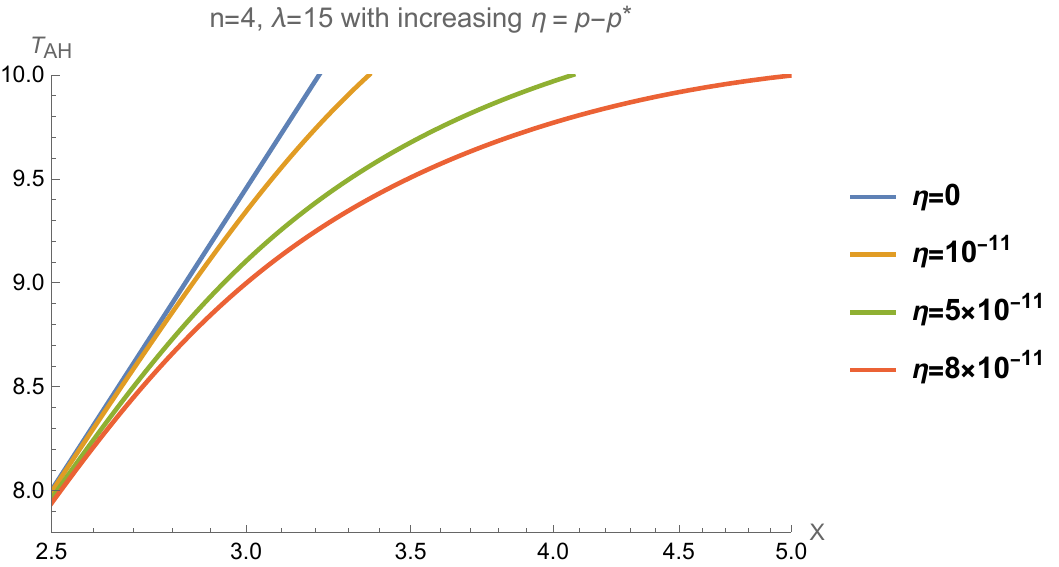}
    \caption{
    Exterior MOTS branch for positive classical detuning $\eta=p-p^\ast\ge 0$, where we consider very small such detuning since the effective classical mode is stronger.  The $\eta=0$ curve is included for comparison.  As $\eta$ increases on the supercritical side, the MOTS curve shifts to smaller $T_{\rm AH}$ at fixed $X$, showing that trapping occurs earlier.  Equivalently, on a fixed $T$ slice the trapped region extends to larger $X$.  This is the expected behavior when the classical growing mode reinforces the quantum tendency toward horizon formation.
    }
    \label{fig:Garfinkle_supercritical_eta}
\end{figure}

For $\eta>0$, the exterior horizon structure is less quantum driven.  The one-loop mode already enhances horizon formation at $\eta=0$, so a positive classical detuning does not compete with the quantum effect but reinforces it.  In this regime the MOTS branch should be viewed as the supercritical classical branch, smoothly deformed by the one-loop correction.  As shown in Figure~\ref{fig:Garfinkle_supercritical_eta}, increasing $\eta>0$ shifts the branch to smaller $T_{\rm AH}$ at fixed $X$, meaning that trapping occurs earlier at that exterior location; equivalently, on a fixed $T$ slice the trapped region extends to larger $X$.  This provides an important consistency check: the quantum-supported branch found at $\eta=0$ connects smoothly to the ordinary supercritical exterior horizon and is not an isolated feature.

The cleanest test of the quantum shielding mechanism is obtained on the classically subcritical side, $\eta<0$.  Classically, negative detuning pushes the solution away from black hole formation.  Quantum mechanically, however, the vacuum polarization can compensate the subcritical classical perturbation over a finite range of negative $\eta$.  In this range the exterior still contains a controlled MOTS branch, even though the classical perturbation alone would not produce one.  This is the exterior analog of the quantum-shifted threshold found in the interior analysis~\cite{Tomasevic:2025kqy,Tomasevic:2025clf}, illustrated schematically in Figure~\ref{fig:phase_quantum}.  The relevant threshold is therefore no longer the classical value $\eta=0$, but a shifted value $\eta_q^\ast<0$, which separates horizon-forming data from data with no controlled exterior trapped branch.  Operationally, we define
\be
\eta_q^\ast(n,\lambda)
\equiv
\inf\left\{
\eta<0\;:\;\text{a controlled exterior MOTS exists at $(T_{\rm AH}, x_{\rm AH})$}
\right\}.
\label{eq:Garfinkle_etaq_def}
\ee
For $\eta_q^\ast\leq\eta<0$, the exterior contains a quantum-supported MOTS branch. For $\eta<\eta_q^\ast$, no controlled exterior MOTS is found.  What is shown here is that, along the connected branch traced in this construction, the first exterior MOTS approaches a finite size as $\eta$ approaches $\eta_q^\ast$ from the horizon-forming side, rather than shrinking continuously to zero, while the other side has no exterior horizon. \eqref{eq:Garfinkle_etaq_def} therefore realizes the possibility that the quantum-shifted threshold lies on the classically subcritical side. We do not attempt to classify every possible one-loop critical solution or to exclude an additional branch outside the perturbative construction.

\begin{figure}[t]
    \centering
    \begin{minipage}{0.49\textwidth}
        \centering
        \includegraphics[width=\linewidth]{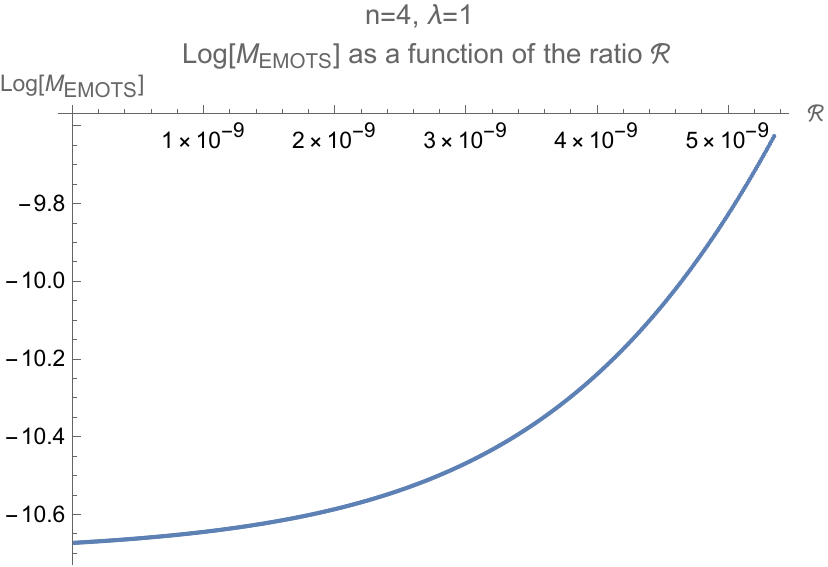}
    \end{minipage}
    \hfill
    \begin{minipage}{0.49\textwidth}
        \centering
        \includegraphics[width=\linewidth]{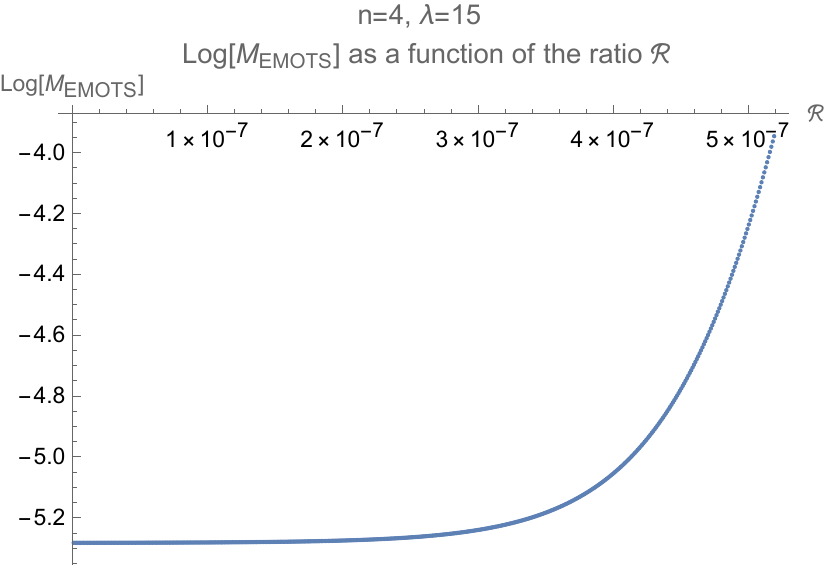}
    \end{minipage}

    \vspace{0.35cm}

    \begin{minipage}{0.52\textwidth}
        \centering
        \includegraphics[width=\linewidth]{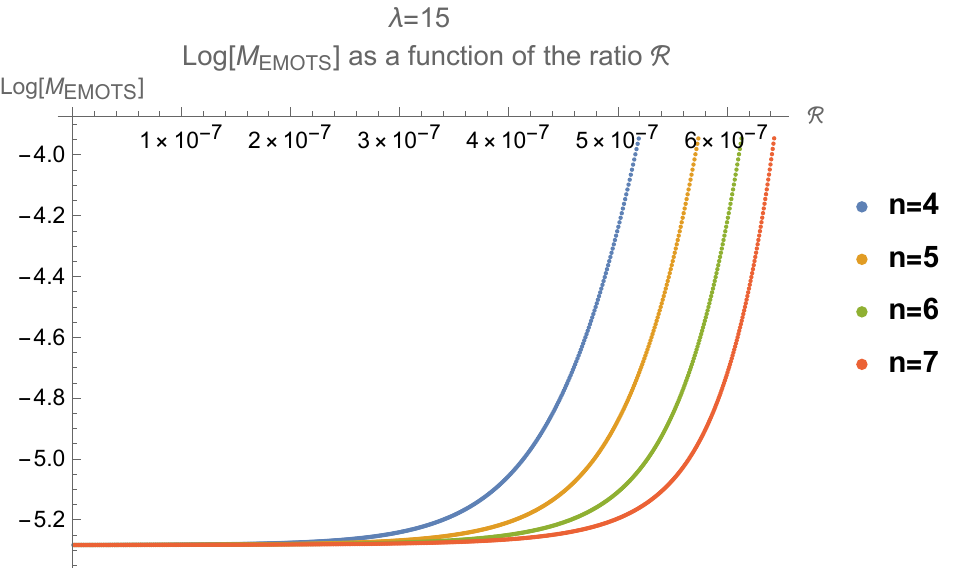}
    \end{minipage}

    \caption{
    Exterior mass gap diagnostic on the classically subcritical side.  The top-left and top-right panels show $\log M_{\rm EMOTS}$ as a function of the branch parameter $\mathcal R$ for the same $n=4$ Garfinkle exterior, with $\lambda=1$ and $\lambda=15$, respectively.  The lower panel fixes $\lambda=15$ and compares several values of $n$.  In all cases, approaching the shifted threshold from the horizon-forming side corresponds to $\mathcal R\to0^+$, and the exterior mass approaches a finite plateau rather than running to zero.  Increasing $\lambda$ raises the quantum-supported exterior mass scale, indicating that the black hole mass need not be microscopic, while changing $n$ modifies the numerical value and crossover behavior but preserves the finite-gap structure.
    }
    \label{fig:Garfinkle_subcritical_massgap}
\end{figure}

To display the approach to the shifted threshold, we parametrize the exterior branch by the ratio between the dressed classical and quantum amplitudes evaluated on the horizon. Since $\Lambda$ plays the role of the effective background, the competition relevant for the shifted threshold is therefore only between the classical and quantum modes.  We define
\be
\mathcal R
\equiv
\left.
\frac{
(\eta-\eta_q^\ast)e^{\omega_c T}
}{
\lambda e^{T}
}
\right|_{\rm EMOTS}.
\label{eq:Garfinkle_ratio_R}
\ee
Note that $\mathcal R$ is not a bare control parameter, it is a branch parameter evaluated at the exterior marginal surface, whose location $(T_{\rm AH},x_{\rm AH})$ changes as $\eta$ is varied.  This distinction is important.  The mass function plotted is not the pointwise function $M(T,x,\eta)$ at fixed coordinates, but the Hawking mass evaluated on that controlled exterior branch,
\be
M_{\rm EMOTS}(\eta)
\equiv
M_{\rm H}^{(3)}
\big(
T_{\rm AH}(\eta),x_{\rm AH}(\eta)
\big),
\label{eq:Garfinkle_MEMOTS_def}
\ee
where we use \eqref{eq:Hawking_mass_3d} on the marginal surface.   As $\mathcal R\to0^+$, the exterior mass approaches a finite plateau rather than decreasing to zero, as shown in Figure~\ref{fig:Garfinkle_subcritical_massgap}.  For $\mathcal R<0$, equivalently $\eta<\eta_q^\ast$, no controlled exterior MOTS is found.  This is the same qualitative hallmark as a Type I threshold: one either disperses, or forms a finite object.  In the present context this finite jump removes the zero-mass naked endpoint and therefore the associated Cauchy-horizon interpretation.

The analytic Garfinkle example therefore provides a sharp demonstration of the shielding mechanism.  Smooth matching and absence of artificial quantum flux select a vacuum polarization state, and the semiclassical Einstein equations turn this state into a quantum trapped region in the classically naked exterior.  Some details are model dependent, especially the special $\Lambda=0$ degeneracy of the null continuation and the precise coefficients of the branch.  The robust lesson is nevertheless that the interior quantum horizon is not lost beyond the past light cone.  The next section tests the same mechanism in the $3+1$ Einstein-scalar problem, where the horizon tracing must be performed numerically.

\section{Quantum critical exterior spacetime in $3+1$ dimensions}
\label{sec:3+1d_Roberts}

We now turn to the $3+1$ dimensional Einstein-scalar problem with Roberts spacetimes, where the exterior construction is more subtle. Unlike the analytic Garfinkle construction, the Roberts model does not admit a closed-form scalar field exterior. We therefore use the Roberts interior as an analytic seed and construct a shell-free exterior strip, where the persistence of the quantum trapped branch is tested numerically. This section thus provides the four-dimensional counterpart of the shielding mechanism.

\subsection{The Roberts interior and its quantum horizon}
\label{sec4:Robertsinterior}

The true $3+1$-dimensional Choptuik critical solution of the spherically symmetric Einstein-scalar system is DSS and is not known in closed analytic form~\cite{Choptuik:1992jv}. The Roberts solution provides a useful analytic laboratory: it is a closed-form CSS solution of the same Einstein-scalar system, supported by a massless scalar field, and captures several structural features of near-critical scalar collapse~\cite{Roberts:1989sk,Brady:1994xfa,Oshiro:1994hd,Brady:1994aq}. It should not be identified with the true Choptuik spacetime, nor used as a quantitative approximation to its geometry or critical exponent. Rather, its role is to provide an analytic CSS background whose perturbations probe how scale-invariant collapse can depart toward DSS critical behavior. As reviewed in Section~5.1 of~\cite{Tomasevic:2025clf}, Roberts admits complex growing modes with oscillatory structure~\cite{Frolov:1997uu,Frolov:1998tq}, while generic growing perturbations drive the solution away from CSS toward the DSS critical solution~\cite{Frolov:1999fv}. In this sense, the Roberts solution provides a qualitative analytic link between the breaking of CSS and the emergence of DSS behavior.

The purpose of the present section is correspondingly more structural. Roberts provides an exact model of scalar collapse and cosmic censorship in which we can test whether the semiclassical mechanism found in the $2+1$-dimensional case survives in four-dimensional Einstein-scalar gravity. In particular, the present construction asks whether the quantum source can be consistently propagated through a genuine scalar-supported, shell-free exterior, rather than through a metric-only cut-and-paste completion. This makes the Roberts spacetime a useful intermediate step toward the considerably more ambitious problem of semiclassical backreaction in the true $3+1$-dimensional DSS Choptuik spacetime.

The Roberts solution is a two-parameter CSS family. In double-null coordinates $(u,v)$, with $u\in(-\infty,0]$ and $v\in[0,\infty)$, it takes the form
\be
ds^2_{\rm int}
=
-2dudv+r^2(u,v)d\Omega_2^2,
\qquad
r^2(u,v)
=
-\alpha v^2+\beta u^2-uv .
\label{eq:Roberts_general_metric}
\ee
The scalar field is
\be
f
=
\frac{\sqrt{2}}{2}
\ln
\left[
-
\frac{
2\alpha v+u(1-\sqrt{1+4\alpha\beta})
}{
2\alpha v+u(1+\sqrt{1+4\alpha\beta})
}
\right]
+
f_0(\alpha,\beta),
\label{eq:Roberts_general_scalar}
\ee
where $f_0(\alpha,\beta)$ is a normalization constant chosen so that $f_{v=0}=0$, allowing the region $v<0$ to be smoothly matched to Minkowski space. We will restrict to $\beta=1$, where we take
\be
f_0(\alpha)=-\frac{\sqrt{2}}{2} \ln{\bigg[\frac{\sqrt{1+4 \alpha}-1}{\sqrt{1+4 \alpha}+1} \bigg]},
\ee
The critical spacetime corresponds to further taking $\alpha \to 0$, resulting in the following background profiles
\be
\bar{r}^2=u^2-uv,
\qquad
\bar{f}=\frac{\sqrt{2}}{2}\ln\left(1-\frac{v}{u}\right).
\label{eq:Roberts_critical_uv}
\ee
The scalar field is switched on at $v=0$, and the critical Roberts region describes the local implosion of scalar radiation from past null infinity.

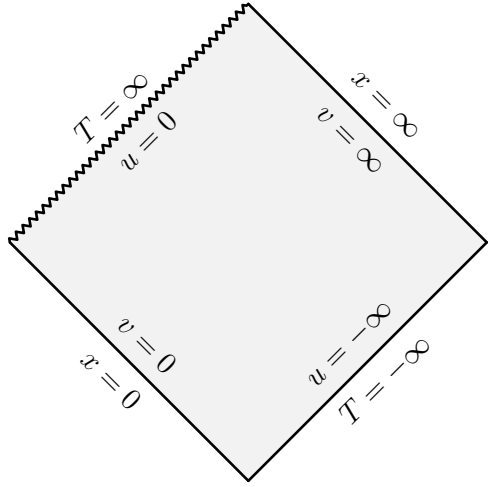
\begin{figure}[t]
\centering
\resizebox{0.42\textwidth}{!}{%
\begin{tikzpicture}[
  line cap=round,
  line join=round,
  every node/.style={inner sep=1.5pt},
  boundary/.style={draw=black, line width=1.25pt},
  singularity/.style={
    draw=black,
    line width=1.25pt,
    decorate,
    decoration={zigzag, segment length=4.5pt, amplitude=1.6pt}
  }
]

\coordinate (L) at (0,0);    
\coordinate (U) at (4,4);    
\coordinate (R) at (8,0);    
\coordinate (D) at (4,-4);   

\coordinate (mLU) at ($(L)!0.5!(U)$);
\coordinate (mUR) at ($(U)!0.5!(R)$);
\coordinate (mDR) at ($(D)!0.5!(R)$);
\coordinate (mLD) at ($(L)!0.5!(D)$);

\def\din{0.3}
\def\dout{0.3}

\fill[gray!10] (L) -- (U) -- (R) -- (D) -- cycle;

\draw[singularity] (L) -- (U);   
\draw[boundary]    (U) -- (R);
\draw[boundary]    (R) -- (D);
\draw[boundary]    (D) -- (L);


\node[font=\Large, rotate=45]
  at ($(mLU)+(\din,-\din)$)
  {$u=0$};

\node[font=\Large, rotate=-45]
  at ($(mUR)+(-\din,-\din)$)
  {$v=\infty$};

\node[font=\Large, rotate=45]
  at ($(mDR)+(-\din,\din)$)
  {$u=-\infty$};

\node[font=\Large, rotate=-45]
  at ($(mLD)+(\din,\din)$)
  {$v=0$};


\node[font=\Large, rotate=45]
  at ($(mLU)+(-\dout,\dout)$)
  {$T=\infty$};

\node[font=\Large, rotate=-45]
  at ($(mUR)+(\dout,\dout)$)
  {$x=\infty$};

\node[font=\Large, rotate=45]
  at ($(mDR)+(\dout,-\dout)$)
  {$T=-\infty$};

\node[font=\Large, rotate=-45]
  at ($(mLD)+(-\dout,-\dout)$)
  {$x=0$};


\end{tikzpicture}%
}
\caption{
Penrose diagram of the Roberts interior region, showing both the null coordinates
$(u,v)$ and the adapted coordinates $(T,x)$. The upper-left null boundary is a
curvature singularity and is drawn as a jagged line.
}
\label{fig:RobertsInteriorOnly}
\end{figure}

It is again useful to introduce adapted self-similar coordinates
\be
T=-\ln(-u),
\qquad
x=\frac12\ln\left(1-\frac{v}{u}\right),
\qquad
u=-e^{-T},
\qquad
v=e^{-T}(e^{2x}-1).
\label{eq:Roberts_Tx_def}
\ee
In these coordinates, the critical solution becomes
\be
ds^2_{\rm int}
=
2e^{-2T}e^{2x}
\left[
(1-e^{-2x})dT^2-2dTdx+\frac{1}{2}d\Omega_2^2
\right],
\qquad
\bar{f}=\sqrt{2}x,
\qquad
x\in[0,\infty).
\label{eq:Roberts_critical_Tx}
\ee
The physical areal radius is $\bar r=e^{-T}e^x$. The critical Roberts singularity is null: in double-null coordinates it lies at $u=0$ with $v>0$, and extends all the way to future null infinity. At fixed finite $x$, taking $T\to\infty$ approaches the endpoint $u=v=0$ of this null singularity; reaching a point with finite $v>0$ on the singular line requires taking $x\to\infty$ together with $T\to\infty$. See Figure~\ref{fig:RobertsInteriorOnly} for a Penrose diagram of the Roberts interior. This causal structure is different from a pointlike naked singularity, and is one of the reasons Roberts is not itself the true Choptuik critical spacetime. The spacetime has no marginal surface in the unperturbed background, as the horizon tracing function remains positive.

The classical Roberts solution also has important limitations as a global spacetime. At first sight, fixing $\beta=1$ and varying $\alpha$ from $\alpha<0$ to $\alpha>0$ appears to give an analytic one-parameter model of dispersal, criticality, and black hole formation. This interpretation turns out to be incorrect. As reviewed in~\cite{Tomasevic:2025clf}, the perturbation spectrum shows that Roberts is not the true intermediate attractor of scalar critical collapse. Furthermore, in the supercritical branch $\alpha>0$, the black hole mass grows without bound as $v\to\infty$, and the global CSS continuation is not asymptotically flat. A typical way to address such problems is to cut the self-similar spacetime at a finite null surface and glue it to an exterior region. 

Such a construction was attempted in~\cite{Wang:1996xh}, where the Roberts region was matched to an outgoing Vaidya exterior. However, that exterior is supported by null dust rather than by the original scalar field, and issues about having a shell at the junction must be treated carefully. For our purposes, the desired analog would be a scalar field supported, shell-free, sufficiently smooth exterior continuation, ideally with weak-field or asymptotically flat behavior. As we will discuss in Section~\ref{sec4:exteriorstrip}, since there is no genuine $3+1$ dimensional pure scalar analog of outgoing Vaidya, the Roberts exterior problem is structurally different from the Garfinkle null continuation and cannot be treated by the same analytic approach. The exterior must be constructed instead as a finite strip and evolved numerically.

Let us now recall the interior one-loop result~\cite{Tomasevic:2025clf}. Applying the $s$-wave one-loop formalism to the critical Roberts background, regularity of the semiclassical backreaction again fixes the choice of state, and the resulting four-dimensional quantum stress tensor is
\begin{alignat}{2}
\langle T^{(4)}_{uu}\rangle_{\rm int}
&=
\frac{
-uv+(u-v)^2\ln(\frac{u}{u-v})
}{
64\pi^2 u^3 (u-v)^3
},
&\qquad
\langle T^{(4)}_{vv}\rangle_{\rm int}
&=
\frac{
uv-2v^2-(u-v)^2\ln(\frac{u}{u-v})
}{
64\pi^2 u v^2 (u-v)^3
}
\label{eq:Roberts_Tuu_vv}
\\
\langle T^{(4)}_{uv}\rangle_{\rm int}
&=
-\frac{v}{64\pi^2 u^2 (u-v)^3},
&\qquad
\langle T^{(4)}_{\theta\theta}\rangle_{\rm int}
&=
\frac{\langle T^{(4)}_{\varphi\varphi}\rangle_{\rm int}}{\sin^2\theta}
=
0.
\label{eq:Roberts_Tuv_angular}
\end{alignat}
These expressions are regular at the incoming surface $v=0$, vanish asymptotically at $u\to-\infty$ and $v\to\infty$, and describe a Boulware-like vacuum polarization state of the collapsing scalar field.

In adapted coordinates, the stress tensor becomes
\begin{alignat}{2}
\langle T^{(4)}_{TT}\rangle_{\rm int}
&=
-
e^{2T}
\frac{
e^{-6x}(e^{2x}-1)^2
}{
16\pi^2
},
&\quad
\langle T^{(4)}_{xT}\rangle_{\rm int}
&=
e^{2T}
\frac{
e^{-4x}\left[2-5e^{2x}-e^{4x}(2x-3)\right]
}{
32\pi^2(e^{2x}-1)
}
\label{eq:Roberts_TTT_Tx}
\\
\langle T^{(4)}_{xx}\rangle_{\rm int}
&=
e^{2T}
\frac{
3-e^{-2x}+2e^{2x}(x-1)
}{
16\pi^2(e^{2x}-1)^2
},
&\quad
\langle T^{(4)}_{\theta\theta}\rangle_{\rm int}
&=
\frac{\langle T^{(4)}_{\varphi\varphi}\rangle_{\rm int}}{\sin^2\theta}
=
0.
\label{eq:Roberts_Txx_Tx}
\end{alignat}
Thus
\be
\langle T^{(4)}_{\mu\nu}\rangle_{\rm int}
=
e^{2T}F_{\mu\nu}(x),
\label{eq:Roberts_quantum_stress_scaling}
\ee
which is the $D=4$ instance of the general scaling $\langle T^{(D)}_{\mu\nu}\rangle\sim e^{(D-2)T}$ discussed around~\eqref{eq:quantum_growing_mode_general}. The quantum Lyapunov exponent is therefore
\be
\omega_q=2.
\ee
The components of $F_{\mu\nu}(x)$ are real analytic in the interior domain $x\in[0,\infty)$. Their growth with $T$ is again the signal that quantum effects become important as the self-similar singular regime is approached.

The corresponding one-loop backreaction can be solved analytically in the interior. We put back an overall $\ell^2$ and use the metric ansatz
\be
ds^2_{\rm int}
=
2\ell^2 e^{-2T}e^{2x}
\left[
e^{2\rho_1}(1-e^{-2x})dT^2
-2e^{2\rho_2}dTdx
\right]
+
 r^2 d\Omega_2^2,
\ee
with
\be
\rho_1
=
\bar{\rho}+\frac{N \hbar}{\ell^2}F_q(x)e^{2T},
\quad
\rho_2
=
\bar{\rho}+\frac{N \hbar}{\ell^2}W_q(x)e^{2T}, 
\quad 
 r
=
\bar{r}
\left[
1+\frac{N \hbar}{\ell^2}r_q(x)e^{2T}
\right],
\ee
\be
\bar{\rho}=0,
\qquad
\bar{r}=\ell e^{-T}e^x .
\ee
Solving the semiclassical Einstein equation with vanishing backreaction at $x=0$ (since the past is Minkowski) and regularity as $x\to\infty$ gives
\be
\label{eq:Roberts_quantum_backreaction1}
F_q(x)
=
-r_q(x)
=
\frac{
6-6e^{-2x}+\pi^2+12x^2
-12x\ln(1-e^{2x})
-6{\rm Li}_2(e^{2x})
}{
768\pi^2
},
\ee
\be
W_q(x)=0.
\label{eq:Roberts_quantum_backreaction2}
\ee
Although the individual terms contain functions with branch cuts for $x>0$, the imaginary parts cancel in the physical combination. The resulting semiclassical geometry is real analytic on the interior domain at finite $T$ and satisfies all components of the semiclassical Einstein equation to $O(\hbar)$.

The interior horizon tracing analysis then parallels the Garfinkle case but with a different scaling. To linear order in perturbations, including both the classical growing mode and the quantum backreaction, the horizon tracing function takes the form
\be
(\nabla r)^2_{\rm int}
\approx
\bar{h}(x)
+
(p-p^\ast)h_c(T,x)
+
\frac{N \hbar}{\ell^2}e^{2T}h_q(x),
\label{eq:Roberts_interior_horizon_function}
\ee
where the unperturbed Roberts background contribution is
\be
\bar{h}(x)=\frac12 e^{-2x}(1+e^{2x}).
\label{eq:Roberts_f0}
\ee
This background function never vanishes, so the exact critical Roberts background contains no apparent horizon. The classical perturbation spectrum around Roberts has been studied in detail in~\cite{Frolov:1997uu,Frolov:1998tq,Frolov:1999fv}; see also Section~5.3 of~\cite{Tomasevic:2025clf} for a review in the present conventions. The function $h_c(T,x)$ denotes the corresponding contribution of the dominant classical mode to the horizon function, which has a classical Lyapunov exponent $\omega_c\sim 1$. We will return to this perturbation sector in Section~\ref{sec4:Robertsperturbations}, where it provides part of the data propagated into the exterior strip. The quantum contribution is
\be
h_q(x)
=
\frac{
e^{-4x}\left(1-e^{2x}+2xe^{2x}\right)
}{
64\pi^2(e^{2x}-1)
}.
\label{eq:Roberts_fq}
\ee
The quantum term is localized near the small-$x$ region where the analytic Roberts approximation and the classical growing mode are both under control. Including the dominant classical perturbation together with the quantum term shifts the threshold and produces a finite mass at the earliest MOTS in the interior analysis. In this sense, the Roberts interior realizes the same qualitative mechanism as the Garfinkle interior. Vacuum polarization produces a quantum growing mode, competes with the classical departure from criticality, and replaces the classical zero-mass endpoint by a quantum-modified finite-mass threshold.

For the purposes of the exterior analysis, the important point is not the full global Roberts continuation, but the local data it supplies on a matching cut. We will use critical Roberts data on a finite null surface $v=v_0$ as the analytic seed for the exterior background. 

\subsection{Shell-free exterior strip and boundary data}
\label{sec4:exteriorstrip}

Before constructing the exterior, it is useful to explain why the four-dimensional problem cannot be treated the same way as the null continuation used in the Garfinkle case.  In $2+1$ dimensions, the exterior null continuation is effectively Vaidya-like. The exterior is organized by outgoing null layers, and the relevant matching problem can be solved analytically.  In the $3+1$ case, however, there is no regular pure scalar analog of an outgoing dust Vaidya exterior.

To see this, consider the general spherically symmetric double-null ansatz
\be
ds^2
=
-2e^{2\rho(u,v)}du\,dv
+
r^2(u,v)d\Omega_2^2,
\qquad
f=f(u,v).
\label{eq:Roberts_ext_double_null}
\ee
In what follows, subscripts $u$ and $v$ denote partial derivatives with respect to the corresponding double-null coordinates.  The Einstein-scalar equations are
\bea
r_{uu}-2\rho_u r_u+\frac12 r f_u^2 &=&0,
\label{eq:Roberts_ext_uu}
\\
r_{vv}-2\rho_v r_v+\frac12 r f_v^2 &=&0,
\label{eq:Roberts_ext_vv}
\\
r_{uv}+\frac{r_u r_v}{r}+\frac{e^{2\rho}}{2r} &=&0,
\label{eq:Roberts_ext_uv}
\\
\rho_{uv}+\frac{r_{uv}}{r}+\frac12 f_u f_v &=&0,
\label{eq:Roberts_ext_rhoeq}
\eea
together with the scalar wave equation
\be
f_{uv}
+
\frac{r_u f_v}{r}
+
\frac{r_v f_u}{r}
=
0.
\label{eq:Roberts_ext_scalar}
\ee
If one tries to impose a purely outgoing scalar exterior,
\be
f=f_+(u),
\qquad
f_v=0,
\ee
then \eqref{eq:Roberts_ext_scalar} gives
\be
\frac{r_v f_u}{r}=0.
\ee
For a nontrivial scalar pulse, $f_u\neq0$, this forces $r_v=0$.  But then \eqref{eq:Roberts_ext_uv} reduces to
\be
\frac{e^{2\rho}}{2r}=0,
\ee
which is impossible for a regular metric with finite areal radius. Thus a scalar exterior cannot be described by a one-function Vaidya family.  It must be a genuine two-variable solution $r(u,v),\rho(u,v),f(u,v)$ of the full Einstein-scalar system.  This is precisely why the analytic surgery used in~\cite{Wang:1996xh}, where the Roberts interior is cut and glued to an outgoing Vaidya dust exterior, cannot be adopted for the present problem.  Instead, we construct a finite shell-free exterior strip. It agrees with Roberts data near the matching cut, is smoothly deformed toward a weak-field outer boundary, and is evolved numerically using the full Einstein-scalar equations.

We now formulate the exterior as a characteristic boundary value problem on a finite double-null strip.  The computational domain is
\be
\mathcal D_{\rm ext}
=
\left\{
(u,v):
u_\ast\leq u\leq u_{\max}<0,
\quad
v_0\leq v\leq v_{\max}
\right\}.
\label{eq:Roberts_ext_strip}
\ee
The lower null boundary $v=v_0$ is the matching cut inherited from the Roberts interior.  On this cut we impose exact critical Roberts data,
\be
r(u,v_0)=r_{\rm R}(u,v_0)
=
\sqrt{u^2-uv_0},
\qquad
\rho(u,v_0)=\rho_{\rm R}(u,v_0
)=0,
\label{eq:Roberts_inner_rho_data}
\ee
\be
f(u,v_0)=f_{\rm R}(u,v_0)
=
\frac{\sqrt2}{2}
\ln\left(1-\frac{v_0}{u}\right).
\label{eq:Roberts_inner_f_data}
\ee
This gives complete analytic control at the matching surface.  In particular, all Taylor jets of the background fields, the classical Roberts growing mode, and the interior quantum state can be computed on the cut before any numerical evolution is performed.

The outer null boundary $u=u_\ast$ is chosen to replace the bad large-$v$ behavior of the global Roberts continuation by a weak-field exterior.  We choose numbers
\be
v_0<v_1<v_1+L<v_{\max},
\ee
and a smooth cutoff $\Xi(s)$ satisfying
\be
\Xi(s)=1\quad(s\leq0),
\qquad
\Xi(s)=0\quad(s\geq1),
\qquad
s\equiv \frac{v-v_1}{L}.
\label{eq:Roberts_cutoff}
\ee
On $u=u_\ast$ we prescribe
\bea
f(u_\ast,v)
&=&
\Xi(s)f_{\rm R}(u_\ast,v),
\label{eq:Roberts_outer_f}
\\
r(u_\ast,v)
&=&
\Xi(s)r_{\rm R}(u_\ast,v)
+
\left[1-\Xi(s)\right]r_{\rm flat}(u_\ast,v),
\label{eq:Roberts_outer_r}
\eea
where
\be
r_{\rm flat}(u,v)
=
-u+\frac{v}{2}.
\label{eq:Roberts_flat_radius}
\ee
This is the flat space areal radius in the convention $ds^2=-2du\,dv+r^2d\Omega_2^2$, since it gives
\be
r_u=-1,
\qquad
r_v=\frac12,
\qquad
r_{uv}=0,
\ee
and hence satisfies \eqref{eq:Roberts_ext_uv} with $\rho=0$ and $f=0$.  The outer boundary is therefore exactly Roberts for $v\leq v_1$, smoothly transitions for $v_1<v<v_1+L$, and becomes exactly Minkowski-like for $v\geq v_1+L$.

The remaining boundary datum $\rho(u_\ast,v)$ is determined from the $vv$ constraint.  Along the outer boundary,
\be
\rho_v(u_\ast,v)
=
\frac{
r_{vv}(u_\ast,v)
+
\frac12 r(u_\ast,v)f_v(u_\ast,v)^2
}{
2r_v(u_\ast,v)
}.
\label{eq:Roberts_outer_rho_constraint}
\ee
We fix the corner normalization
\be
\rho(u_\ast,v_0)=0,
\ee
and integrate \eqref{eq:Roberts_outer_rho_constraint} upward in $v$.  Since the boundary is exactly Roberts for $v\leq v_1$, one has $\rho=0$ in the initial Roberts wedge.  Furthermore, the far part of the boundary is exactly flat-like, with
\be
r_{vv}=0,
\qquad
f_v=0,
\qquad
r_v=\frac12,
\ee
one also has $\rho_v=0$ there.  Thus the outer boundary contains no artificial incoming scalar flux from the weak-field region.

The shell-free requirement is implemented at the matching cut $v=v_0$.  The surface $v=v_0$ is not a physical null shell, but the artificial boundary between the analytical interior seed and the exterior strip.  We therefore require the metric and scalar field data to match the Roberts solution there, and we do not allow discontinuities in the first derivatives that would generate a Barrab\`es-Israel null shell.  Equivalently, the $uu$ and $vv$ constraints should not develop distributional violations at the cut. 

For numerical evolution it is convenient to introduce the first-derivative variables
\be
p\equiv r_u,
\qquad
q\equiv r_v,
\qquad
a\equiv f_u,
\qquad
b\equiv f_v,
\qquad
\alpha\equiv \rho_u,
\qquad
\beta\equiv \rho_v.
\label{eq:Roberts_first_order_vars}
\ee
The definitions \eqref{eq:Roberts_first_order_vars} are used to reconstruct
$r,f,\rho$ from their derivatives, while the nontrivial mixed evolution equations are
\be
p_v=q_u=s_R,
\qquad
a_v=b_u=s_F,
\qquad
\alpha_v=\beta_u=s_\rho,
\label{eq:Roberts_mixed_first_order}
\ee
with
\be
s_R
=
-\frac{pq}{r}
-
\frac{e^{2\rho}}{2r},
\qquad
s_F
=
-\frac{pb+qa}{r},
\qquad
s_\rho
=
\frac{pq}{r^2}
+
\frac{e^{2\rho}}{2r^2}
-
\frac{ab}{2}.
\ee
The $uu$ and $vv$ equations are treated as null constraints,
\bea
\mathcal C_u
&\equiv&
p_u
-
2\alpha p
+
\frac12 r a^2
=
0,
\label{eq:Roberts_Cu}
\\
\mathcal C_v
&\equiv&
q_v
-
2\beta q
+
\frac12 r b^2
=
0.
\label{eq:Roberts_Cv}
\eea
They are imposed on the characteristic boundaries through the data construction and then monitored throughout the strip.

\subsection{Classical and quantum perturbations in the exterior}
\label{sec4:Robertsperturbations}

We now describe the perturbation sectors propagated on the shell-free exterior strip.  The background fields constructed in Section~\ref{sec4:exteriorstrip} will be denoted by $(\bar r,\bar\rho,\bar f)$. On top of this background we keep two independent linear sectors.  The first is the classical growing perturbation from near-critical Roberts spacetime.  The second is the quantum perturbation sourced by the one-loop vacuum polarization.  We write
\be
r
=
\bar r
+
\epsilon_{\rm c} r_{\rm c}
+
\epsilon_{\rm q} r_{\rm q},
\qquad
\rho
=
\bar\rho
+
\epsilon_{\rm c}\rho_{\rm c}
+
\epsilon_{\rm q}\rho_{\rm q},
\label{eq:Roberts_ext_pert_expansion_metric}
\ee
and
\be
f
=
\bar f
+
\epsilon_{\rm c}f_{\rm c}
+
\epsilon_{\rm q}\psi_{\rm q}.
\label{eq:Roberts_ext_pert_expansion_scalar}
\ee
Here $\psi_{\rm q}$ is introduced only as a bookkeeping variable for the induced scalar response in the quantum sector.  We keep the notation distinct as it is not an independent quantum scalar mode and carries no new physical amplitude.  The only genuinely new source at order $\epsilon_{\rm q}$ is the renormalized one-loop stress tensor.  Nevertheless, after the geometry is corrected by the one-loop source, the classical mean scalar field must still obey the classical wave equation on the corrected geometry:
\be
\Box_{\bar g+\epsilon_{\rm q} \delta g_{\rm q}}
\left(\bar f+\epsilon_{\rm q} \psi_{\rm q}\right)=0,
\ee
which is understood as the scalar sector consistency condition of the linearized system.  Equivalently, in the sourced semiclassical Einstein equations, the $O(\epsilon_{\rm q})$ variation of the classical stress tensor belongs to the linearized operator on the left-hand side, while the right-hand side contains only the renormalized quantum stress tensor. Schematically, the quantum perturbation satisfies
\be
\delta
\left(
G_{\mu\nu}
-
T_{\mu\nu}
\right)
\bigg|_{(\bar g,\bar f)}
=
\langle T_{\mu\nu}\rangle .
\label{eq:Roberts_quantum_sourced_linear_eq}
\ee
In the Roberts interior, the induced scalar response equation can be solved analytically on the quantum-corrected geometry, and the regular solution is precisely $\psi_{\rm q}=0$. For the present problem, we use this vanishing interior solution only as the matched jet on the cut $v=v_0$, and then solve for the induced response in the exterior subject also to the requirement that no incoming homogeneous scalar perturbation be injected from the weak-field outer boundary.   Although $\psi_q$ is not an independent mode, it must be included in a complete linear-order analysis because it enters the metric perturbation equations and can affect the corrected horizon function indirectly.\footnote{In the analytic Garfinkle exterior, this scalar response need not be introduced as a separate variable. This is because the full semiclassical Einstein equations are solved analytically, and stress tensor conservation together with the Bianchi identity makes the corrected scalar equation an integrability condition.  In the numerical Roberts exterior, we instead monitor the scalar residual and the null constraints explicitly.}

For a general linear perturbation $(\delta r,\delta\rho,\delta f)$, we introduce
\be
\delta p\equiv(\delta r)_u,
\quad
\delta q\equiv(\delta r)_v,
\quad
\delta a\equiv(\delta f)_u,
\quad
\delta b\equiv(\delta f)_v,
\quad
\delta\alpha\equiv(\delta\rho)_u,
\quad
\delta\beta\equiv(\delta\rho)_v .
\ee
As for the background, these definitions are used to reconstruct the perturbed fields from their first derivatives.  The nontrivial linearized mixed equations are
\be
(\delta p)_v=(\delta q)_u=\delta s_R,
\qquad
(\delta a)_v=(\delta b)_u=\delta s_F,
\qquad
(\delta\alpha)_v=(\delta\beta)_u=\delta s_\rho,
\label{eq:Roberts_ext_linearized_mixed_system}
\ee
where
\bea
\delta s_R
&=&
-\frac{\bar q\,\delta p+\bar p\,\delta q}{\bar r}
+
\left(
\frac{\bar p\bar q}{\bar r^2}
+
\frac{e^{2\bar\rho}}{2\bar r^2}
\right)\delta r
-
\frac{e^{2\bar\rho}}{\bar r}\,\delta\rho,
\label{eq:Roberts_delta_sR}
\\
\delta s_F
&=&
-\frac{
\bar b\,\delta p
+
\bar p\,\delta b
+
\bar a\,\delta q
+
\bar q\,\delta a
}{\bar r}
+
\frac{\bar p\bar b+\bar q\bar a}{\bar r^2}\,\delta r,
\label{eq:Roberts_delta_sF}
\\
\delta s_\rho
&=&
\frac{\bar q\,\delta p+\bar p\,\delta q}{\bar r^2}
-
\frac{2\bar p\bar q+e^{2\bar\rho}}{\bar r^3}\,\delta r
+
\frac{e^{2\bar\rho}}{\bar r^2}\,\delta\rho
-
\frac{\bar b\,\delta a+\bar a\,\delta b}{2}.
\label{eq:Roberts_delta_srho}
\eea
These are the homogeneous linearized Einstein-scalar equations on the exterior background. The corresponding linearized null constraints are
\bea
\delta\mathcal C_u
&\equiv&
(\delta p)_u
-
2\bar\alpha\,\delta p
-
2\bar p\,\delta\alpha
+
\frac12 \bar a^2\,\delta r
+
\bar r\bar a\,\delta a
=
0,
\label{eq:Roberts_linearized_Cu}
\\
\delta\mathcal C_v
&\equiv&
(\delta q)_v
-
2\bar\beta\,\delta q
-
2\bar q\,\delta\beta
+
\frac12 \bar b^2\,\delta r
+
\bar r\bar b\,\delta b
=
0.
\label{eq:Roberts_linearized_Cv}
\eea
As in the background evolution, these constraints are imposed through consistent characteristic data and monitored in the strip.

\subsubsection*{Classical perturbation sector}

The classical perturbation is obtained by solving the homogeneous system
\eqref{eq:Roberts_ext_linearized_mixed_system}--\eqref{eq:Roberts_delta_srho} with 
\be
(\delta r,\delta\rho,\delta f)
=
(r_{\rm c},\rho_{\rm c},f_{\rm c}).
\ee
Its boundary data on the matching cut $v=v_0$ are inherited from the dominant Roberts growing mode.  We refer to Section~5.3 of~\cite{Tomasevic:2025clf} for a detailed review of the Roberts perturbation analysis, and to~\cite{Frolov:1997uu,Frolov:1998tq,Frolov:1999fv} for the original treatment.

Let us briefly recall the ingredients needed here. The classical Roberts perturbations are expanded in modes labelled by the complex scaling exponent $k$, which controls the dependence of the mode on the self-similar scale. In the numerical construction we use the endpoint mode with 
\be 
k_0=1+i\sqrt2. 
\ee 
Since this eigenmode is complex, the physical classical perturbation is taken to be a real linear combination of the $k_0$ mode and its complex conjugate. Equivalently, the real part of the complex mode gives a representative seed, while the relative phase corresponds to the phase of the oscillatory perturbation.  This is the sense in which we use the real part of the mode as the classical growing perturbation on the cut.

Following the standard Roberts perturbation analysis, we write a general spherically symmetric metric perturbation as
\be
\delta g_{\mu\nu}dx^\mu dx^\nu
=
k_{uu}du^2
+
2k_{uv}du\,dv
+
k_{vv}dv^2
+
r_c^2K\,d\Omega_2^2,
\label{eq:Roberts_metric_pert_k}
\ee
and use the gauge-invariant scalar perturbation satisfying
\be
2u(u-v) f_{{\rm c},uv}
+
(2u-v) f_{{\rm c},v}
-
u f_{{\rm c},u}
-
2f_{\rm c}
=
0.
\label{eq:Roberts_classical_mode_scalar_eq}
\ee  
After constructing the gauge-invariant perturbation variables, one may choose a convenient gauge.  In the field gauge
\be
K=0,
\qquad
k_{vv}=0,
\ee
the scalar gauge-invariant variable coincides with the scalar perturbation used in \eqref{eq:Roberts_classical_mode_scalar_eq}, and the metric perturbations satisfy the simple relations
\be
k_{uv}=2f_{\rm c},
\qquad
k_{uu,v}=-\frac{4f_{\rm c}}{u}.
\label{eq:Roberts_Frolov_field_gauge_relations}
\ee
These relations are exact on the Roberts background.  Here they are used only to construct analytic cut data on $v=v_0$, where the background is exactly Roberts.  They are not imposed as global formulas in the deformed exterior strip.

The classical perturbation is then propagated into the shell-free exterior by solving the full linearized system \eqref{eq:Roberts_ext_linearized_mixed_system}--\eqref{eq:Roberts_delta_srho}.  On the outer boundary $u=u_\ast$, we impose no incoming classical perturbation from the weak-field region.  Operationally, the perturbation is kept Roberts-like in the exact Roberts wedge, smoothly tapered through the same transition interval used for the background, and set to zero in the far weak-field portion of the outer boundary.  This prevents the outer boundary from injecting an artificial classical growing mode into the strip.

It is useful to keep in mind the distinction between two uses of the classical mode.  Propagating only the scalar part $f_{\rm c}$ on the fixed exterior background is a useful diagnostic of the seed and boundary conditions.  However, the horizon condition is geometric, so the final perturbative data needed for horizon tracing are the induced metric perturbations $r_{\rm c}$ and $\rho_{\rm c}$ obtained from the full linearized Einstein-scalar system.

\subsubsection*{Quantum perturbation sector}

The quantum sector is sourced by the same vacuum polarization state selected in the interior Roberts analysis. We use the same localized anomaly-induced effective action introduced in Section~\ref{sec:critical collapse and nakedness}.  On the exterior strip, the reduced two-dimensional background is
\be
\bar g^{(2)}_{ab}dx^a dx^b
=
-2e^{2\bar\rho(u,v)}du\,dv,
\qquad
\bar\phi=-\ln\bar r .
\ee
The auxiliary fields obey
\be
\Box^{(2)}_{\bar g}\chi_1
=
\lambda_1 R^{(2)}[\bar g]
+
\lambda_2
\left(\nabla^{(2)}\bar\phi\right)^2,
\qquad
\Box^{(2)}_{\bar g}\chi_2
=
-\mu_1 R^{(2)}[\bar g]
-
\mu_2
\left(\nabla^{(2)}\bar\phi\right)^2 .
\label{eq:Roberts_ext_auxiliary_eqs}
\ee
Here $\Box^{(2)}_{\bar g}$, $R^{(2)}[\bar g]$, and $\left(\nabla^{(2)}\bar\phi\right)^2$ are two-dimensional quantities evaluated on the reduced exterior background. The homogeneous parts of $\chi_1$ and $\chi_2$ encode the quantum state.  On the matching cut $v=v_0$ they are fixed by the interior quantum state. On the outer boundary $u=u_\ast$ they are chosen so that the weak-field exterior does not inject spurious incoming quantum flux.  Once the auxiliary fields are solved, the renormalized two-dimensional stress tensor is obtained from the same one-loop functional as before,
\be
\langle T_{ab}^{(2)}\rangle
=
\langle T_{ab}^{(2)}\rangle
[
\chi_1,\chi_2;\bar g^{(2)},\bar\phi
],
\ee
and is then uplifted to the effective four-dimensional $s$-wave source $\langle T_{\mu\nu}\rangle$. Note that the one-loop stress tensor is evaluated on the exterior background. This is sufficient at linear order in the quantum amplitude, since varying $\langle T_{\mu\nu}^{(4)}\rangle$ with respect to the quantum-corrected geometry would give higher-order terms in the semiclassical expansion.  The metric response itself, however, must be computed with the full linearized Einstein-scalar operator, as in~\eqref{eq:Roberts_quantum_sourced_linear_eq}.

For the quantum sector we use the same linearized operator defined above, now evaluated on
\be
(\delta r,\delta\rho,\delta f)
=
(r_{\rm q},\rho_{\rm q},\psi_{\rm q}) .
\ee
The full linearized semiclassical system is obtained by adding the one-loop source only to the Einstein equations.  In terms of the linearized quantities defined in
\eqref{eq:Roberts_ext_linearized_mixed_system}--\eqref{eq:Roberts_linearized_Cv}, this means
\be
\delta s_R
\longrightarrow
\delta s_R+\mathcal J_R^{(q)},
\qquad
\delta s_\rho
\longrightarrow
\delta s_\rho+\mathcal J_\rho^{(q)},
\label{eq:Roberts_q_sourced_mixed}
\ee
while the scalar equation is unchanged $\delta s_F
\longrightarrow
\delta s_F$. Equivalently, the sourced null constraints are
\be
\delta\mathcal C_u
=
\mathcal J_u^{(q)},
\qquad
\delta\mathcal C_v
=
\mathcal J_v^{(q)} .
\label{eq:Roberts_q_sourced_constraints}
\ee
The source functions $\mathcal J_R^{(q)}$, $\mathcal J_\rho^{(q)}$, $\mathcal J_u^{(q)}$, and $\mathcal J_v^{(q)}$ are the projections of $\langle T_{\mu\nu}\rangle$ appearing in the corresponding double-null Einstein equations. With our conventions, the projections are
\begin{alignat}{2}
\mathcal J_R^{(q)}
&=
-\frac12\,\bar r\,\langle T_{uv}\rangle,
\qquad&
\mathcal J_\rho^{(q)}
&=
-\frac{e^{2\bar\rho}}{2\bar r^2}
\langle T_{\theta\theta}\rangle,
\\
\mathcal J_u^{(q)}
&=
-\frac12\,\bar r\,\langle T_{uu}\rangle,
\qquad&
\mathcal J_v^{(q)}
&=
-\frac12\,\bar r\,\langle T_{vv}\rangle .
\end{alignat}
The overall factor $\epsilon_{\rm q}$ is kept outside these definitions.  No source appears in the scalar equation because the one-loop correction enters through the renormalized stress tensor in the Einstein equations, not through a direct correction to $\Box f=0$.

The perturbative data propagated through the strip are therefore
\be
(r_{\rm c},\rho_{\rm c},f_{\rm c})
\qquad
\text{and}
\qquad
(r_{\rm q},\rho_{\rm q},\psi_{\rm q}),
\ee
with amplitudes $\epsilon_{\rm c}$ and $\epsilon_{\rm q}$. In the next subsection these data are combined into the geometric horizon function and used to search for exterior MOTS branches.

\subsection{Horizon tracing and quantum trapped branch}
\label{sec4:Roberts_horizontracing}

We now use the data constructed in
Sections~\ref{sec4:exteriorstrip} and~\ref{sec4:Robertsperturbations}
to trace MOTS in the Roberts
exterior.

\subsubsection*{Geometric horizon condition}

The MOTS condition is imposed directly on the invariant quantity 
\be
(\nabla r_{\rm})^2
=
-2e^{-2\rho_{\rm}}
(r_{\rm})_u
(r_{\rm})_v .
\label{eq:Roberts_full_horizon_condition}
\ee
The corrected exterior geometry is written as~\eqref{eq:Roberts_ext_pert_expansion_metric}.
To linear order,
\be
(\nabla r)^2
=
(\nabla\bar r)^2
+
\epsilon_{\rm c}
\delta_{\rm c}\!\left[(\nabla r)^2\right]
+
\epsilon_{\rm q}
\delta_{\rm q}\!\left[(\nabla r)^2\right],
\label{eq:Roberts_linear_horizon_expansion}
\ee
where
\be
(\nabla\bar r)^2
=
-2e^{-2\bar\rho}\bar p\,\bar q,
\qquad
\bar p=\bar r_u,
\qquad
\bar q=\bar r_v .
\ee
For either perturbative sector $i={\rm c},{\rm q}$,
\be
\delta_i\!\left[(\nabla r)^2\right]
=
-2e^{-2\bar\rho}
\left(
\bar q\,p_i
+
\bar p\,q_i
-
2\bar p\,\bar q\,\rho_i
\right),
\label{eq:Roberts_delta_horizon_general}
\ee
with $p_i=(r_i)_u$ and $q_i=(r_i)_v$.  The scalar perturbations $f_c$ and $\psi_q$ do not enter
\eqref{eq:Roberts_delta_horizon_general} directly.  They enter indirectly
through the linearized Einstein-scalar equations that determine
$r_i,\rho_i,p_i,q_i$.

A zero of \eqref{eq:Roberts_full_horizon_condition} is not automatically the
outgoing branch. For the full geometry, we explicitly check that the selected zero satisfies
\be
p<0,
\qquad
q\simeq0 .
\label{eq:Roberts_outer_MOTS_orientation}
\ee
Equivalently, for the expansions
\be
\theta_u
=
\frac{2p}{r}
<0,
\qquad
\theta_v
=
\frac{2q}{r}
\simeq0 .
\ee
The Hawking mass evaluated at the selected EMOTS is then given by~\eqref{eq:Hawking_mass_4d}.

\subsubsection*{Numerical workflow and controlled regime}

The evolution is implemented with a double-null characteristic
predictor-corrector scheme.  We first evolve the shell-free exterior
background $(\bar r,\bar\rho,\bar f)$ on a finite strip.  We then propagate
the classical and quantum perturbations on
this background. Finally,
we combine the two perturbative sectors in
\eqref{eq:Roberts_linear_horizon_expansion} and search for controlled sign
changes of $(\nabla r)^2$.

The production strip is
\be
u_\ast=-40,
\qquad
u_{\max}=-3,
\qquad
v_0=1,
\qquad
v_1=4,
\qquad
L=4,
\qquad
v_{\max}=15,
\label{eq:Roberts_production_strip_parameters}
\ee
resolved on a $320\times160$ grid.  The region
$v_0\leq v\leq v_1$ is kept as an exact Roberts wedge, the region
$v_1<v<v_1+L$ is a smooth transition, and the region
$v_1+L<v<v_{\max}$ is a weak-field exterior buffer.  These parameters define a fiducial finite exterior patch large enough
to contain the relevant EMOTS region while keeping the outer boundary
weak-field.  We do not push the evolution arbitrarily close to $u=0$, where
the local continuation becomes increasingly sensitive. 

Several background checks are performed before adding perturbations.  With an
exact Roberts boundary choice, the code reproduces the Roberts solution to the
expected grid accuracy.  In the shell-free weak-field completion, we monitor
the null constraints on the full strip and near the candidate EMOTS.  For the production runs, the full-strip residuals are of order
\be
\max|\mathcal C_u|=O(10^{-1}),
\qquad
\max|\mathcal C_v|=O(10^{-2}),
\ee
with smaller residuals in the local EMOTS region used below.  These full-strip
norms are used as diagnostics rather than sharp local acceptance tests,
because their maxima can be dominated by boundary regions or by places where
a denominator in a relative diagnostic is small.

The control conditions are imposed locally at each candidate crossing.  The invariant criteria stated in Section~\ref{sec2:horizontracing} are conditions on the full truncated geometry.  In the numerical Roberts exterior we implement practical diagnostics adapted to the double-null evolution.

For the weak curvature diagnostic we use the background Ricci scalar scale evaluated at the EMOTS,
$\epsilon_{\rm curv}
\equiv
|\bar{R}||_{\rm EMOTS}$.
This is a proxy for the invariant curvature control. The reason is that  we will simultaneously require the classical and quantum corrections to remain perturbative at the same point.  Under these local linearity conditions, the curvature invariants of the truncated geometry are expected to differ from the background curvature scale only perturbatively. Similarly, for the source-to-curvature control we use a component-wise diagnostic built from the stress tensor components that actually enter the evolution and constraint equations:
\be
\epsilon_{\rm src}
\equiv
\left.
\frac{
|\langle T_{uu} \rangle|
+
|\langle T_{uv}\rangle|
+
|\langle T_{vv}\rangle|
+
|\langle T_{\theta\theta}\rangle|/r^2
}{
|\bar{R}|
}
\right|_{\rm EMOTS}.
\label{eq:Roberts_source_curvature_diagnostic}
\ee
This is the numerical counterpart of the invariant stress tensor norm, which directly checks that the one-loop source terms driving the Roberts exterior equations remain small compared with the local background curvature scale near the selected EMOTS.  In the production runs we impose
\be
\epsilon_{\rm curv}<0.06,
\qquad
\epsilon_{\rm src}<0.2 .
\label{eq:Roberts_curv_src_cutoffs}
\ee
The most constraining local linearity diagnostics are the ratios involving the $v$-derivative of the areal radius,
\be
\eta_{\rm c}
\equiv
\left.
\left|
\frac{q_{\rm c}}{\bar q}
\right|
\right|_{\rm EMOTS},
\qquad
\eta_{\rm q}
\equiv
\left.
\left|
\frac{q_{\rm q}}{\bar q}
\right|
\right|_{\rm EMOTS}.
\label{eq:Roberts_control_eta_def}
\ee
These ratios are local perturbative-size diagnostics for the classical and quantum corrections in the $r_v$ sector.  We impose
\be
\eta_{\rm c}<1,
\qquad
\eta_{\rm q}<0.3 .
\label{eq:Roberts_eta_cutoffs}
\ee
The two cutoffs are not chosen symmetrically because the two perturbations
play different roles.  The classical perturbation is the tuning direction away
from the critical solution, and near the shifted threshold it is
allowed to produce an $O(1)$ deformation of the most sensitive derivative
component, provided the selected root still satisfies the other local
linearity, curvature, source, and outer orientation checks.  By contrast, the quantum
sector is a one-loop correction to the background geometry, so we
require it to remain parametrically smaller in the same component.  The
$v$-derivative ratio is also especially sensitive because $\bar q$ can become
small in parts of the exterior strip.  For this reason, global maxima of
relative quantities such as $|\delta q/\bar q|$ are treated only as warning
diagnostics.  The final
acceptance criterion uses the local values at the selected EMOTS.  The
remaining ratios $|r_i/\bar r|$ and $|p_i/\bar p|$, for
$i={\rm c},{\rm q}$, are also monitored and remain within the corresponding
production tolerances along the branch.

The induced scalar response is nonzero in the exterior bulk.  Near the selected EMOTS, however,
its derivative ratios remain moderate,
\be
\left|
\frac{a_{\rm q}}{\bar a}
\right|_{\rm EMOTS}
\sim
10^{-1},
\qquad
\left|
\frac{b_{\rm q}}{\bar b}
\right|_{\rm EMOTS}
\sim
10^{-2},
\ee
and the scalar residuals remain small compared with the geometric scales monitored in the
same region.

\subsubsection*{Quantum-shifted threshold and finite mass gap}

A scan over the amplitudes $(\epsilon_{\rm c},\epsilon_{\rm q})$ shows that
raw exterior roots of $(\nabla r)^2$ are not organized into a single
universal surface.  Several raw crossing families can appear in the finite
strip, and many fail one or more local control tests.  For each amplitude pair
we therefore select the earliest trapped surface that satisfies the local control conditions and the outer-trapped orientation
\eqref{eq:Roberts_outer_MOTS_orientation}. This is the quasi-local analog of the
earliest horizon onset in a collapse spacetime.

The EMOTS branch relevant for the quantum transition forms a
narrow threshold ribbon in the $(\epsilon_{\rm c},\epsilon_{\rm q})$ plane,
\be
\epsilon_{\rm low}(\epsilon_{\rm q})
\leq
\epsilon_{\rm c}
\leq
\epsilon_{\rm high}(\epsilon_{\rm q}) .
\ee
Its center is
\be
\epsilon_{\rm center}(\epsilon_{\rm q})
=
\frac{
\epsilon_{\rm low}(\epsilon_{\rm q})
+
\epsilon_{\rm high}(\epsilon_{\rm q})
}{2}.
\ee
The production scan follows this centerline as a one-parameter near-threshold
family.  For $3.0\leq\epsilon_{\rm q}\leq3.25$, the controlled ribbon has a
typical half-width of order $5\times10^{-5}$ in $\epsilon_{\rm c}$.  This
narrow band is the numerical realization of the quantum-shifted threshold.

To parametrize the approach to the shifted threshold, we follow the same logic
as in Section~\ref{sec3:horizontracing} and in our earlier interior
analyses~\cite{Tomasevic:2025clf,Tomasevic:2025kqy}, adapted to the numerical
exterior problem.  Along the EMOTS ribbon, we compare the classical
and quantum contributions to the horizon function locally, and use the resulting ratio as an operational coordinate on the branch.
After shifting this coordinate so that the quantum onset is located
at the origin, we denote it by $\mathcal R$:
\be
\mathcal R
=\left.
\frac{
|\epsilon_{\rm c} \delta_{\rm c}[(\nabla r)^2]|
}{
|\epsilon_{\rm q} \delta_{\rm q}[(\nabla r)^2]|
}
\right|_{\rm EMOTS},
\label{eq:Roberts_shifted_ratio}
\ee
Thus $\mathcal R\to0$ denotes the approach to the quantum-shifted threshold in
the sampled production window. In evaluating this ratio numerically, a small denominator floor is used only to avoid meaningless spikes when the quantum contribution to the horizon function is accidentally very small at isolated points.  This regulator has no role in the horizon tracing itself.  The EMOTS are selected directly from the zeroes of $(\nabla r)^2$, together with the local control criteria and the outer orientation check.  The regulator therefore affects only the plotting coordinate used to display the approach to the shifted threshold, not the existence, location, or mass of the selected EMOTS.

\begin{figure}[t]
\centering
\begin{subfigure}{0.49\textwidth}
\includegraphics[width=\linewidth]{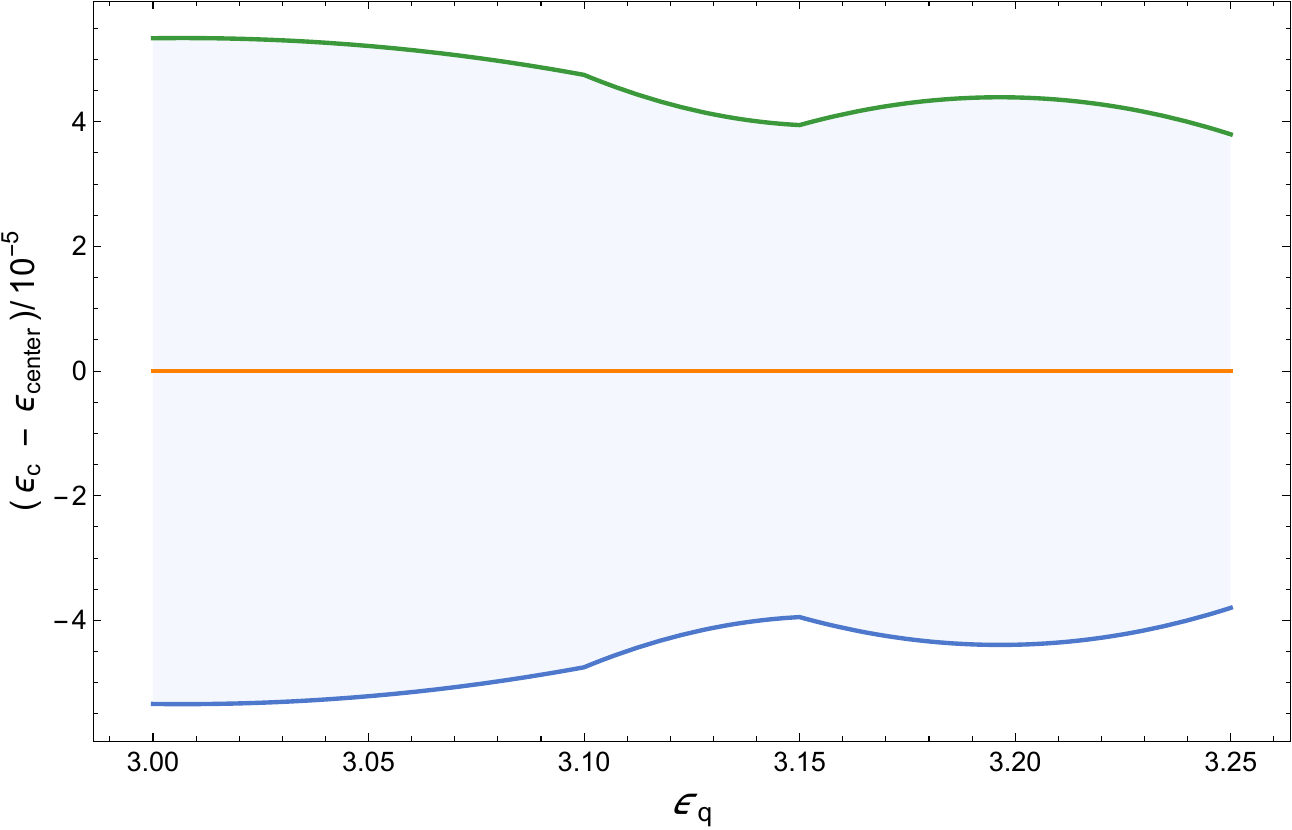}
\caption{}
\end{subfigure}
\hfill
\begin{subfigure}{0.49\textwidth}
\includegraphics[width=\linewidth]{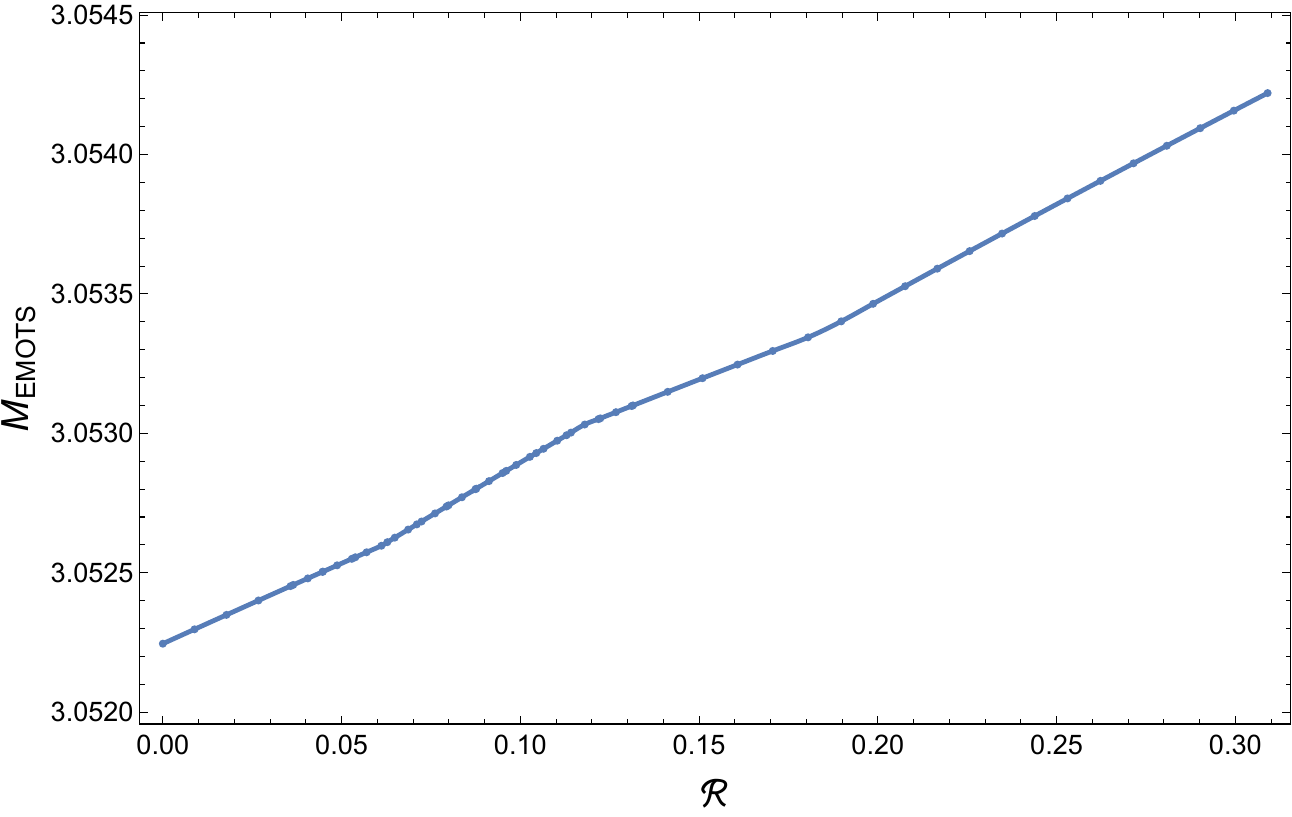}
\caption{}
\end{subfigure}
\caption{
Quantum-shifted threshold ribbon and finite exterior mass gap in the Roberts production run. The left plot shows the EMOTS ribbon in the $(\epsilon_{\rm c},\epsilon_{\rm q})$ plane. The vertical axis is the offset from the ribbon center in units of $10^{-5}$, illustrating that the accepted threshold family forms a narrow but resolved ribbon rather than an isolated point. The right plot shows the Hawking mass of the EMOTS along the ribbon centerline, plotted against the shifted ratio $\mathcal R$. The mass approaches a finite plateau, instead of scaling to zero as $\mathcal R\to0$. Only points satisfying the local curvature, source-to-curvature, constraint, scalar residual, and perturbative control cuts are included.}
\label{fig:Roberts_ribbon_massgap}
\end{figure}

\begin{figure}[hbt!]
    \centering
    \includegraphics[
        width=0.85\textwidth,
        height=0.50\textheight,
        keepaspectratio
    ]{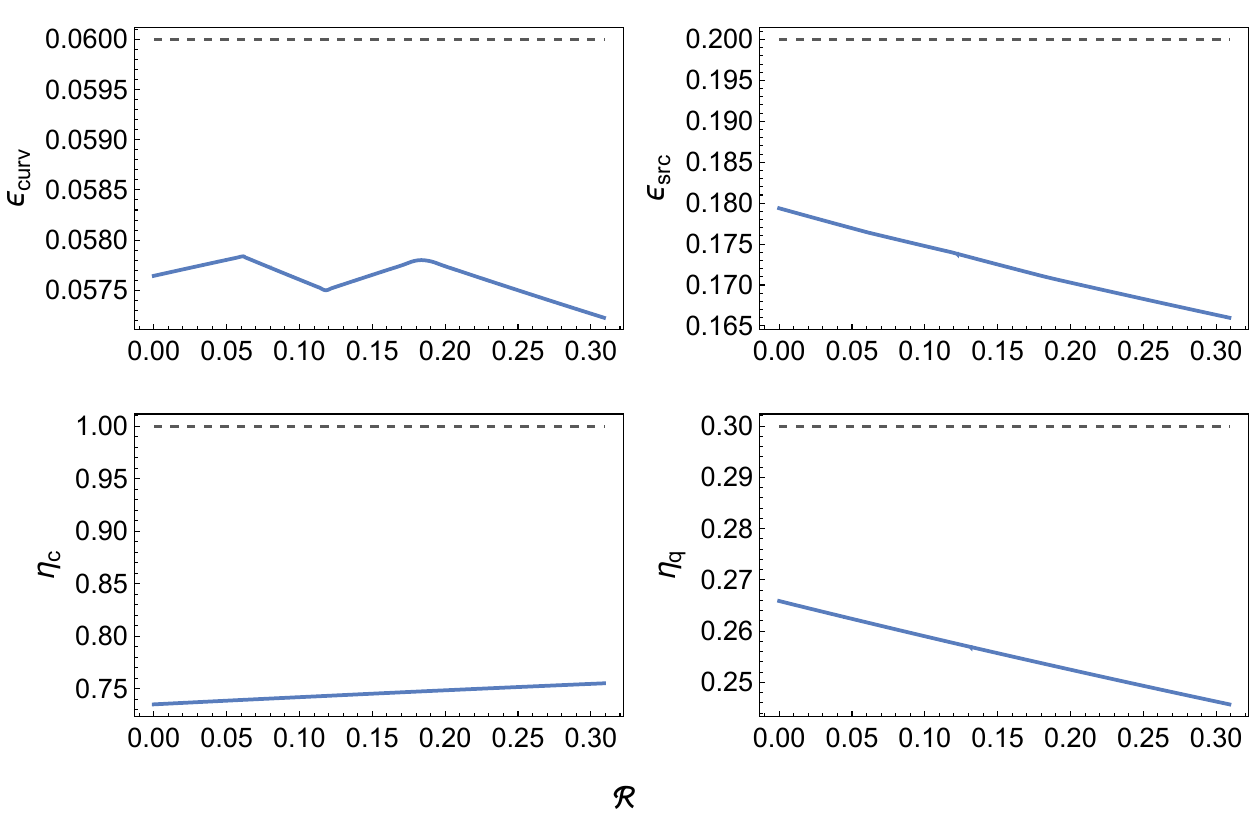}
    \caption{
    Local control diagnostics along the sampled threshold ribbon centerline in the Roberts exterior.  The curves show the curvature diagnostic $\epsilon_{\rm curv}$, the source-to-curvature diagnostic $\epsilon_{\rm src}$, and the dominant linearity diagnostics $\eta_{\rm c}$ and $\eta_{\rm q}$ evaluated along the candidate EMOTS branch.  Dashed horizontal lines indicate the corresponding production cutoffs.  All accepted points remain within the local semiclassical and perturbative control window used in the horizon tracing analysis.
    }
    \label{fig:Roberts_controlpanel}
\end{figure}

\begin{figure}[t]
    \centering
    \includegraphics[width=0.55\textwidth,
        height=0.31\textheight,
        keepaspectratio]{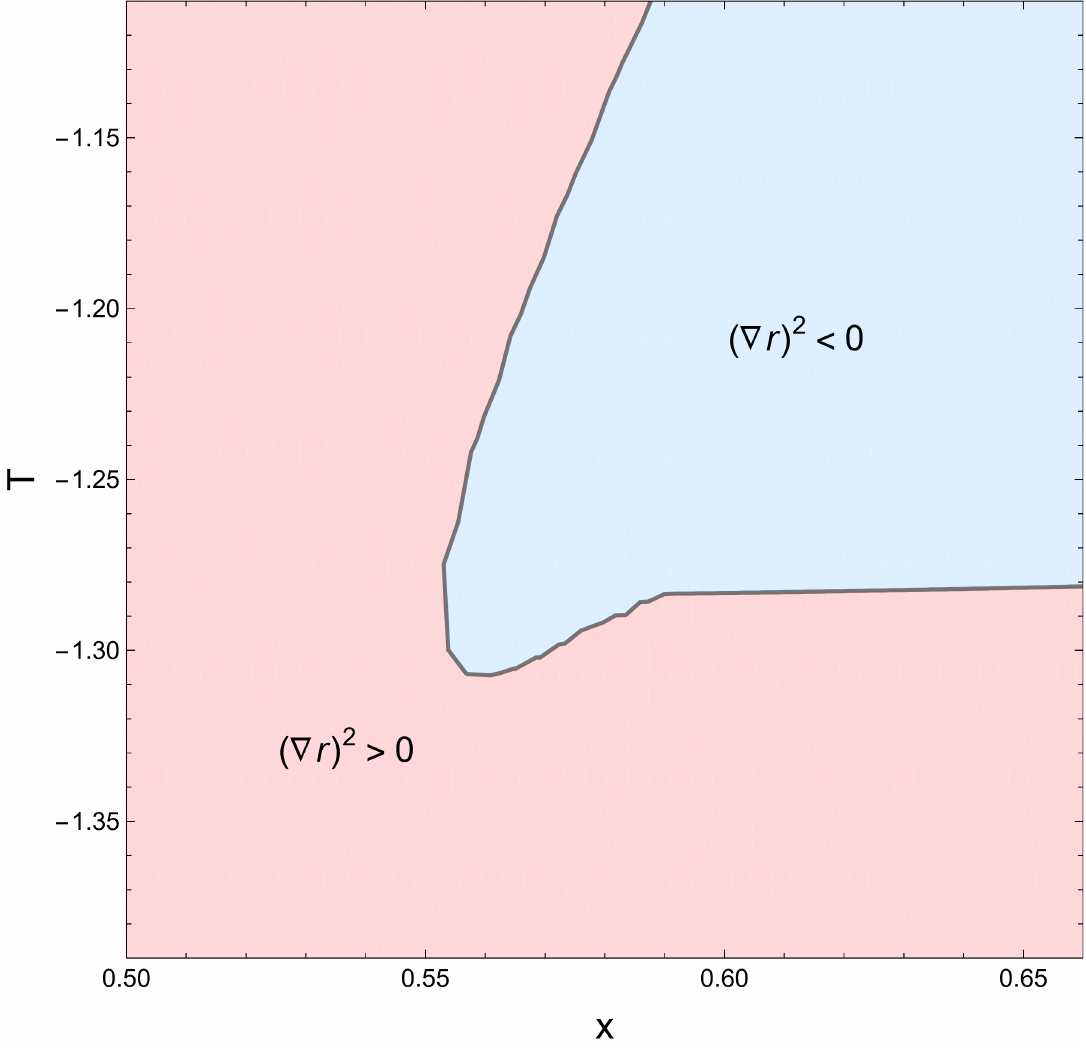}
    \caption{
    Representative sign plot of the horizon function in the $(x,T)$ plane.  The black curve is the zero contour of $(\nabla r)^2$ and gives the MOTS branch.  The figure illustrates that the accepted root is not merely an isolated zero of a one-dimensional scan, but the boundary of a trapped region in the finite exterior strip.
    }
    \label{fig:Roberts_signplot}
\end{figure}

The main numerical result is shown in Figure~\ref{fig:Roberts_ribbon_massgap}.
Along the centerline, the EMOTS mass does not scale to zero
as $\mathcal R\to0$.  Instead it approaches a finite plateau $M_{\rm EMOTS}
\simeq
3.05$, in the units used for the Roberts exterior strip, with the total variation only of order $10^{-3}$ across the displayed range of $\mathcal R$. The significance is not merely the limiting value at the endpoint, but the persistence of a nearly constant finite mass over an extended portion of the branch. Similarly, in Figure~\ref{fig:Roberts_controlpanel}, we show the
control quantities evaluated at all sampled EMOTS points along the threshold, where they remain within the admissible regime.\footnote{The selected EMOTS must lie in the exterior portion of the
finite strip, away from the matching cut where the exact Roberts wedge is
attached to the smoothed weak-field completion.  This is part of
the selection criteria, because roots too close to the cut can be sensitive to
the interpolation between the analytic Roberts data and the numerical exterior
completion.  Along the production centerline we find
\be
x_{\rm EMOTS}
\simeq
0.557,
\qquad
\Delta x_{\rm EMOTS}
\sim
10^{-4}.
\ee
The selected EMOTS is therefore well inside the exterior region, rather than in the near-cut region where matching artifacts
would be most likely to occur.} Figure~\ref{fig:Roberts_signplot} shows the corresponding sign
structure of $(\nabla r)^2$ in the $(x,T)$ plane, displaying the local boundary of the
trapped region.  Thus the exterior retains the same qualitative signature as the
interior picture, where the zero-mass Roberts endpoint is replaced by a
finite-mass horizon onset.

It is important to distinguish the robust features from details that depend on the exterior modeling. The strip is a finite completion of the Roberts self-similar
region, constructed within the $s$-wave and one-loop approximation, and the
horizons we trace are quasi-local.  These limitations affect numerical
coefficients, the precise location of the threshold ribbon, and the behavior
of raw roots in the strip.

Furthermore,  varying
the transition profile, the size of the strip, or the details of the
boundary smoothing changes the precise amplitude window and the coordinate
locations of MOTS roots.  In shifted completions, many
raw roots persist but fail one or more of the local control tests, most often
the weak curvature or source-to-curvature condition.  This sensitivity is
expected, since changing $v_1$ or $L$ changes the model rather than
merely changing gauge.  We therefore do not require the amplitude pair,
the coordinate location, or the raw root pattern to survive
unchanged under such variations.

These issues do not invalidate the central conclusion. The
robust features are the existence of a quantum-shifted exterior trapped
branch, the nonzero Hawking mass of the controlled EMOTS, and the
continuity with the interior quantum picture. We checked these in several
ways, including changes of grid resolution, exterior strip size, boundary
smoothing, threshold ribbon definition, and control cutoffs. The model-dependent details are the
exact numerical value of the mass plateau, the precise position and width of
the ribbon in the $(\epsilon_{\rm c},\epsilon_{\rm q})$ plane, and the detailed
shape of raw roots in alternative finite completions.

The exterior numerics are therefore consistent with the physical picture
developed in the interior analysis.  The quantum contribution behaves as a
growing deformation of the horizon function and competes with the classical
departure from criticality.  Instead of tuning to the old classical Roberts
threshold, the exterior EMOTS defines a shifted threshold ribbon.
Approaching this ribbon along its centerline, the EMOTS
does not shrink to zero mass.  It remains a finite-mass trapped
surface.

This is precisely the role of the Roberts model in the present analysis.  It
provides maximal analytic control over the self-similar background and the
one-loop source, while still allowing a numerical finite-patch test of
exterior EMOTS formation. This is a concrete step toward simulating quantum effects in
true critical collapse.

\section{Discussion and Outlook}
\label{sec:Discussion}

We have addressed the exterior part of the quantum critical collapse problem.  The interior analysis showed that quantum backreaction of the collapsing scalar field generates a universal quantum growing mode, shifts the critical threshold, and produces a finite mass gap~\cite{Tomasevic:2025kqy, Tomasevic:2025clf}. However, the global status of the Choptuik naked singularity cannot be decided from the interior region alone. The present work shows that
this effect is not confined to the self-similar interior. In both the analytic Garfinkle
continuation and the numerical Roberts exterior strip, the corrected geometry develops
a finite-mass quasi-local trapped branch.

This suggests a global quantum picture of critical collapse.  The Choptuik naked singularity, understood broadly as the zero-mass naked endpoint of Type II critical collapse, does not appear as a visible singularity in the semiclassical regime.  Instead, one-loop effects drive the exterior geometry toward the structure of an ordinary black hole, and the classical loss of predictability is converted into the familiar problem of black hole evaporation~\cite{Hawking:1975vcx,Hawking:1976ra}.  

This is a conceptual reversal relative to ordinary black hole formation.  There, classical evolution preserves predictability outside the horizon, while quantum evaporation introduces the problem of information loss.  In critical collapse, by contrast, the classical theory already threatens predictability through a visible singularity, while one-loop effects hide
that singularity and leave only the familiar evaporation problem.  The issue is not eliminated, but it is not enlarged into a new form of globally naked quantum evolution. 

This conclusion should not be confused with a proof of classical cosmic censorship. The classical critical spacetime still contains a naked singularity at the threshold of black hole formation. The role of quantum effects is not to restore the classical energy conditions or to remove the singularity by hand, but to modify the near-threshold and exterior horizon structure. This gives a concrete form of \emph{quantum censorship}: the relevant exterior Cauchy horizon region is cloaked by the backreaction of the same matter field that drives the collapse.

This also clarifies the relation to recent work on global visibility and evaporating black holes.  Classical obstructions to global visibility can be formulated using Raychaudhuri evolution and energy conditions along outgoing null generators \cite{Jay:2026huw}. Our mechanism does not rely on the classical null energy condition. The semiclassical stress tensor depends on the quantum state and can violate classical energy conditions. The common lesson is that local nakedness is not the same as global visibility, and that the exterior structure is decisive. Similarly, recent quantum singularity theorems show that standard evaporating black hole models remain singular, in the sense of future null geodesic incompleteness, when classical energy conditions and global hyperbolicity are replaced by generalized entropy assumptions such as the generalized second law and quantum trappedness \cite{Engelhardt:2026qqs}. In a related direction, the classical notion of a trapped surface must itself be refined in semiclassical gravity, since violations of the null convergence condition can invalidate the usual focusing, area, and censorship arguments. This motivates stronger notions such as sufficiently trapped surfaces \cite{Kontou:2026kej}. Our analysis addresses the complementary question of how a classically naked critical endpoint becomes cloaked in the first place. The subsequent loss of predictability is the standard one associated with black hole evaporation.

There are several important limitations to keep in mind.  The horizons we identify are quasi-local trapped surfaces, and our analysis remains perturbative, one-loop, and tied to controlled exterior models. Quantitative details such as the precise location of the trapped branch, the numerical value of the quantum-shifted threshold, and the smoothness class of the exterior continuation are therefore model dependent. The growth of the quantum source is not tied to these details and follows from its scaling in a self-similar geometry. The horizon-forming consequence is established by the explicit horizon-tracing calculation. The controlled models instead make this mechanism sharp.

It is important, however, that the introduction of a quantum scale and a growing quantum mode does not by itself guarantee Type-I-like behavior. The sign and spatial profile of the backreaction in the horizon equation are essential, see the detailed analysis in~\cite{Tomasevic:2025clf}. The mass gap found in the present scalar models is therefore a nontrivial result of the explicit backreaction and horizon-tracing calculation, rather than a consequence of dimensional analysis alone.

Finally, let us discuss several open problems suggested by this work.  A broader set of open questions related to quantum critical collapse was discussed in \cite{Tomasevic:2025kqy,Tomasevic:2025clf}.

\begin{itemize}
    \item \textbf{DSS critical collapse.}
    The present work uses CSS models because they provide analytic advantages.  In $2+1$ dimensions this is well motivated by the CSS nature of the numerical critical solution, while in $3+1$ dimensions the Roberts solution serves as a CSS proxy.  A complete treatment should extend the one-loop formalism to DSS critical collapse, either analytically or numerically, and test explicitly how the quantum growing mode and exterior trapped branch behave.

    \item \textbf{Large-$D$ expansion.}
The scaling of the quantum Lyapunov exponent, $\omega_q=D-2$, makes the large-$D$ limit a particularly interesting setting for semiclassical critical collapse. Recent progress has made both sides of the problem analytically accessible. CSS Einstein-scalar critical solutions, generalizing the Roberts spacetime, and their perturbations have been constructed in the large-$D$ expansion~\cite{Clark:2025tqi}, while exact DSS Einstein-scalar solutions have also been obtained recently~\cite{Ecker:2026akf, Ecker:2026joz}. The latter provide a particularly promising analytic setting for extending the present analysis beyond CSS. It would be particularly interesting to understand whether the rapid growth implied by $\omega_q=D-2$ translates into an enhanced semiclassical effect, or whether it is accompanied by a corresponding large-$D$ localization of the quantum response. More generally, the large-$D$ expansion may make the quantum stress tensor and its backreaction tractable while retaining the essential self-similar structure, providing a natural intermediate step before attempting the full four-dimensional DSS Choptuik problem.
    
    \item \textbf{Global exterior completion and late-time evolution.}
    Our analysis establishes quasi-local trapped branches in exterior spacetimes.  A natural next step is to construct a more complete quantum-corrected exterior development and understand how the first trapped surface connects to the later evaporating black hole region.  In a fully specified global spacetime, an event horizon can in principle be located numerically by evolving null surfaces backward from the asymptotic future, but such a construction requires the global future completion that is beyond the local exterior analysis performed here.  The more immediate problem is therefore to extend the construction far enough to connect the near-critical trapped branch to an evaporation regime.

    \item \textbf{From vacuum polarization to Hawking evaporation.}
    The quantum state used in the near-critical analysis is Boulware-like. It captures genuine self-energy of the collapsing scalar field and does not impose an outgoing Hawking flux by hand.  This is the appropriate state for the earliest pre-horizon regime and for locating the first marginally trapped surface.  Once a global future horizon has formed, however, an Unruh-like description with outgoing Hawking radiation at future null infinity may become the appropriate late-time effective description.  Understanding the transition between these two descriptions is essential.

    On the other hand, the present separation does not exclude the more subtle possibility of pre-Hawking radiation before a trapped region forms~\cite{Barcelo:2006uw, Barcelo:2010pj}. Whether such radiation is present in the causal collapse state~\cite{Chen:2017pkl, Unruh:2018jlu}, and whether it can compete with the vacuum polarization growing mode in determining the first marginally trapped surface, remains an important open question. See the discussion in~\cite{Tomasevic:2025clf} and the references therein.

    \item \textbf{Quantum focusing and generalized entropy.}
    The horizons traced in this paper are geometric trapped surfaces.  In a fully quantum setting, especially near an evaporating horizon, the more appropriate notions may involve quantum expansions, generalized entropy, and quantum trapped surfaces.  It would be valuable to reformulate the exterior critical collapse problem in this language and determine whether the quantum trapped branch found here satisfies an appropriate generalized entropy criterion.

    \item \textbf{Other matter models and spacetime dimensions.}
    The Einstein-scalar system is the canonical arena for critical collapse, but critical phenomena occur in many other matter systems and spacetime dimensions~\cite{Gundlach:2025yje}.  In our previous work we showed that the leading quantum growing mode is kinematical, arising from the scaling of the one-loop stress tensor in a self-similar background~\cite{Tomasevic:2025kqy, Tomasevic:2025clf}.  This suggests that the quantum growing mode may be universal in the self-similar domain. Whether it produces shielding, and whether its exterior manifestation is equally universal, remain to be understood.

    \item \textbf{Primordial black holes.}
    A finite quantum mass gap at the critical threshold can have direct implications for primordial black hole formation, especially for the low-mass tail of the mass spectrum.  The phenomenological consequences of the interior mass gap were discussed in \cite{Tomasevic:2025kqy}. A realistic treatment requires simulations of radiation-fluid collapse with quantum corrections

    \item \textbf{Quantum censorship beyond critical collapse.} It is natural to ask whether the mechanism found here is part of a broader quantum or stringy form of cosmic censorship. In addition to critical collapse, several classical or semiclassical evolutions appear to produce visible singular endpoints: the Gregory-Laflamme instability drives a black string toward a horizon pinch-off, and Hawking evaporation can push an ordinary black hole toward a regime where semiclassical evolution itself may threaten censorship~\cite{Gregory:1993vy,Horowitz:2001cz,Figueras:2022zkg,Hod:2019wcw}. These examples are physically different, and there is no reason to expect a single universal resolution.  Nevertheless, existing results suggest a common pattern:  naked singularities may not define a genuinely new endpoint, but may instead share the fate of black hole singularities. 

    In semiclassical three-dimensional examples, quantum backreaction can dress a naked singularity with an event horizon, and recent holographic constructions similarly suggest that quantum matter can build horizons acting as quantum censors~\cite{Casals:2016ioo,Frassino:2025buh}.  In the Gregory-Laflamme problem, the classical pinch-off may be cut off by string-scale physics and replaced by a string star~\cite{Horowitz:1996nw,Horowitz:1997jc,Ceplak:2023afb,Seitz:2025wpc,Emparan:2024mbp}. Although these mechanisms are different, they all suggest that classical violations of predictability may be converted into black hole, stringy, or evaporating quantum states rather than surviving as visible singular evolutions.  The critical collapse mechanism studied here gives a precise semiclassical realization of this idea.
\end{itemize}

\paragraph{Acknowledgements} The author is grateful to Gustavo J. Turiaci for useful discussions, to Marija Toma\v{s}evi\'c for collaboration on previous papers, and to Roberto Emparan for insightful feedback that motivated this work. This work was supported by the DOE Early Career Award DE-SC0026287.

\addcontentsline{toc}{section}{References}
\bibliographystyle{JHEP}
\bibliography{bibliography}

\end{document}